\documentclass[letterpaper,11pt]{article}
\pdfoutput=1

\usepackage{jheppub}

\usepackage{multirow}

\usepackage{subfig}
\usepackage{xspace}
\usepackage[countmax]{subfloat}

\newcommand{\Nsub}[2]{\tau_{#1}^{(#2)}}

\newcommand{\GeV}{\textrm{GeV}}

\DeclareRobustCommand{\Sec}[1]{Sec.~\ref{#1}}

\DeclareRobustCommand{\App}[1]{App.~\ref{#1}}
\DeclareRobustCommand{\Tab}[1]{Table~\ref{#1}}

\DeclareRobustCommand{\Fig}[1]{Fig.~\ref{#1}}
\DeclareRobustCommand{\Figs}[2]{Figs.~\ref{#1} and \ref{#2}}
\DeclareRobustCommand{\Eq}[1]{Eq.~(\ref{#1})}

\DeclareRobustCommand{\Ref}[1]{Ref.~\cite{#1}}
\DeclareRobustCommand{\Refs}[1]{Refs.~\cite{#1}}

\newcommand{\pythia}[1]{\textsc{Pythia\xspace #1}}

\newcommand{\fastjet}[1]{\textsc{FastJet\xspace #1}}

\preprint{MIT--CTP 4472}

\title{Unsafe but Calculable: \\ Ratios of Angularities in Perturbative QCD}

\author{Andrew J. Larkoski}
\author{and Jesse Thaler}

\affiliation{Center for Theoretical Physics, Massachusetts Institute of Technology, Cambridge, MA 02139, USA}

\emailAdd{larkoski@mit.edu}
\emailAdd{jthaler@mit.edu}

\abstract{Infrared- and collinear-safe (IRC-safe) observables have finite cross sections to each fixed-order in perturbative QCD.  Generically, ratios of IRC-safe observables are themselves not IRC safe and do not have a valid fixed-order expansion.  Nevertheless, in this paper we present an explicit method to calculate the cross section for a ratio observable in perturbative QCD with the help of resummation.  We take the IRC-safe jet angularities as an example and consider the ratio formed from two angularities with different angular exponents.  While the ratio observable is not IRC safe, it is ``Sudakov safe'', meaning that the perturbative Sudakov factor exponentially suppresses the singular region of phase space.   At leading logarithmic (LL) order, the distribution is finite but has a peculiar expansion in the square root of the strong coupling constant, a consequence of IRC unsafety.   The accuracy of the LL distribution can be further improved with higher-order resummation and fixed-order matching.  Non-perturbative effects can sometimes give rise to order one changes in the distribution, but at sufficiently high energies $Q$, Sudakov safety leads to non-perturbative corrections that scale like a (fractional) power of $1/Q$, as is familiar for IRC-safe observables.  We demonstrate that Monte Carlo parton showers give reliable predictions for the ratio observable, and we discuss the prospects for computing other ratio observables using our method.}

\begin{document} 
\maketitle

\section{Introduction}

Observables are fundamental quantities that probe the structure of quantum field theories.  Ever since it was realized that quantum chromodynamics (QCD) was a perturbative gauge theory at high energies \cite{Gross:1973id,Politzer:1973fx}, numerous observables have been designed to verify and study the jetty nature of QCD \cite{Farhi:1977sg,Georgi:1977sf,Basham:1977iq,Basham:1978bw,Basham:1978zq,Parisi:1978eg,Donoghue:1979vi,Rakow:1981qn,Ellis:1986ig,Berger:2003iw,Almeida:2008yp,Stewart:2010tn}, to define jet algorithms \cite{Sterman:1977wj,Bethke:1988zc,Catani:1993hr,Ellis:1993tq,Dokshitzer:1997in,Wobisch:1998wt,Wobisch:2000dk,Cacciari:2008gp}, and to reveal the substructure of jets themselves \cite{Abdesselam:2010pt,Altheimer:2012mn}.  To date, infrared and collinear (IRC) safety has been a key guiding principle for constructing observables that can be analyzed in perturbative QCD (and other perturbative field theories).  IRC safety ensures that virtual and real diagrams will be consistently combined according to the KLN theorem \cite{Kinoshita:1962ur,Lee:1964is} to yield physical cross sections order-by-order in perturbation theory.

The standard definition of IRC safety is simply a rule-of-thumb for determining whether the cross section for an observable will be finite at any fixed order \cite{Ellis:1991qj}:\footnote{Issues with this definition of IRC safety as a precise mathematical statement has been discussed in the literature \cite{Banfi:2004yd}.}
\begin{quote}
An observable is IRC safe if it is insensitive to infinitesimally soft emissions or exactly collinear splittings.
\end{quote}
This requirement guarantees that divergences associated with real emission of soft or collinear particles will be canceled exactly by infrared (IR) divergences in virtual diagrams.  Here, we will focus on the jet angularities $e_\alpha$ which are IRC safe for $\alpha>0$ \cite{Berger:2003iw,Almeida:2008yp,Ellis:2010rwa}:\footnote{This definition differs from \Ref{Berger:2003iw} where the angularities are defined in a hemisphere of the event, and from \Ref{Almeida:2008yp} where the jet angularities are normalized with respect to the jet mass.  Our definition of the  jet angularities is closer to that of \Ref{Ellis:2010rwa}, except we choose a different angular behavior (which agrees in the $\theta \to 0$ limit) to simplify the resulting cross section formulae.  This definition of jet angularites is appropriate for $e^+e^-$ collisions; for a hadron collider, energy and angle would be replaced by $p_T$ and $R$, respectively.}
\begin{equation}
\label{eq:ang_def}
e_\alpha = \frac{1}{E_J}\sum_{i\in J}E_i \theta_i^\alpha \ ,
\end{equation}
where $E_J$ is the total jet energy, the sum runs over all particles $i$ in the jet with energy $E_i$, and the angle $\theta_i$ is measured with respect to an appropriately chosen jet axis.  Angularities belong to a broad class of IRC-safe observables which are linear in the energy of each particle, symmetric under particle exchange, and weighted by positive powers of angles between particles; this class includes thrust \cite{Farhi:1977sg} and jet mass.  While IRC safety is certainly a sufficient condition for an observable to be finite order-by-order in the strong coupling constant $\alpha_s$, it is not yet established whether more general observables might still be tractable using alternative approaches to perturbative QCD.

In this paper, we present a case study of a non-IRC-safe observable whose cross section can nevertheless be calculated using perturbative QCD with the help of resummation.  We consider a ratio observable formed from two different angularities measured on the same jet:
\begin{equation}
\label{eq:ratio_def}
r_{\alpha,\beta} \equiv \frac{e_\alpha}{e_\beta},
\end{equation}
where $r_{\alpha,\beta} \in [0,1]$ for the choice of angular exponents $\alpha > \beta$ and jet radius $R_0 = 1$.  We will sometimes drop the subscripts $r_{\alpha, \beta} \to r$ for readability.  While $r_{\alpha,\beta}$ is not IRC safe, it belongs to a category of observables we call ``Sudakov safe'', where the singular region of phase space is exponentially suppressed due to Sudakov factors.  We conclude from this example that the set of observables computable in a perturbative quantum field theory is larger than just the set of IRC-safe observables.\footnote{Of course, there are many other observables which can be calculated with the help of non-perturbative objects such as parton distribution functions and fragmentation functions.  Here, we are referring to observables that are finite and calculable even without considering non-perturbative effects.}

More broadly considered, ratio observables are an interesting class of observables which have been used, for example, to reduce experimental uncertainties in the
measurement of the strong coupling constant \cite{Abazov:2012lua,ATLAS:2013lla,Chatrchyan:2011wn,Chatrchyan:2013txa} and probe the substructure of jets \cite{Thaler:2010tr,Thaler:2011gf,Larkoski:2013eya,Almeida:2008yp,Field:2012rw,Jankowiak:2011qa}. It is therefore imperative to know whether the distributions for ratio observables can be predicted from first principles.  As noted in \Ref{Soyez:2012hv}, however, observables formed from the ratio of two IRC-safe observables are generically themselves not IRC safe.  The logic for the case of the angularity ratio $r_{\alpha,\beta}$ is as follows.  Both $e_\alpha$ and $e_\beta$ go to zero in the region of phase space where radiation is soft and/or collinear with respect to the jet axis, but the ratio $r_{\alpha,\beta}$ in this region can be arbitrary.  This implies that the virtual contribution to the ratio observable would have to be divergent for all values of $r_{\alpha,\beta}$ in order to cancel divergences from the real emission diagrams, but this is impossible.  Of course, one can deform $r_{\alpha,\beta}$ to make it IRC safe by, say, applying a cut on the denominator to avoid the singular region $e_\beta \to 0$.  But naively, this lack of IRC safety prevents generic ratio observables from being computed in perturbative QCD without some kind of non-perturbative input.   

\begin{figure}
\begin{center}
\includegraphics[width=7cm]{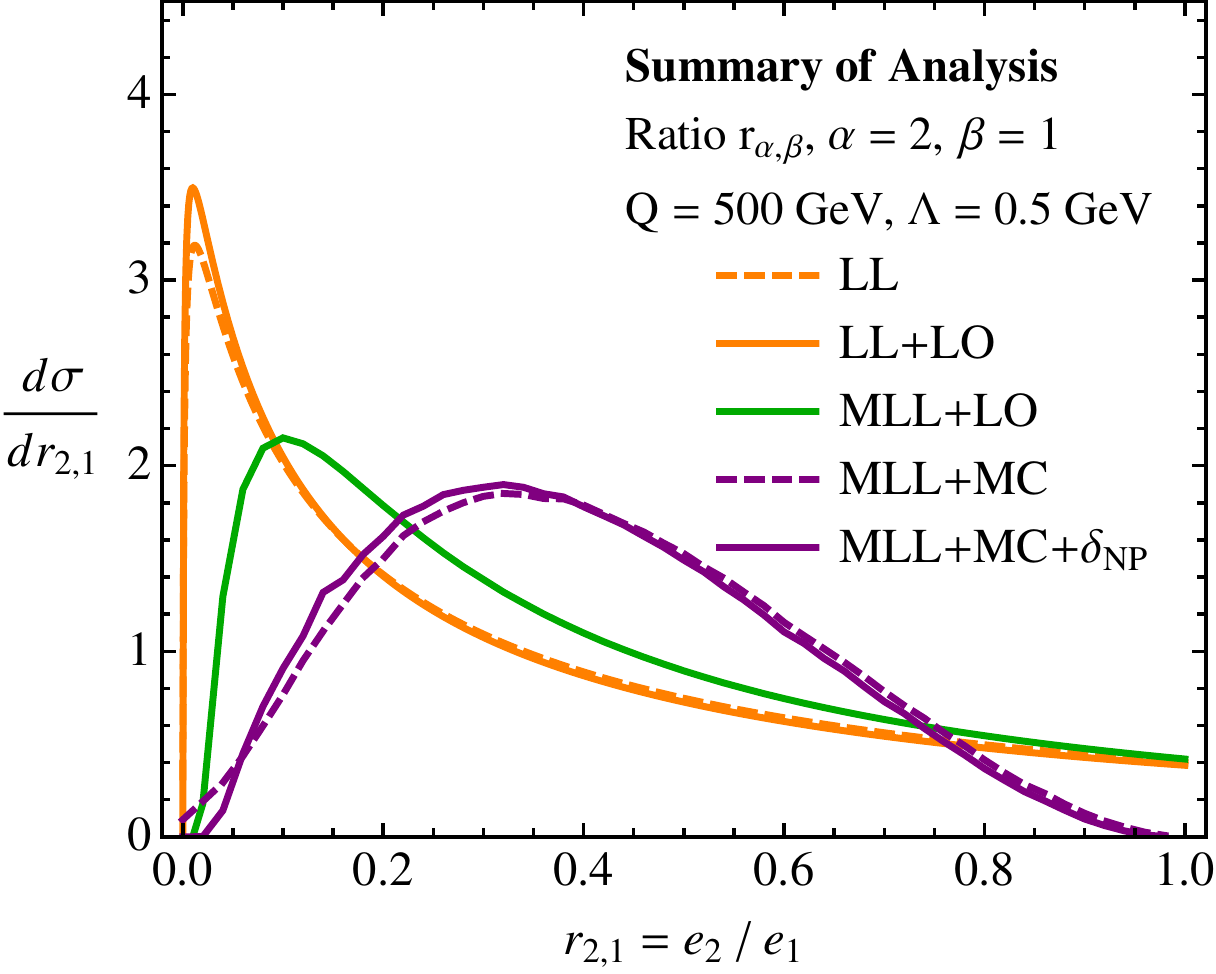}
\end{center}
\caption{Summary of results from this paper.  Plotted is the differential cross section for the angularity ratio $r_{2,1} \equiv e_2 / e_1$.  The curves appear in increasing order of theoretical accuracy, with each subsequent curve including all previous effects.  ``LL'' is the baseline leading logarithmic result (with the scaling $\alpha_s \log^2 r \simeq 1$) that demonstrates Sudakov safety.  ``LL+LO''  includes $\mathcal{O}(\alpha_s)$ fixed-order matching in the Log-R scheme.  ``MLL+LO'' includes modified leading log resummation which has running $\alpha_s$.  ``MLL+MC'' is Monte Carlo resummation which includes multiple emissions.  Finally, ``MLL+MC+$\delta_{\rm NP}$'' includes an estimate of non-perturbative corrections through a simple shape function.  We note that most parton shower Monte Carlo programs include all of these effects.  
}
\label{fig:summary_plot}
\end{figure}

The key insight of this paper is that while the angularity ratio $r_{\alpha,\beta}$ is not IRC safe and therefore not defined order-by-order in perturbation theory, it is well-defined in resummed perturbative QCD when one accounts for logarithmic effects to all orders in $\alpha_s$.  The logic and outline of this paper is as follows, with the results summarized in \Fig{fig:summary_plot}.
\begin{itemize}
\item {\bf IRC Unsafety at Fixed Order}.  In \Sec{sec:rat_ex}, we attempt to compute the differential cross section for $r_{\alpha,\beta}$ in fixed-order QCD.  As expected, we find that the fixed-order cross section is IR divergent because the ratio is sensitive to the singular region of phase space for all values of $r_{\alpha,\beta}$.

\item {\bf Resummed Cross Section of the Ratio}. In \Sec{sec:LL_double}, we compute the leading-logarithmic (LL) resummed differential cross section of the ratio $r_{\alpha,\beta}$ by marginalizing the resummed double differential cross section of angularities:
\begin{equation}
\label{eq:ratio_dist_def}
\frac{d \sigma}{d r_{\alpha,\beta}} \equiv \int d e_\alpha \, d e_\beta \,  \frac{d^2 \sigma}{d e_\alpha d e_\beta } \, \delta\left( r_{\alpha,\beta} -\frac{e_\alpha}{e_\beta} \right).
\end{equation}
The Sudakov factor in the double differential distribution provides exponential suppression of the singular region of phase space, resulting in a finite cross section for the ratio.  We call this feature ``Sudakov safety''.  We expect that this feature will generalize to a wide class of ratio observables.

\item {\bf Incorporating Higher-Order Corrections}.  Because the cross section for the ratio $r_{\alpha,\beta}$ is defined from the double differential cross section of angularities $e_\alpha$ and $e_\beta$, the accuracy of the cross section can be systematically improved by matching to fixed-order or by resumming to higher logarithmic order.  We discuss applications of these corrections in \Sec{sec:highord}.

\item {\bf Monte Carlo Resummation of Ratios}.  It is well-known that Monte Carlo parton showers formally resum IRC-safe observables to LL order.  In \Sec{sec:MC_resum}, we provide strong evidence that Monte Carlos also resum the ratio observable correctly to LL order.  In addition, the parton shower includes important multiple emission effects which arise at next-to-leading logarithmic (NLL) order.

\item {\bf Evidence for Small Non-Perturbative Corrections}.  Although the ratio of angularities is not IRC safe, we show in \Sec{sec:nonpert} that the non-perturbative corrections to the perturbative cross section are small in the high energy limit.  For small values of the energy $Q$ the non-perturbative effects can be order one, but for $Q \gtrsim 100\text{--}1000~\GeV$ the power corrections scale like a (fractional) power of $1/Q$.  This is a consequence of  Sudakov safety, since the perturbative Sudakov factor exponentially suppresses the non-perturbative region of phase space at sufficiently high energies.
\end{itemize}
For this final point, we must assume the existence of a shape function for the double differential cross section of angularities.  A proof of this assumption lies beyond the scope of this work.  We conclude in \Sec{sec:conc} where we discuss the calculation of other phenomenologically well-motivated ratio observables formed from, e.g., $N$-subjettiness \cite{Stewart:2010tn,Thaler:2010tr,Thaler:2011gf} and energy correlation functions \cite{Banfi:2004yd,Larkoski:2013eya}.
 
\section{Ratio of Angularities and IRC Unsafety}\label{sec:rat_ex}

We begin this section with a brief discussion of the appropriate axis to use for defining the jet angularities.  We then describe the phase space for two different angularities $e_\alpha$ and $e_\beta$ in order to make some general statements about the region of support relevant for the ratio.  We compute the fixed-order double differential cross section $d^2 \sigma/ d e_\alpha d e_\beta$ and attempt to use \Eq{eq:ratio_dist_def} to define the ratio distribution $d \sigma/ d r_{\alpha,\beta}$.  Because the ratio observable is IRC unsafe, the differential cross section for $r_{\alpha,\beta}$ is not defined at fixed-order, precisely because it is sensitive to the singular region of phase space for all values of $r_{\alpha,\beta}$.  This will set the stage for the resummed calculations in the remainder of the paper.

Throughout this paper, we will focus on jets initiated by an energetic quark.  At lowest non-trivial order, the jet angularities $e_\alpha$ can be found by accounting for a single gluon emission from that quark.  The case of gluon jets would be similar, except there is an additional complication arising  because the splitting $g \to gg$ has two soft singularities, whereas $q \to q g$ has only one soft singularity.

\subsection{Angularities with Respect to the Broadening Axis}
\label{sec:broadening}

\begin{figure}
\begin{center}
\includegraphics[width=6cm]{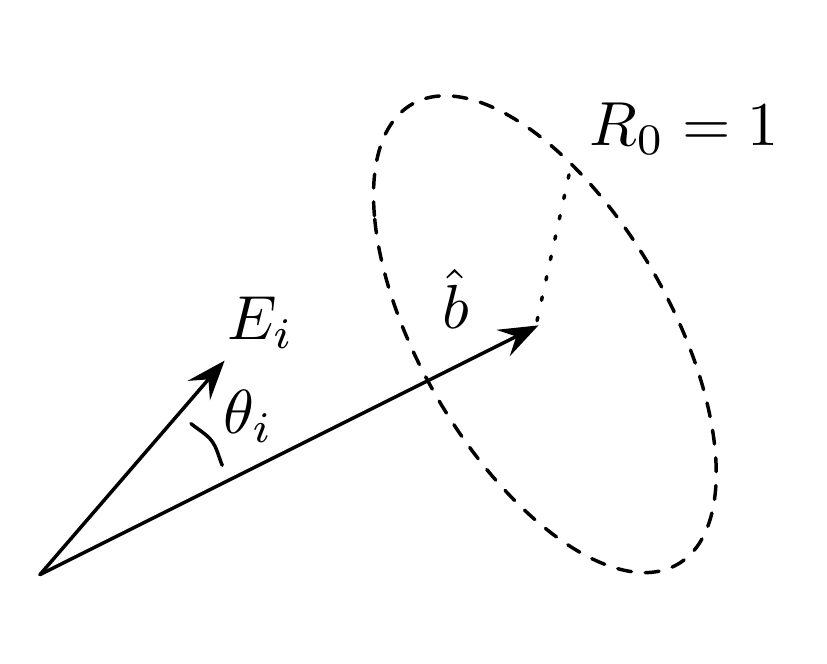}
\end{center}
\caption{Jets and angles defined with respect to the broadening axis $\hat{b}$ defined in \Eq{eq:broadening_axis}.  For two particles, the broadening axis coincides with the direction of the most energetic particle.  Throughout this paper, we focus on jets initiated by an energetic quark, and take the jet radius to be $R_0 = 1$.}
\label{fig:simplejet}
\end{figure}

In order to define the jet angularities $e_\alpha$ in \Eq{eq:ang_def}, we have to choose the appropriate axis from which to measure angles.  While the natural choice would seem to be the jet momentum axis, this choice suffers from the effect of recoil for small values of $\alpha$ \cite{Larkoski:2013eya,broadening}.  That is, instead of measuring the soft radiation pattern about the hard central jet core, recoil-sensitive observables are dominated simply by the displacement of the jet axis caused by soft radiation.

To avoid recoil effects, we instead measure angles with respect to the broadening axis of the jet shown in \Fig{fig:simplejet}.  The broadening axis is defined as the axis which minimizes the $\beta=1$ measure of $N$-subjettiness \cite{Thaler:2010tr,Thaler:2011gf} (which minimizes broadening \cite{Rakow:1981qn,Ellis:1986ig} with respect to that axis).  This corresponds to finding the axis $\hat{b}$ which minimizes the scalar sum of the momentum transverse to that axis:
\begin{equation}
\label{eq:broadening_axis}
\min_{\hat{b}} \left[ \sum_{i \in J} E_i \theta_{i \hat{b}}   \right] \ ,
\end{equation}
where the sum runs over all particles in the jet and $\theta_{i \hat{b}}$ is the angle from the axis $\hat{b}$ to particle $i$.  For a jet with two constituents, the broadening axis coincides with the direction of the most energetic particle.  More generally, the broadening axis corresponds quite closely to the direction of the hard central jet core as desired.

For consistency, we also define the jet region in \Eq{eq:ang_def} as all particles within a radius $R_0$ of the broadening axis (and not the momentum axis).  For simplicity, we take the jet radius to be $R_0 = 1$ such that the angles of particles within the jet obey $\theta_i \in [0,1]$.  When considering the ratio of angularities $r_{\alpha,\beta}$ in \Eq{eq:ratio_def}, we always take $\alpha > \beta$ which then implies $e_\alpha < e_\beta$ and $r_{\alpha, \beta}\in [0,1]$ for every jet configuration.

\subsection{Allowed Phase Space}
\label{sec:ps}

\begin{figure}
\centering
    \includegraphics[width=7
   cm]{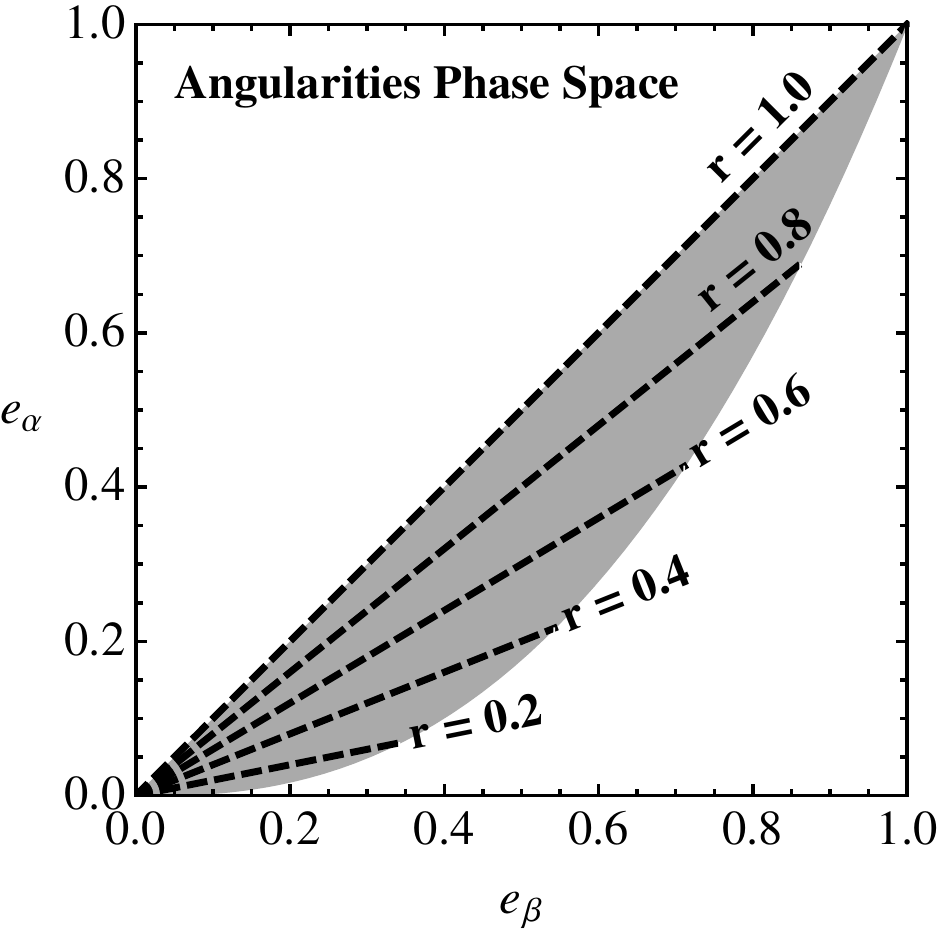}
\caption{Allowed phase space region for the double differential cross section in the $(e_\alpha,e_\beta)$ plane.  The allowed phase space is gray, lines of constant ratio $r_{\alpha,\beta} \equiv e_\alpha/e_\beta$ are illustrated by dashed lines, and the forbidden regions are white.  The boundary of the lower forbidden region has zero slope only at $e_\beta=0$, such that $r_{\alpha,\beta} = 0$ is only possible at the origin.
}\label{region2}
\end{figure}

In order to find the double differential cross section $d^2 \sigma/ d e_\alpha d e_\beta$, it is helpful to first determine the region in the $(e_\alpha,e_\beta)$ plane over which it has support.  The allowed phase space region is summarized in \Fig{region2}, where dashed lines emanating from the origin correspond to constant values of the ratio $r_{\alpha,\beta} \equiv e_\alpha/e_\beta$.  As already mentioned, if $\alpha> \beta$, then $e_\alpha < e_\beta$ on the physical phase space, such that $e_\beta \to 0$ implies $e_\alpha \to 0$.  Less obvious is that $e_\alpha\to0$ implies $e_\beta \to 0$, which can be understood because $e_\alpha$ and $e_\beta$ are first non-zero at the same order in perturbation theory.  So in addition to the upper boundary on $e_\alpha$, there must be a lower boundary on $e_\alpha$ (for fixed value of $e_\beta$).  Because the angularities are independent observables for $\alpha \neq \beta$, there is no fixed constant $k$ such that $e_\alpha = k e_\beta$; this implies that that the ratio $r_{\alpha,\beta} \to 0$ can only be achieved as both $e_\alpha$ and $e_\beta$ approach zero.  In fact, by the above arguments, the only point in phase space with $r_{\alpha,\beta} = 0$ is $e_\alpha = e_\beta = 0$.\footnote{\label{footnote:zero_slope}This also implies that the lower boundary on $e_\alpha$ has zero slope on the $(e_\alpha, e_\beta)$ plane at $e_\beta = 0$.}  Note that all values $r_{\alpha,\beta} \in [0,1]$ are achievable somewhere in phase space.

For a jet with two constituents arising from the splitting $q \to q g$, these properties can be made concrete.  The variables that describe the phase space are the energy fraction $z$ taken by the gluon and the splitting angle $\theta$ between the two particles.  Conservation of energy requires $0\leq z \leq 1$, and $0\leq \theta \leq 1$ is required for both particles to be in a jet of radius $R_0 = 1$.  Thus, the matrix element necessarily contains the phase space restrictions
\begin{equation}
\Theta(1-z)\Theta(1-\theta) \ ,
\end{equation}
where we have implicitly assumed that $z,\theta > 0$.  In these phase space coordinates, the recoil-free angularity $e_\alpha$ in the soft emission limit is\footnote{\label{footnote:angdefsubtle}Strictly speaking, the recoil-free angularities at this order in perturbation theory are $e_\alpha = \min[z,1-z]\theta^\alpha$.  The definition in \Eq{eq:angdef} is correct for $z < 1/2$, and since soft quarks ($z \to 1$) have no associated singularities in QCD, we are free to apply \Eq{eq:angdef} to the whole range of $z$ for the purposes of resummation.  For the fixed order corrections, this will lead to a (small) difference in the distribution.  For the \pythia{8} plots in \Fig{fig:ang_MC}, we use the full expression for the angularities.}
\begin{equation}\label{eq:angdef}
e_\alpha = z \theta^\alpha \ ,
\end{equation}
which ranges from $0$ to $1$.  To determine the phase space constraints on the $(e_\alpha, e_\beta)$ plane, we can simply invert \Eq{eq:angdef} and express $z$ and $\theta$ in terms of the angularities $e_\alpha$ and $e_\beta$:
\begin{equation}\label{eq:ps_irc}
z=e_\alpha^{-\frac{\beta}{\alpha-\beta}}e_\beta^{\frac{\alpha}{\alpha-\beta}} \ , \qquad \theta = e_\alpha^{\frac{1}{\alpha-\beta}}e_\beta^{-\frac{1}{\alpha-\beta}} \ .
\end{equation}
The phase space restrictions written in terms of $e_\alpha$ and $e_\beta$ are
\begin{equation}\label{eq:ps_consts}
\Theta(1-z)\Theta\left(1-\theta\right)\quad \Rightarrow \quad \Theta\left( e_\alpha^\beta - e_\beta^\alpha \right) \Theta\left( e_\beta - e_\alpha  \right).
\end{equation}
The upper boundary  in \Fig{region2} ($e_\alpha < e_\beta$)  corresponds to the requirement that the two particles are clustered in the jet.  The lower boundary in \Fig{region2} ($e_\beta^{\alpha/\beta} < e_\alpha$) comes from energy conservation.\footnote{Because $\beta< \alpha$, this lower boundary has 0 slope at $e_\beta = 0$, consistent with the comment made in footnote~\ref{footnote:zero_slope}.} 

Without doing a calculation, \Fig{region2} already illustrates why the ratio of two angularities is IRC unsafe.  Lines for every value of $r_{\alpha,\beta}$ pass through the singular region of the phase space at the origin $e_\alpha = e_\beta = 0$.  There is no way that the virtual contribution to the observable at a given order in perturbation theory can cancel the real contribution because the real contribution is singular for all values of $r_{\alpha,\beta}$.  We will now see this explicitly by computing the fixed-order cross section for $r_{\alpha,\beta}$.

\subsection{Fixed-Order Distributions}\label{sec:ang_double}

To calculate the double differential cross section of angularities to ${\cal O}(\alpha_s)$, we will use the $q \to q g$ QCD splitting function as representative of the matrix element for a narrow quark jet.  This only differs from the full QCD matrix element at ${\cal O}(\alpha_s)$ by non-singular terms. 

The quark splitting function is
\begin{equation}
\label{split}
S_q(z,\theta) \,d\theta \,dz = \frac{\alpha_s}{\pi} C_F\, \frac{1}{\theta}\,  \frac{1+(1-z)^2}{z} \, d\theta \, dz \,\Theta(1-z)\Theta(1-\theta)\ ,
\end{equation}
where $C_F = 4/3$ is the quark color factor, $z$ is the energy fraction of the emitted gluon, and $\theta$ is the splitting angle between the quark and the gluon.  As discussed in \Sec{sec:ps}, the variable $z$ ranges from $0$ to $1$ and the angle $\theta$ ranges from $0$ to the jet radius $R_0 = 1$.  Using the expressions for $z$ and $\theta$ in terms of $e_\alpha$ and $e_\beta$ from \Eq{eq:ps_irc}, we can simply perform a change of variables to rewrite \Eq{split} in terms of $e_\alpha$ and $e_\beta$ to determine the double differential cross section.  Including the appropriate Jacobian factor, we obtain the leading order (LO) cross section
\begin{equation}
\label{doubledist}
\frac{d^2\sigma^{\rm LO}}{de_\alpha\, de_\beta} = 2\frac{\alpha_s}{\pi} \frac{C_F}{\alpha - \beta} \left( \frac{1}{e_\alpha e_\beta} - e_\alpha^{-\frac{\alpha}{\alpha-\beta}}e_\beta^{\frac{\beta}{\alpha-\beta}}+\frac{e_\alpha^{-\frac{\alpha+\beta}{\alpha-\beta}}e_\beta^{\frac{\alpha+\beta}{\alpha-\beta}} }{2} \right)\Theta\left(e_\beta -  e_\alpha \right) \Theta\left( e_\alpha^\beta - e_\beta^\alpha \right)  \ .
\end{equation}

Armed with the double differential distribution, we can attempt to calculate the differential cross section for the ratio observable $r_{\alpha, \beta}=e_\alpha/e_\beta$ using \Eq{eq:ratio_dist_def}.   Dropping the subscripts $r_{\alpha, \beta} \to r$ for readability, we make the change of variables $e_\alpha = r e_\beta$ and integrate the double differential cross section over $e_\beta$.  The $\Theta$-function constraints with this change of variables becomes
\begin{equation}
\Theta\left(e_\beta -  e_\alpha \right) \Theta\left( e_\alpha^\beta - e_\beta^\alpha \right) \quad \Rightarrow \quad \Theta\left(1-r \right) \Theta\left( r^{\frac{\beta}{\alpha-\beta}} - e_\beta \right) \ .
\end{equation}
The differential cross section for $r$ is then
\begin{align}\label{eq:rxsec}
\frac{d\sigma^{\rm LO}}{dr} &= \int_0^1 de_\beta \int_0^1 de_\alpha \, \frac{d^2\sigma^{\rm LO}}{de_\alpha\, de_\beta} \, \delta\left( r - \frac{e_\alpha}{e_\beta}  \right) \nonumber \\
&=\int_0^{r^{\frac{\beta}{\alpha - \beta}}} de_\beta \, e_\beta \, \left. \frac{d^2\sigma^{\rm LO}}{de_\alpha\, de_\beta}\right|_{e_\alpha = r e_\beta} \ .
\end{align}
Integrating, we find
\begin{equation}\label{eq:fordist}
\frac{d\sigma^{\rm LO}}{dr} = -\frac{3}{2}\frac{\alpha_s}{\pi}\frac{C_F}{\alpha-\beta}\frac{1}{r} + 2\frac{\alpha_s}{\pi}\frac{C_F}{\alpha-\beta}\frac{1}{r}\int_0^{r^{\frac{\beta}{\alpha - \beta}}}\frac{de_\beta}{e_\beta} \ ,
\end{equation}
for $0<r<1$.  This expression manifests the IR-unsafeness of the ratio observable.  Because the singularity of the remaining integral is unregulated, the ratio observable is not defined in fixed-order perturbation theory.

To be able to compute the ratio observable in fixed-order perturbation theory, we need to regulate the remaining integral somehow.  The simplest prescription is to impose a lower limit on $e_\beta$, which prohibits the denominator of the ratio observable from becoming arbitrarily small and removes the phase space region where $e_\alpha$ and $e_\beta$ are both small but their ratio is arbitrary.  For a cut of $e_\beta > \epsilon$, the cross section in \Eq{eq:fordist} becomes
\begin{equation}
\label{eq:ircsaferwithepsilon}
\left.\frac{d\sigma^{\rm LO}}{dr}\right|_{e_\beta > \epsilon} = -\frac{3}{2}\frac{\alpha_s}{\pi}\frac{C_F}{\alpha-\beta}\frac{1}{r} + 2\frac{\alpha_s}{\pi}C_F \frac{\beta}{(\alpha-\beta)^2}\frac{\log r}{r} - 2\frac{\alpha_s}{\pi}\frac{C_F}{\alpha-\beta}\frac{\log \epsilon}{r}  + \mathcal{O}(\epsilon) \ .
\end{equation}
As this cut is lowered, the distribution becomes unbounded because of the $\log \epsilon$ term and is properly IR-unsafe as the cut is removed.  Another possible regularization is to deform the definition of the ratio observable to be
\begin{equation}
r_\delta = \frac{e_\alpha}{e_\beta^{1-\delta}} \ ,
\end{equation}
where $\delta > 0$.  Now, the singular region is regulated by an explicit $\Theta$-function that becomes trivial when $\delta\to 0$.  However, all of these fixes are arbitrary, and change the definition of the observable.  In the next section, we will see how to make the ratio observable well-defined by resumming the double differential cross section.

\section{Leading-Log Resummed Double Differential Distribution}
\label{sec:LL_double}

While the ratio of angularities $r_{\alpha,\beta}$ is not an IRC-safe observable, we now argue that it is a ``Sudakov safe'' observable.  That is, the Sudakov factor arising from the resummation of large logarithms acts as a natural regulator for the double differential cross section $d^2 \sigma/ d e_\alpha d e_\beta$.   We then marginalize appropriately to determine the resummed differential cross section of the ratio observable $r_{\alpha,\beta}$, which has a number of interesting and unfamiliar properties.  In particular, $d\sigma/dr_{\alpha,\beta}$ has a series expansion in $\sqrt{\alpha_s}$ (not $\alpha_s$) and ``leading logarithmic'' resummation for $r_{\alpha,\beta}$ corresponds to summing a tower of terms of the form $(\alpha_s \log^2 r)^{n/2}$ (instead of $(\alpha_s \log^2 r)^{n}$).  We emphasize that throughout this paper, we define LL to capture the leading logarithms $L$ with the scaling $\alpha_s L^2 \sim 1$.

\subsection{The Strongly-Ordered Limit}
\label{sec:strongorder}

For usual IRC-safe observables, one can perform LL resummation by considering the all-orders cross section in the strongly-ordered limit.  In this limit, there are multiple emissions from an eikonal Wilson line, but the value of the observable is determined by the leading emission(s).  Since the double differential cross section is IRC safe, we can easily determine the LL distribution for $d^2 \sigma/ d e_\alpha d e_\beta$ and use \Eq{eq:ratio_dist_def} to find the ratio distribution.  The meaning of ``LL'' for $d\sigma/dr_{\alpha,\beta}$ is more subtle, and we save that discussion for \Sec{sec:LL}.

\begin{figure}
\begin{center}
\subfloat[]{\label{fig:one_emis}
\includegraphics[width=6.85cm]{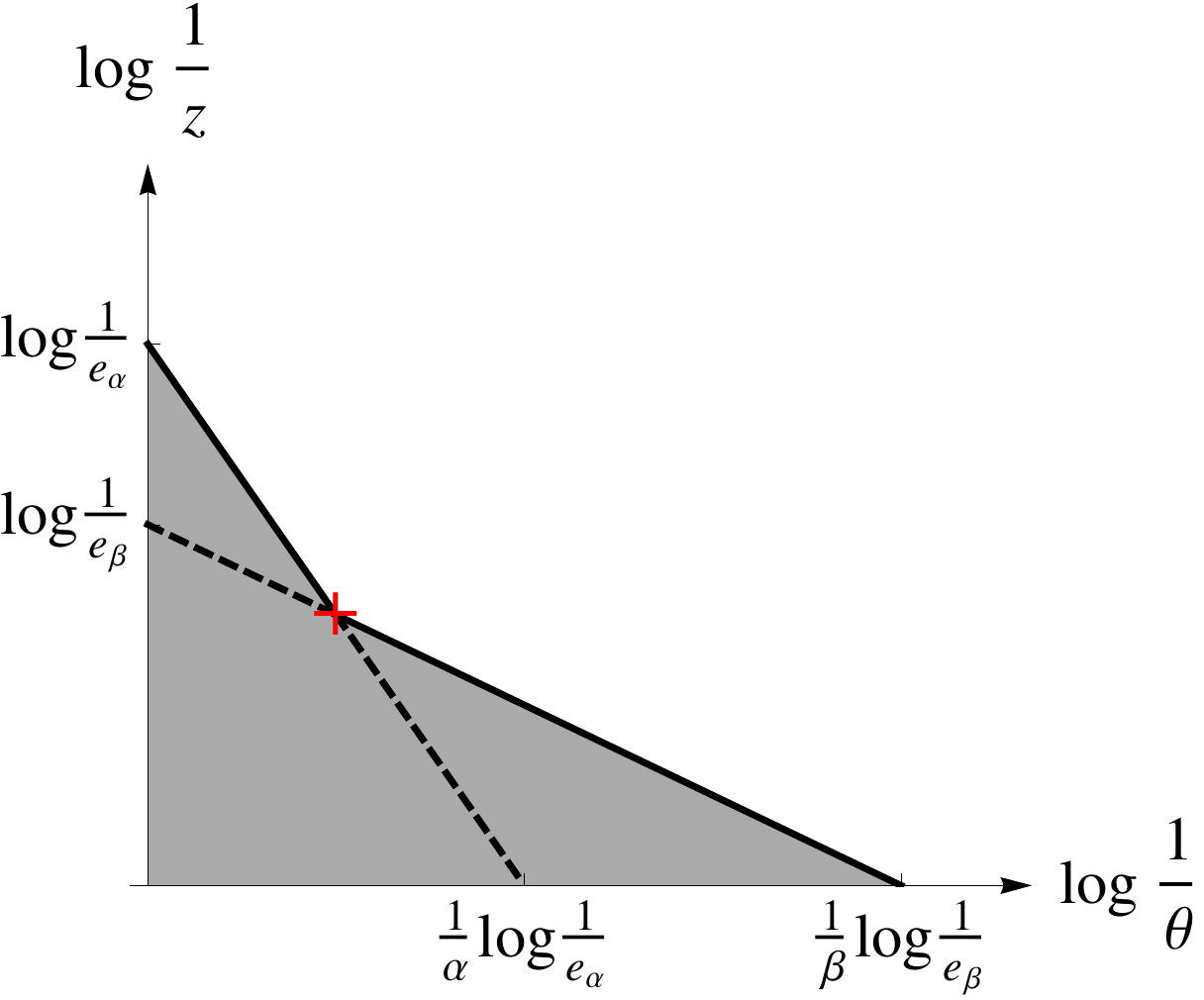}
}
$\qquad$
\subfloat[]{\label{fig:two_emis} 
\includegraphics[width=6.85cm]{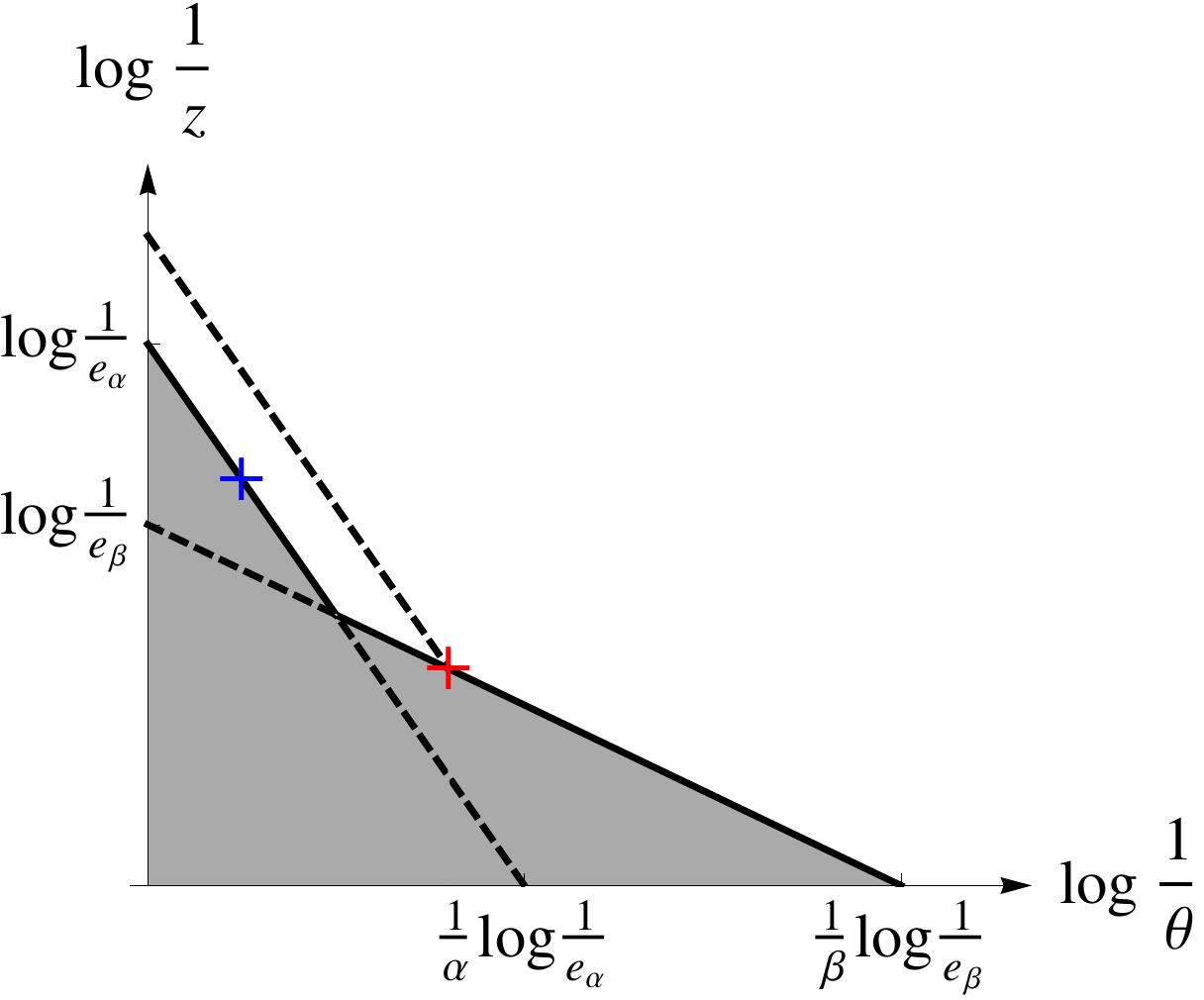}
}
\end{center}
\caption{Phase space for strongly-ordered emissions in the $\left(\log1/\theta,\log1/z\right)$ plane.  Left:  a single emission dominates the value of both $e_\alpha$ and $e_\beta$.  Right:  one emission dominates the value of $e_\alpha$ in blue, while a second emission dominates the value of $e_\beta$ in red.  In both cases, further emissions are forbidden in the gray region below the solid black line, and the area of the gray forbidden region determines the Sudakov factor.  The singular region of phase space is up and to the right.  
}
\label{fig:ps_emis}
\end{figure}

We first need to determine the region of phase space that contributes in the strongly-ordered limit.  This is best understood in $\left(\log1/\theta,\log1/z\right)$ space, where $\theta$ and $z$ are the emission angle and energy fraction, respectively.\footnote{We thank Gavin Salam for helpful discussions on the LL resummation.}  At LL order, we can ignore subleading terms in the splitting functions, and treat emissions as having uniform probability in the $\left(\log1/\theta,\log1/z\right)$ plane.  This phase space is illustrated in \Fig{fig:ps_emis}, where constant values of angularity $e_\alpha$ correspond to straight lines
\begin{equation}
\log\frac{1}{e_\alpha} = \log\frac{1}{z} + \alpha\log\frac{1}{\theta} \ ,
\end{equation}
and the singular region extends up and to the right.  In the strongly-ordered limit, a single leading emission determines the value of the angularity $e_\alpha$.

For the double differential cross section, a single emission may or may not determine both $e_\alpha$ and $e_\beta$ simultaneously.  In general, there are two possibilities for the leading emission(s):  either there is a single emission which determines the value of both $e_\alpha$ and $e_\beta$ (shown in \Fig{fig:one_emis}), or there is one emission that dominates the value of $e_\alpha$ and a different emission that dominates $e_\beta$ (shown in \Fig{fig:two_emis}).   Both possibilities contribute at LL order.  This is shown visually in \Fig{fig:ps_emis}, where the dominant emissions (denoted by the crosses) fix the values of $e_\alpha$ and $e_\beta$, such that no emissions can occur below the solid black line.  Additional emissions can then fill out the region above the solid black line, toward the singular region at infinity, but at LL order, these subdominant emissions do not modify the value of the angularities.

Given this phase space, we can easily determine the Sudakov factor in the strongly-ordered limit.  The Sudakov factor is just the probability that there were no emissions between two given scales.  The observation of $e_\alpha$ and $e_\beta$ introduces explicit scales and defines the area of phase space in which emissions are forbidden. The area of the gray forbidden region under the solid black line happens to be the same for both cases considered in \Fig{fig:ps_emis}:
\begin{equation}
\text{Area} = \frac{1}{2} \left(\frac{1}{\beta}\log^2e_\beta + \frac{1}{\alpha-\beta}\log^2\frac{e_\beta}{e_\alpha} \right) \ .
\end{equation}
Restoring the quark color factor and strong coupling constant and then exponentiating, the Sudakov factor at LL is
\begin{equation}\label{eq:sudakov}
\Delta(e_\alpha,e_\beta) =e^{- \frac{\alpha_s}{\pi}C_F\left( \frac{1}{\beta}\log^2 e_\beta +\frac{1}{\alpha-\beta}\log^2\frac{e_\alpha}{e_\beta} \right)} \ .
\end{equation}

From the Sudakov factor, we can determine the  resummed double differential cross section by differentiating with respect to $e_\alpha$ and $e_\beta$:
\begin{align}\label{eq:resum_double}
\frac{d^2\sigma^\text{LL}}{de_\alpha \, de_\beta} =& \ \frac{\partial}{\partial e_\alpha}\frac{\partial}{\partial e_\beta}\Delta(e_\alpha,e_\beta) \nonumber \\
=&\  \left( \frac{2\alpha_s}{\pi} \frac{C_F}{\alpha-\beta}\frac{1}{e_\alpha e_\beta} 
+ \frac{4\alpha_s^2}{\pi^2} \frac{C_F^2}{\beta(\alpha-\beta)^2}\frac{1}{e_\alpha e_\beta} \log\frac{e_\beta}{e_\alpha} \, \log\frac{e_\alpha^\beta}{e_\beta^\alpha}
\right)\Delta(e_\alpha,e_\beta).
\end{align}
The double differential cross section is defined on the physical phase space region from \Eq{eq:ps_consts} with $e_\beta > e_\alpha$,  $e_\alpha^\beta > e_\beta^\alpha $.      Note that the Sudakov factor suppresses the singular region of phase space at $e_\alpha \to 0$, $e_\beta \to 0$.  As a cross check of this calculation, we can integrate over one of the angularities (making sure to impose the proper phase space constraints) to reproduce the LL cross section for the other angularity:
\begin{align}
\frac{d\sigma^\text{LL}}{de_\alpha} &= \int_{e_\alpha}^{e_\alpha^{\beta/\alpha}} de_\beta \, \frac{d^2\sigma^\text{LL}}{de_\alpha \ de_\beta} = -2\frac{\alpha_s}{\pi}\frac{C_F}{\alpha}\frac{\log e_\alpha}{e_\alpha}e^{-\frac{\alpha_s}{\pi}\frac{C_F}{\alpha}\log^2 e_\alpha} \ , \nonumber \\
\frac{d\sigma^\text{LL}}{de_\beta} &= \int_{e_\beta^{\alpha/\beta}}^{e_\beta} de_\alpha \, \frac{d^2\sigma^\text{LL}}{de_\alpha \ de_\beta} = -2\frac{\alpha_s}{\pi}\frac{C_F}{\beta}\frac{\log e_\beta}{e_\beta}e^{-\frac{\alpha_s}{\pi}\frac{C_F}{\beta}\log^2 e_\beta}, \label{eq:LL_ang}
\end{align}
which is indeed correct.\footnote{The area under a curve of constant $e_\alpha$ is $\frac{1}{2 \alpha} \log^2 e_\alpha$, so the Sudakov factor is $\Delta(e_\alpha) = e^{-\frac{\alpha_s}{\pi}\frac{C_F}{\alpha}\log^2 e_\alpha}$.  \Eq{eq:LL_ang} comes from differentiating $\Delta(e_\alpha)$.}

\begin{figure}
\begin{center}
\subfloat[]{\label{fig:rat_distro_2}
\includegraphics[width=7.0cm]{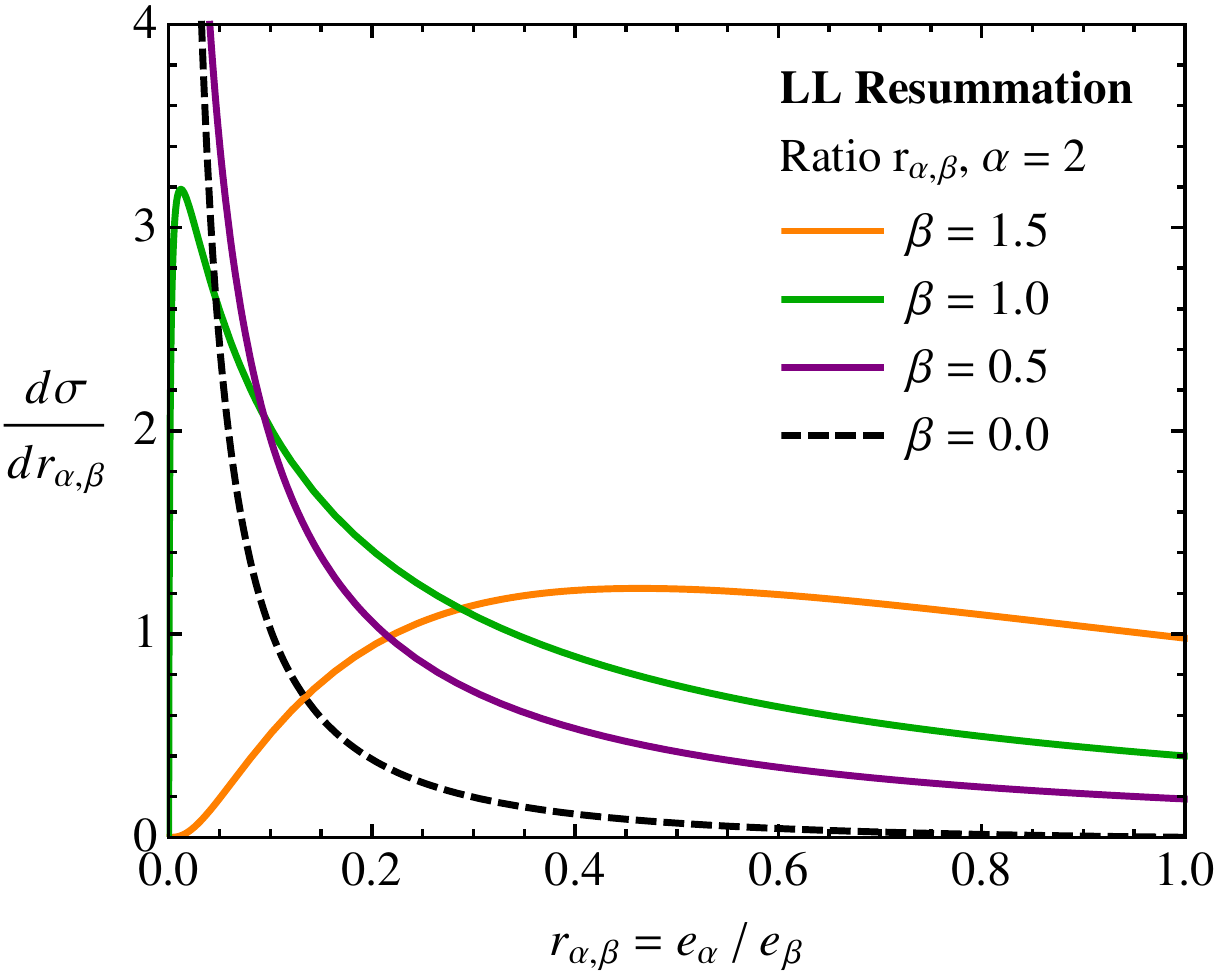}
}
$\quad$
\subfloat[]{\label{fig:rat_distro_1} 
\includegraphics[width=7.0cm]{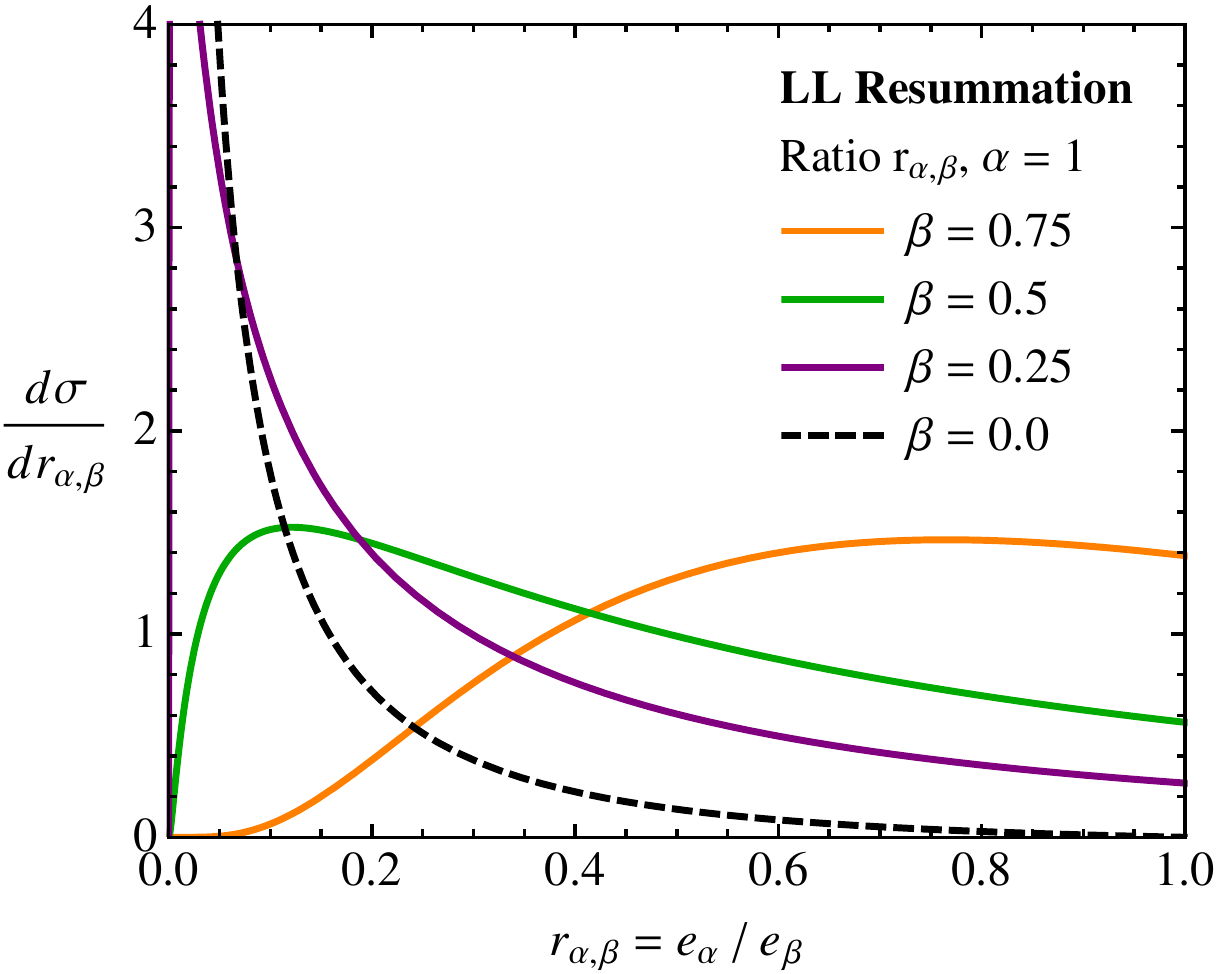}
}
\end{center}
\caption{LL differential cross section for the ratio observable $r_{\alpha,\beta} \equiv e_\alpha/e_\beta$ from \Eq{eq:resumrLL}.  Left:  numerator fixed to $\alpha = 2$ (thrust measure), and denominator sweeping over $\beta$.  Right: $\alpha = 1$ (broadening measure), sweeping $\beta$.  In both cases, the $\beta = 0$ curves give the LL differential cross section for the angularity with the corresponding $\alpha$, i.e.~$r_{\alpha,0} = e_\alpha$.}
\label{fig:rat_distro}
\end{figure}

Finally, from the resummed double differential cross section, we can determine the differential cross section for the ratio observable $r_{\alpha,\beta}=e_\alpha / e_\beta$ using \Eq{eq:ratio_dist_def}.   Dropping the subscripts $r_{\alpha,\beta} \to r$ for clarity,
\begin{align}\label{eq:resumrLL}
\frac{d\sigma^\text{LL}}{dr} =  &\  \frac{\sqrt{\alpha_sC_F\beta} }{\alpha
   -\beta }\frac{1}{r} \left(1-2\frac{\alpha_s}{\pi}\frac{ C_F }{\alpha -\beta}
   \log^2 r  \right)
   \left(\text{erf}\left[\frac{ \sqrt{\alpha_s C_F \beta} }{\sqrt{\pi } (\alpha -\beta )}\log r\right]+1\right)e^{-\frac{\alpha_s}{\pi}\frac{C_F}{\alpha-\beta} \log ^2r}\nonumber \\
   &-2\frac{\alpha_s}{\pi}\frac{C_F}{\alpha - \beta}\frac{\log r}{r} e^{-\frac{\alpha_s}{\pi}C_F\frac{\alpha}{(\alpha-\beta)^2}\log^2r},
\end{align}
where the error function $\text{erf}(x)$ is
\begin{equation}
\text{erf}(x) = \frac{2}{\sqrt{\pi}}\int_0^x dt \, e^{-t^2} \ .
\end{equation}
Because of the Sudakov regularization of the double differential cross section, the ratio distribution is well-defined and finite (i.e.~it is Sudakov safe).  

In \Figs{fig:rat_distro_2}{fig:rat_distro_1}, we plot the cross section of the ratio observable $r_{\alpha, \beta}$, taking a fixed coupling $\alpha_s = 0.12$.  In \Fig{fig:rat_distro_2}, we set the numerator to $\alpha=2$ (thrust) and scan over $\beta = 1.5$, $1.0$, $0.5$, to $0$.  In \Fig{fig:rat_distro_1}, we set the numerator to $\alpha=1$ (recoil-free broadening) and scan over $\beta = 0.75$, $0.5$, $0.25$, to $0$.  

Two limits of the ratio distribution can be easily understood.  As $\beta \to \alpha$, the cross section for the ratio approaches a $\delta$-function located at $r=1$.  As $\beta \to 0$, $e_\beta \to 1$, so the cross section for the ratio should degenerate to the cross section for $e_\alpha$ itself.  This behavior can be seen from \Eq{eq:resumrLL} directly.  With $\alpha$ fixed and taking $\beta \to 0$, all terms with error functions have a coefficient proportional to a positive power of $\beta$, and so vanish in the $\beta \to 0$ limit.  This leaves the last term of the cross section which has a non-zero limit:
\begin{equation}
\left.\frac{d\sigma^\text{LL}}{dr}\right|_{\beta\to 0} =-2\frac{\alpha_s}{\pi}\frac{C_F}{\alpha}\frac{\log r}{r} e^{-\frac{\alpha_s}{\pi}
\frac{C_F}{\alpha}\log^2r} =\frac{d\sigma}{de_\alpha} \ .
\end{equation}

\subsection{The Meaning of ``Leading Log''}
\label{sec:LL}

We have argued that the ratio $r_{\alpha,\beta}$ is a Sudakov safe observable.   Unlike an IRC-safe observable, it is not defined order-by-order in perturbation theory, so it is therefore interesting to ask how the cross section for the ratio behaves in the small $\alpha_s$ limit.  Expanding \Eq{eq:resumrLL}, we find
\begin{equation}\label{eq:r_exp}
\frac{d\sigma^\text{LL}}{dr}=\sqrt{\alpha_s}\frac{ \sqrt{C_F \beta}}{\alpha -\beta }\frac{1}{r} + {\cal O}(\alpha_s) \ .
\end{equation}
Because the expansion starts at $\mathcal{O}(\sqrt{\alpha_s})$, there is not a proper Taylor expansion in $\alpha_s$.  This is not surprising as the fixed-order cross section for the ratio observable does not exist.\footnote{The fact that the differential cross section is proportional to $\sqrt{\alpha_s}$ is reminiscent of the anomalous dimension of fragmentation functions for Mellin moment $j\to 1$ \cite{Bassetto:1979nt,Furmanski:1979jx,Konishi:1979ft,Mueller:1981ex,Bassetto:1982ma,Ellis:1991qj}, corresponding to the hadron multiplicity.  In that case, for $j\neq 1$, there is a sensible Taylor expansion in $\alpha_s$ of the anomalous dimension:
\begin{equation}\label{eq:anom}
\gamma\left(j,\alpha_s\right) = \frac{\alpha_s C_A}{\pi}\frac{1}{j-1}+{\cal O}(\alpha_s^2) \ ,
\end{equation}
which, however, does not exist at $j=1$.  The entire series must be resummed and the Taylor expansion which is valid for $j\neq1$ must be analytically continued outside of its radius of convergence.  It follows that the anomalous dimension at $j=1$ is
\begin{equation}
\gamma\left(j=1,\alpha_s\right) = \sqrt{\frac{\alpha_s C_A}{2\pi}} \ ,
\end{equation}
which is not reproduced by any finite-order expansion of \Eq{eq:anom}.}

This then raises the question as to the formal accuracy of \Eq{eq:resumrLL} and the meaning of ``LL'' for an IRC-unsafe observable.  For an ordinary IRC-safe observable, LL order is defined through the logarithms that appear in the cumulative distribution.  Given the double cumulative distribution $\Sigma(e_\alpha, e_\beta)$, its logarithm has the expansion
\begin{equation}
\log \Sigma(e_\alpha,e_\beta) = \alpha_s L^2 + \alpha_s L + \alpha_s + {\cal O}(\alpha_s^2) \ ,
\end{equation}
where $L$ is the logarithm of $e_\alpha$ or $e_\beta$.  Here, LL order includes all terms in $\log \Sigma$ at order $\alpha_s L^2\sim 1$, all of which are captured in the strongly-ordered limit.\footnote{An alternative definition of LL includes the leading terms with the scaling $\alpha_s L \sim 1$.  These are captured in the MLL procedure of \App{app:mll} which includes running $\alpha_s$.}  Because the ratio observable is IRC-unsafe, though, all values of $r\in[0,1]$ are sensitive to the singular region, and there is no simple correspondence between the singular region of phase space and the existence of large logarithms as there is with IRC-safe observables.

To figure out which logarithms have been resummed in \Eq{eq:resumrLL}, we first find the cumulative distribution $\Sigma(r)$ for the ratio observable:
\begin{align}
\Sigma(r)= \int_0^r dr \, \frac{d\sigma}{dr} &= \sqrt{\alpha_s}\frac{\sqrt{ C_F \beta}}{\alpha-\beta} \log r \,
   \left( 1+ \text{erf}\left[\frac{
   \sqrt{\alpha_s C_F \beta} }{\sqrt{\pi }
   (\alpha -\beta )}\log r\right] \right) e^{-\frac{\alpha_s}{\pi}\frac{ C_F }{\alpha
   - \beta }\log ^2 r} \nonumber\\
   &\qquad ~ +e^{-\frac{\alpha_s}{\pi}C_F\frac{\alpha }{(\alpha
   - \beta)^2 }\log ^2 r} \ .
\end{align}
The expansion of the logarithm of the cumulative distribution in $\alpha_s$ is
\begin{equation}
\log\Sigma(r)=\sqrt{\alpha_s}\frac{\sqrt{C_F \beta} }{\alpha -\beta }\log
   r-\frac{\alpha_s}{2\pi}C_F\frac{ 2
   \alpha -(4-\pi ) \beta  }{(\alpha -\beta )^2}\log ^2 r+{\cal O}(\alpha_s^{3/2}) \ .
\end{equation}
Every term in this expansion is of the form $(\alpha_s \log^2 r)^{n/2}$, where $n$ is a positive integer.  So just as for ordinary IRC-safe observables, LL resummation means that we capture leading terms in the limit $\alpha_s L^2\sim1$ where $L = \log r$, albeit starting at $\mathcal{O}(\sqrt{\alpha_s})$.  Note that LL resummation captures logarithms of $r$ in the $r \to 0$ region.  While finite values of $r$ are also sensitive to the singular region of phase space, those effects are subleading in the logarithmic power counting.  We will see in \Sec{sec:MC_resum} that there are also logarithms of $(1-r)$ which show up beyond LL.

\section{Higher-Order Corrections}
\label{sec:highord}

To increase the accuracy of our LL calculation in \Eq{eq:resumrLL}, we would like to include both fixed-order corrections and higher-order resummation.  Because the fixed-order cross section for the ratio observable does not exist, though, we cannot use standard matching methods.  That said, in the same way as the LL resummation proceeded in \Sec{sec:strongorder}, we can perform the matching procedure on the double differential cross section $d^2 \sigma/ d e_\alpha d e_\beta$ and then marginalize using \Eq{eq:ratio_dist_def} to define the ratio cross section $d\sigma/dr_{\alpha,\beta}$.  Because the Sudakov factor provides a natural cut-off of the singular region of phase space, the ratio observable will still have a finite cross section as higher-order effects are included, as long as the matching procedure does not affect the Sudakov-suppressed region of phase space.

As discussed in \Sec{sec:LL}, it is not entirely straightforward to define the accuracy of a non-IRC-safe distribution.  For this reason, we will subsequently refer to the accuracy of a calculation for $d\sigma/dr_{\alpha,\beta}$ in terms of the accuracy of the double differential distribution $d^2 \sigma/ d e_\alpha d e_\beta$.  For illustrative purposes, we show how to include LO fixed-order information and some effects beyond LL, leaving more accurate calculations to future work.  Throughout this paper, LO means $\mathcal{O}(\alpha_s)$ fixed-order corrections.

\begin{figure}
\begin{center}
\subfloat[]{\label{fig:match_2}
\includegraphics[width=7.0cm]{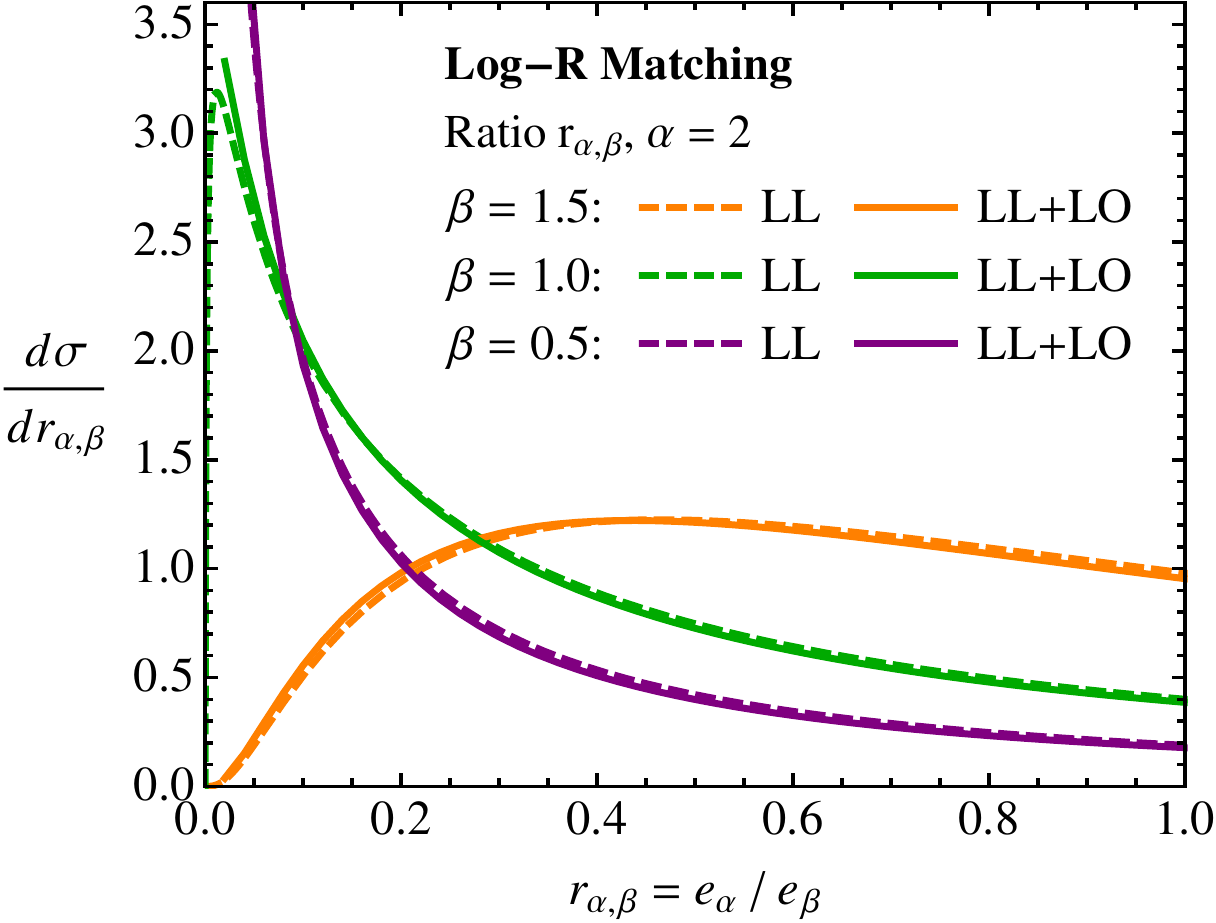}
}
$\quad$
\subfloat[]{\label{fig:match_1} 
\includegraphics[width=7.0cm]{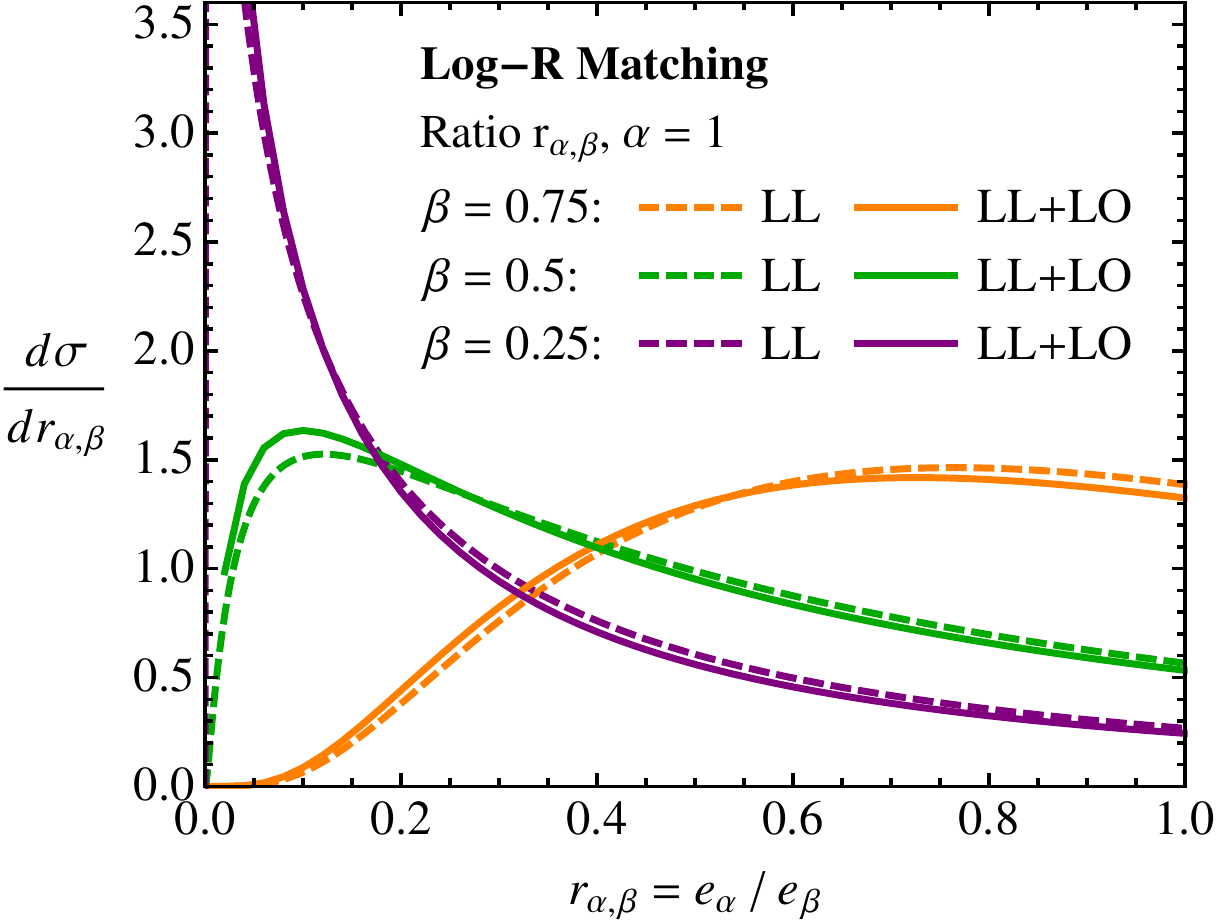}
}
\end{center}
\caption{Cross section for the ratio observable $r_{\alpha,\beta}=e_\alpha/e_\beta$ at LL+LO using the Log-R matching procedure in \App{app:match}.  Shown are $\alpha=2$ (left) and $\alpha=1$ (right), sweeping over $\beta$.  Because the ratio distribution is dominated by the Sudakov peak, there are only small changes in going from LL to LL+LO.}
\label{fig:match_plot}
\end{figure}

To incorporate fixed-order corrections at LO,  we use the Log-R matching procedure \cite{Catani:1992ua} in \App{app:match}, effectively merging \Eq{doubledist} with \Eq{eq:resumrLL} to obtain an LL+LO distribution.  We show the final results in \Fig{fig:match_plot}, which compares the LL cross section of the ratio observable to the LL+LO distribution obtained after fixed-order corrections are incorporated into $d^2 \sigma/ d e_\alpha d e_\beta$.  In general, the effect of matching is quite small over the entire range of $r_{\alpha,\beta}$.  This is expected because the integral defining the ratio observable $r_{\alpha,\beta}$ is dominated by the Sudakov peak region where fixed-order corrections are small.  Nevertheless, matching does formally increase the accuracy of the cross section for the ratio observable.

\begin{figure}
\begin{center}
\subfloat[]{\label{fig:mll_2}
\includegraphics[width=7.0cm]{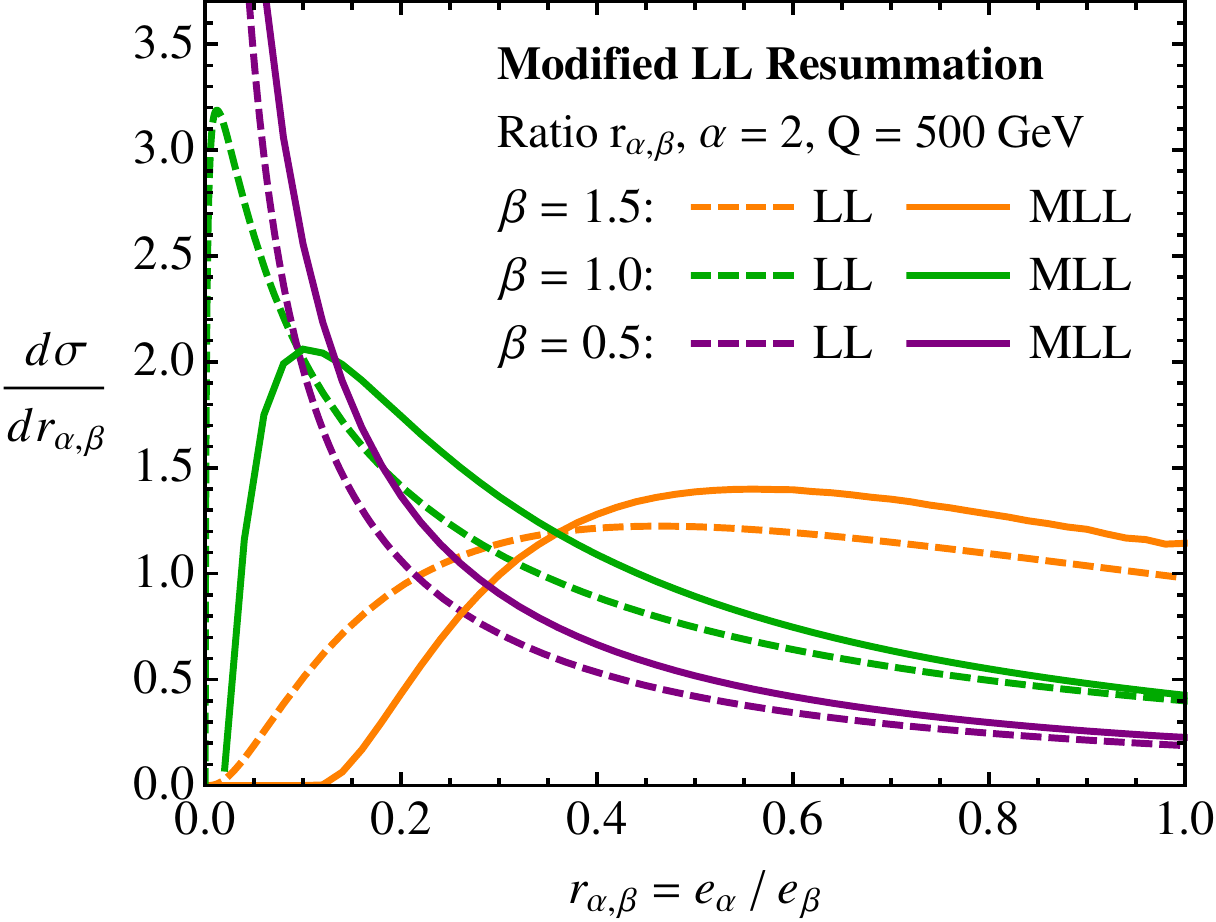}
}
$\quad$
\subfloat[]{\label{fig:mll_1} 
\includegraphics[width=7.0cm]{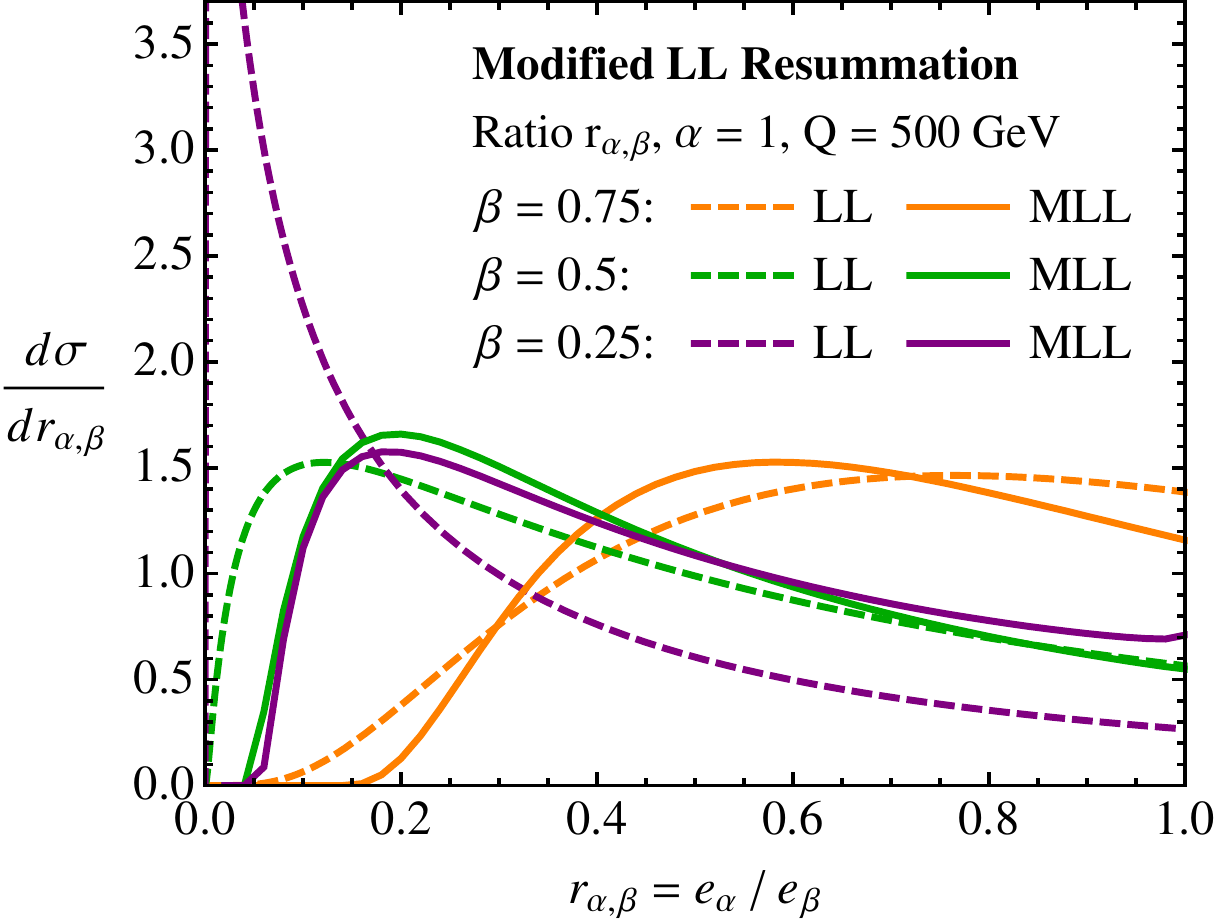}
}
\end{center}
\caption{Cross section for the ratio observable $r_{\alpha,\beta}=e_\alpha/e_\beta$ with MLL resummation described in \App{app:mll}.  This is shown for quark jets with energy $Q = 500$ GeV (to set the scale for $\alpha_s$), with $\alpha=2$ (left) and $\alpha=1$ (right), sweeping over $\beta$.  Because MLL includes the effects of running $\alpha_s$, there are larger changes in the cross section than for the fixed-order corrections in \Fig{fig:match_plot}.
}
\label{fig:mll_plot}
\end{figure}

In addition to including fixed-order corrections, the double differential cross section can be resummed to higher orders as well.  In \App{app:mll}, we compute the modified leading logarithmic (MLL, as in, e.g., \Ref{Seymour:1997kj}) Sudakov factor for $e_\alpha$ and $e_\beta$.  MLL resummation includes one-loop running of $\alpha_s$ and subleading terms in the splitting function, resumming logarithmically enhanced terms of the form $\alpha_s^n L^m$, where $m\geq n$.  However, MLL is not fully accurate at NLL order because there are effects that arise at the same formal order in the logarithmic counting that are not included (namely, multiple emissions and two-loop running coupling).  Because of the inclusion of a running coupling, we expect that MLL is significantly more accurate than LL resummation, especially for the ratio observable $r_{\alpha,\beta}$.  We compare the MLL cross section to the LL cross section for jets of energy $Q = 500$ GeV in \Fig{fig:mll_plot}, where the running coupling effects suppress the cross section at small values of the ratio $r_{\alpha,\beta}$.

\begin{figure}
\begin{center}
\subfloat[]{\label{fig:mlllo_2}
\includegraphics[width=7.0cm]{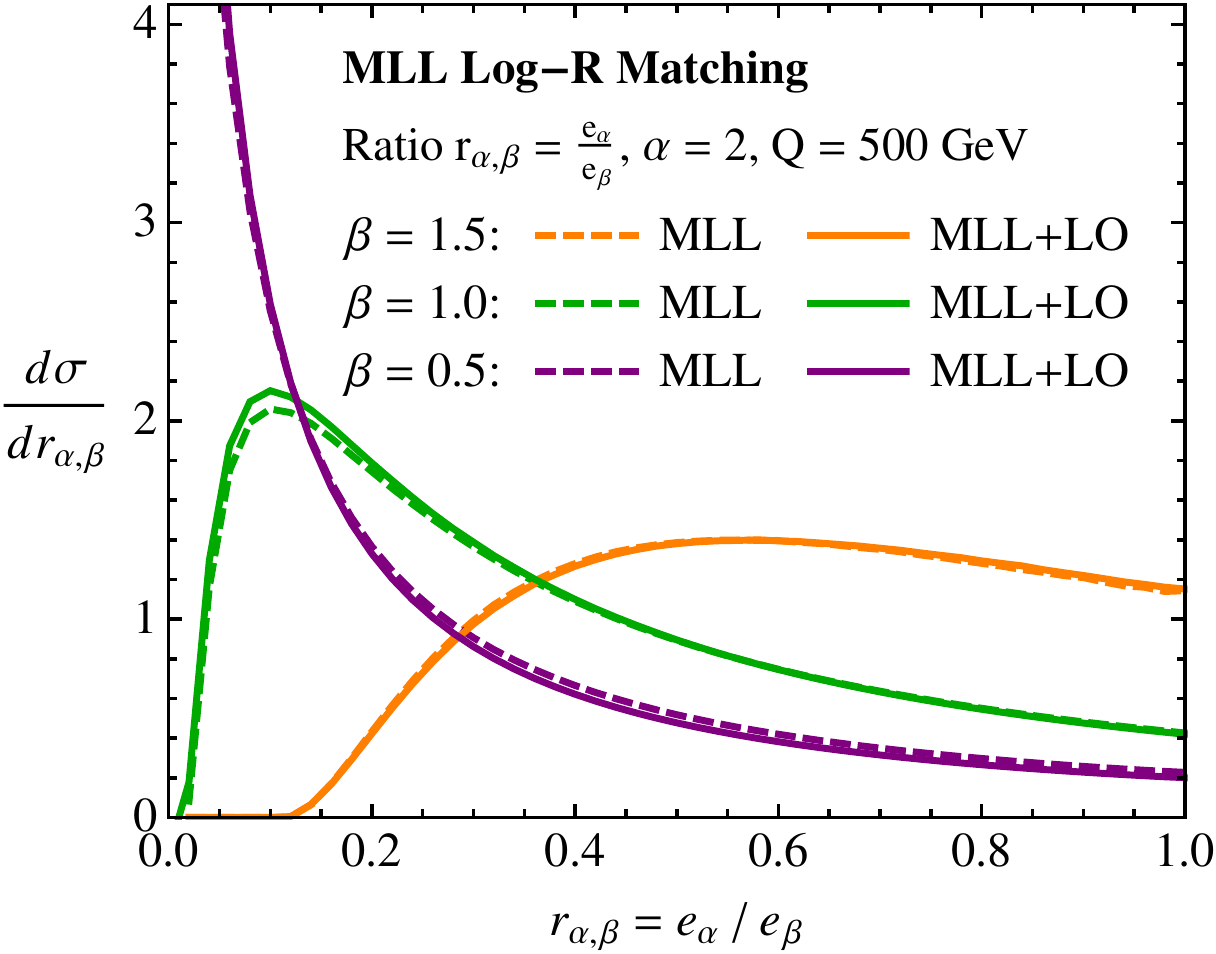}
}
$\quad$
\subfloat[]{\label{fig:mlllo_1} 
\includegraphics[width=7.0cm]{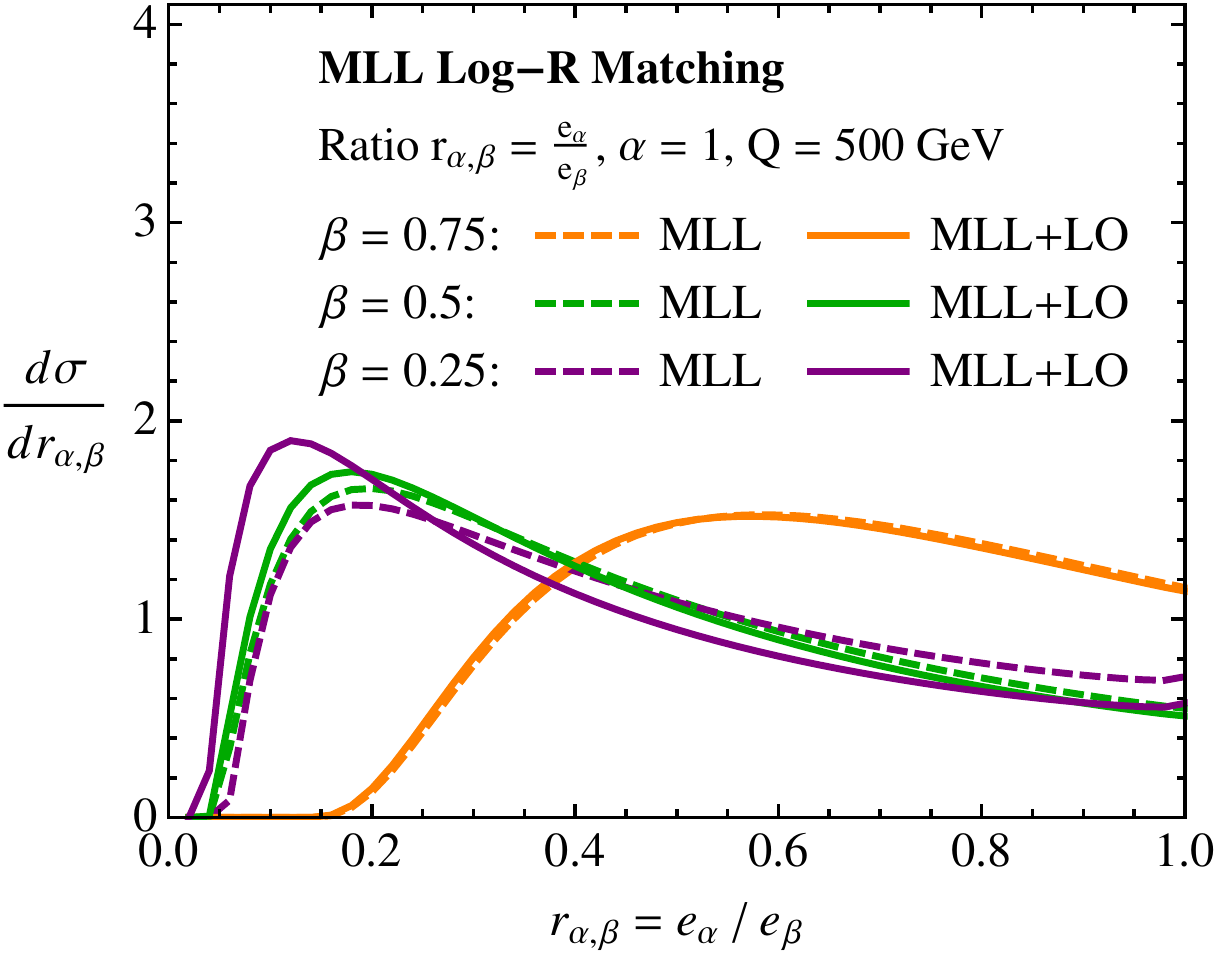}
}
\end{center}
\caption{
Cross section for the ratio observable $r_{\alpha,\beta}=e_\alpha/e_\beta$ with both MLL resummation and LO fixed-order matching (MLL+LO).  Shown are $\alpha=2$ (left) and $\alpha=1$ (right), sweeping over $\beta$.  Because the ratio distribution is dominated by the Sudakov peak, there are only small changes in going from MLL to MLL+LO.  The effect of matching grows as $\beta$ decreases because the Sudakov peak moves to larger values of the angularities where the fixed-order contribution becomes more important.
}
\label{fig:mlllo_plot}
\end{figure}

Finally, the accuracy of the ratio distribution can be further improved by combining fixed-order corrections with higher-order resummation.  In \Fig{fig:mlllo_plot}, we improve the MLL resummation by using the full splitting function in \Eq{eq:splitfull} (instead of just the average splitting function in \Eq{eq:splitave}) to achieve LO matching.  This is formally the same accuracy as applying the Log-R matching procedure to the MLL distribution.  The effect of matching in going from MLL to MLL+LO is small (and comparable to going from LL to LL+LO in \Fig{fig:match_plot}). 
That said, the matching becomes a larger effect as $\beta$ gets small.  This is because the Sudakov peak of the double differential cross section moves to larger values of $e_\alpha$ and $e_\beta$ as $\beta$ gets small, and enters a regime in which the fixed-order contribution becomes a significant part of the cross section.

\section{Monte Carlo vs.\ Analytic Resummation}
\label{sec:MC_resum}

Parton shower Monte Carlo programs are ubiquitous tools for predicting the outcome of particle collisions, so it is important to know whether the parton shower can accurately determine the distribution of the ratio observable $r_{\alpha, \beta}$.  It is well-known that the parton shower formally provides LL resummation of IRC-safe observables \cite{Ellis:1991qj}, but strictly speaking, these arguments assume that the observable in question is well-defined at each order in perturbation theory.  Because the ratio observable is not IRC safe, one might worry whether or not the parton shower could reproduce the LL expression in \Eq{eq:resumrLL}.  In this section, we argue that parton showers not only achieve LL (and even MLL) accuracy, but they also include the important effect of multiple emissions (formally appearing at NLL order) which is particularly relevant near $r_{\alpha,\beta} = 1$.

To study resummation in a Monte Carlo context, we have written a program that resums the leading logarithms of any angularity through a simplified parton shower.  Compared to a full-blown Monte Carlo program, our simplified shower only treats emissions from a single quark jet and does not include effects like energy-momentum conservation or color coherence  which are beyond LL order.  Crucially, the shower does allow for multiple emissions within a single jet.  We have two versions of the simplified shower:  ``LL+MC'' which uses fixed $\alpha_s = 0.12$ and only the leading terms in the splitting function, and ``MLL+MC'' which includes one-loop running $\alpha_s$ and subleading terms in the splitting function (i.e.~the same effects included in MLL in \Sec{sec:highord}).  Because the full splitting functions are used for MLL+MC, it contains all of the physics of MLL+LO from \Sec{sec:highord}.

We can take the evolution variable of our shower to be any angularity $e_\alpha$ for arbitrary $\alpha>0$.  For the LL+MC shower, the probability that no emission has occurred between two scales $e_\alpha^i$ and $e_\alpha^f$ is given by the ratio of Sudakov factors
\begin{equation}\label{eq:Sud_rat}
\frac{\Delta(e_\alpha^f)}{\Delta(e_\alpha^i)} = e^{-\frac{\alpha_s}{\pi}\frac{C_F}{\alpha}\left( \log^2 e_\alpha^f - \log^2 e_\alpha^i  \right)} \ ,
\end{equation}
where this expression is for fixed $\alpha_s$ and $e_\alpha^f<e_\alpha^i$.  By construction, a parton shower that distributes emissions according to \Eq{eq:Sud_rat} will have LL accuracy for $e_\alpha$, and if the parton shower also distributes emissions in energy and angle according to the QCD splitting functions, then it will have LL accuracy for the other angularities $e_\beta$ with $\beta \not= \alpha$.  We have checked that our final results are robust to the choice of evolution variable, but for concreteness,  we take $\alpha = 1$ for our MC studies, corresponding to recoil-free broadening or $k_T$.\footnote{The LL+MC results are independent of the evolution variable.  However, the MLL+MC results depend on the evolution variable because it fixes the scale at which $\alpha_s$ is evaluated.  Using the angularity with $\alpha=1$ for evolution is consistent with analytic resummation procedures, such as we discuss in \App{app:mll}.}

The structure of our LL+MC shower is as follows:
\begin{enumerate}
\item Given an initial scale $e_\alpha^i$, determine the scale of the next splitting $e_\alpha^f$ . To do this, let $R \in [0,1]$ be a random number and set it equal to the ratio of Sudakov factors in \Eq{eq:Sud_rat}.  Invert the expression to find the scale of the next emission:
\begin{equation}
e_\alpha^f = e^{-\left( \log^2 e_\alpha^i - \frac{\pi}{\alpha_s}\frac{\alpha}{C_F}\log R  \right)^{1/2}} \ .
\end{equation}

\item  From the emission scale $e_\alpha^f$, determine the phase space variables $z$ and $\theta$.  From \Eq{eq:angdef}, we have the constraint $e^f_\alpha = z \theta^\alpha$.\footnote{As discussed in footnote~\ref{footnote:angdefsubtle}, this choice is fine for quark jets and only differs from the exact answer by non-singular terms.  A more sophisticated shower would be needed to address $g \to gg$ splittings.}  With a random number $R \in [0,1]$, the phase space variables are
\begin{equation}\label{eq:variablesMC}
z = \left(e_\alpha^f\right)^R, \qquad \theta = \left( e_\alpha^f  \right)^\frac{1-R}{\alpha}.
\end{equation}
This kinematic map generates phase space that is flat in $\log1/\theta$ and $\log 1/z$, and therefore distributes events according to the most singular terms of the splitting function in \Eq{split}.

\item With the phase space coordinates $z$ and $\theta$, compute the contribution of the current emission to any observable of interest.  To LL accuracy, we do not need to include correlations between different emissions; only correlations between each emission and the hard jet core are necessary at this order.

\item Set the current scale $e_\alpha^f$ equal to the new initial scale $e_\alpha^i$ and go to step 1.  Continue the Monte Carlo until the starting scale $e_\alpha^i$ falls below some stopping scale $x_{\text{end}}$.
\end{enumerate}
This LL+MC procedure is guaranteed to resum the leading logarithms from gluon emission off of a quark jet for arbitrary IRC-safe observables.

For the MLL+MC shower, we use the Sudakov veto method (see \Ref{Sjostrand:2006za} for an example) to account for running $\alpha_s$ and subleading terms in the splitting function.  For the MLL+MC shower, we follow steps 1 and 2 of the LL+MC shower, taking a large, fixed value of $\alpha_s$ which we denote as $\hat{\alpha}_s\sim 0.5$.  Once an emission scale and phase space point have been chosen, the running coupling and subleading terms in the splitting function can be included by veto.  We note that, for sufficiently large $\hat{\alpha}_s$, the probability of an emission using the most singular terms of the splitting function is larger than that using the full splitting function on all of phase space:
\begin{equation}
2\frac{\hat{\alpha}_s}{\pi}C_F \frac{1}{\theta}\frac{1}{z} \geq 2\frac{\alpha_s(k_T)}{\pi}C_F \frac{1}{\theta}\frac{1+(1-z)^2}{z}  \ ,
\end{equation}
where $k_T$ is the energy scale of the emission.  Therefore, the running coupling and full splitting functions can be included in the Monte Carlo by vetoing emissions if the ratio of the two probabilities is less than a random number $R\in[0,1]$.  That is, an emission is vetoed if
\begin{equation}
\frac{\alpha_s(k_T)}{\hat{\alpha}_s}\frac{1+(1-z)^2}{2} < R\in[0,1] \ .
\end{equation}
For emissions that are accepted, we can then compute the contribution of that emission to any observable of interest.  The shower terminates when the scale $k_T$ of an emission falls below a low, but still perturbative, scale $\mu$.  We set $\mu=1$ GeV in all of the following plots of the MLL+MC distributions.

\begin{figure}
\begin{center}
\subfloat[]{\label{fig:ang_MC_pdf2}
\includegraphics[width=7.0cm]{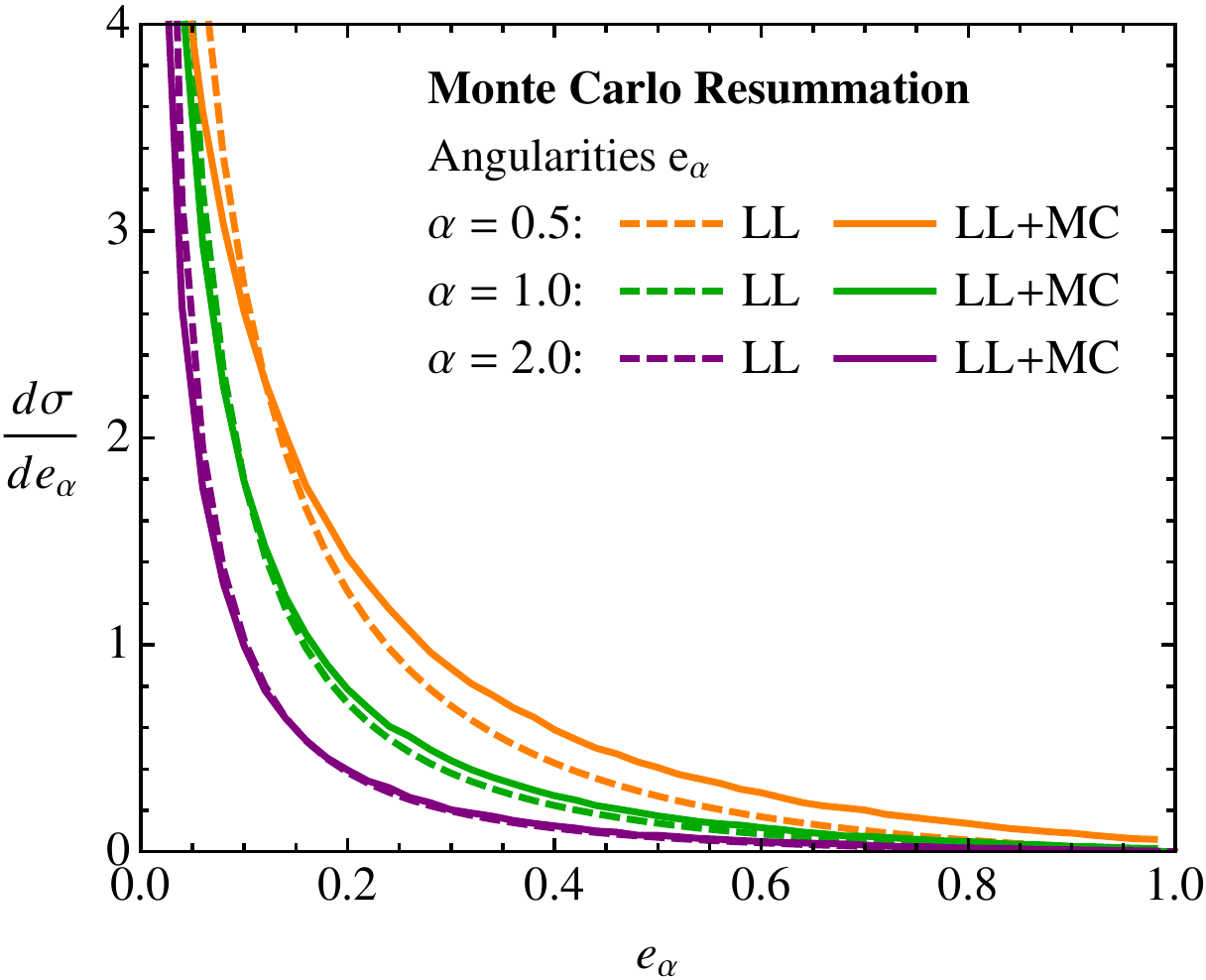}
}
$\quad$
\subfloat[]{\label{fig:ang_MC_pdf1} 
\includegraphics[width=7.0cm]{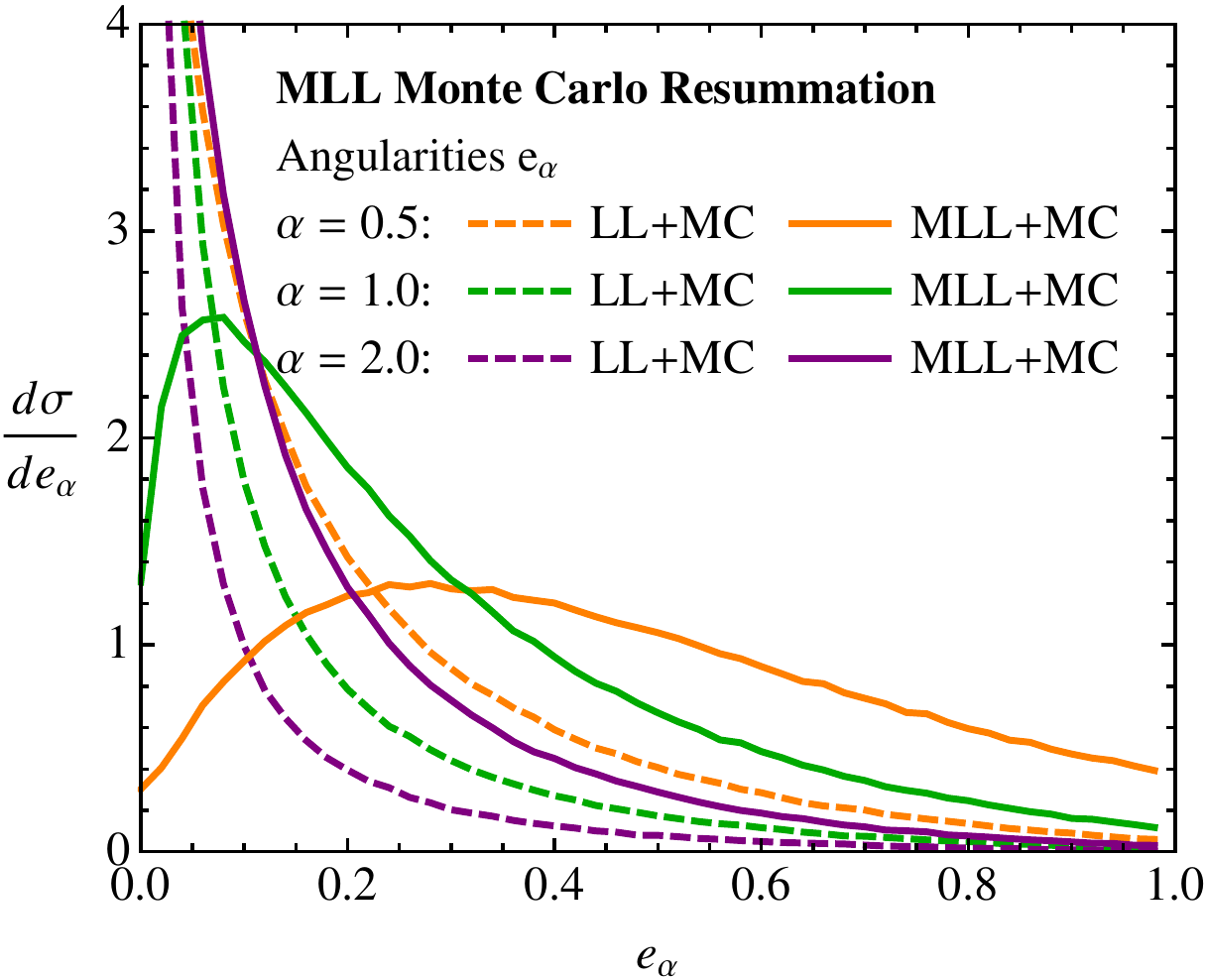}
}
\end{center}
\caption{Left: Comparing the distribution of angularities $e_\alpha$ between the LL resummed distribution in \Eq{eq:LL_ang} (dashed) and the LL+MC shower with fixed $\alpha_s$ (solid).  The distributions agree to LL accuracy, but because LL+MC includes the effect of multiple emissions, there are differences become apparent at small values of the angular exponent $\alpha$.  Right: Comparing the distribution of angularities $e_\alpha$ from the LL+MC to MLL+MC showers.  Because of the running coupling, the MLL+MC is suppressed at small values of $e_\alpha$ with respect to LL+MC.}
\label{fig:ang_MC_pdf}
\end{figure}

As an initial test of our shower, we first verify that the differential cross sections for the individual angularities $e_\alpha$ agree with the analytic expression for the LL resummation in \Eq{eq:LL_ang}.  Overall, the agreement in \Fig{fig:ang_MC_pdf} is quite good, though as the angular exponent $\alpha$ decreases, there is a greater discrepancy between the parton shower and the analytic expression.  This discrepancy can be attributed to the effect of multiple emissions.  Because the shower allows an arbitrary number of emissions to contribute to the observable $e_\alpha$, the region near $e_\alpha \simeq 0$ is suppressed beyond the na\"ive LL Sudakov factor.  The multiple emissions effect formally begins at $\alpha_s^2 L^2$ order in the resummation, and becomes more pronounced as the angular exponent exponent $\alpha$ decreases.  Because this multiple emission effect is physical, we expect the MC result to be a better estimate of the true distribution of angularities than the LL result.  We also compare the distribution of angularities from the LL+MC shower to the MLL+MC shower.  Because of the running coupling, the MLL+MC distributions are further suppressed  with respect to LL+MC at small values of $e_\alpha$.  Note that our toy Monte Carlo does not attempt to conserve energy/momentum, so the distributions have an (unphysical) finite cross section at $e_\alpha = 1$.

\begin{figure}
\begin{center}
\subfloat[]{\label{fig:angrat_MC_pdf2}
\includegraphics[width=7.0cm]{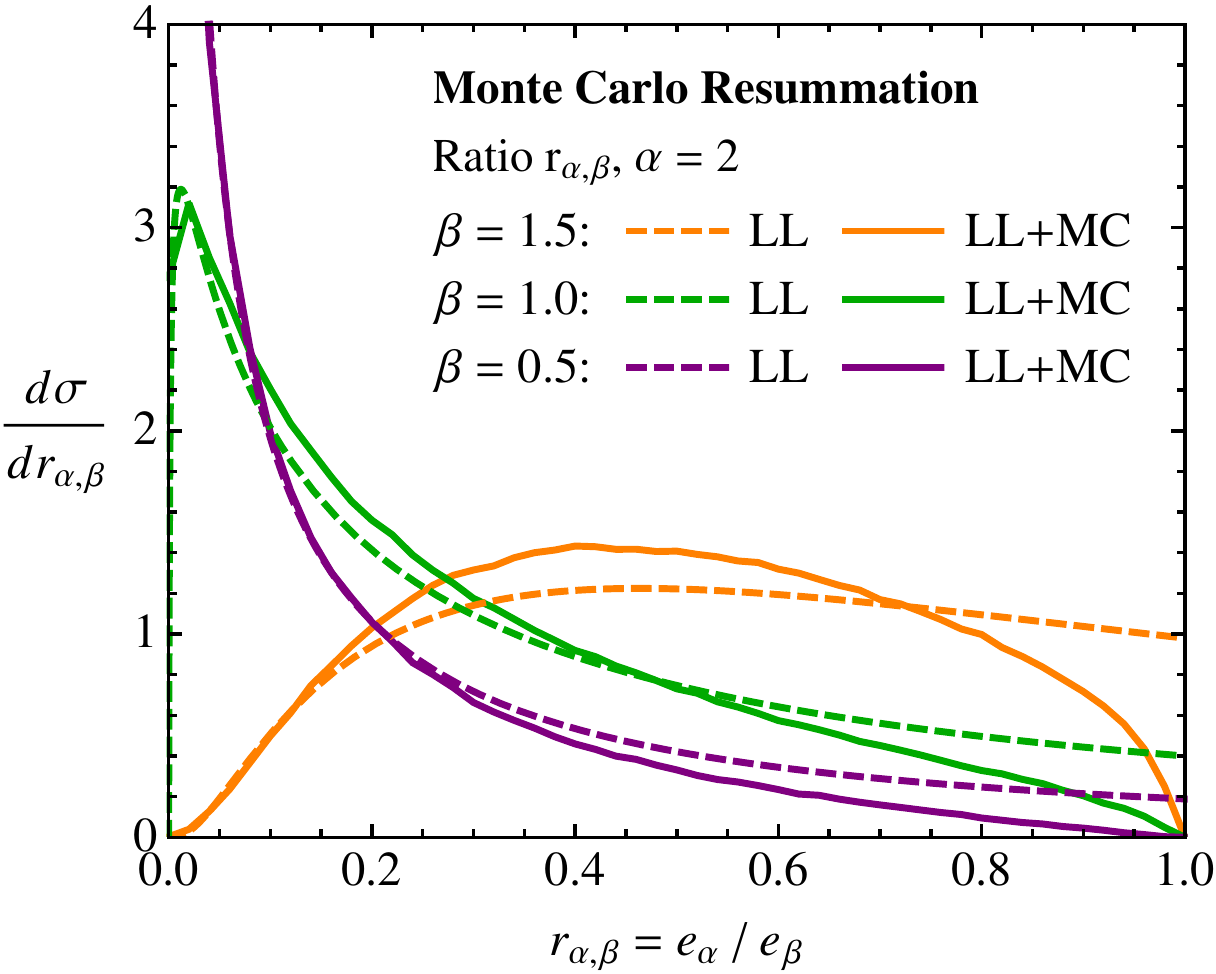}
}
$\quad$
\subfloat[]{\label{fig:angrat_MC_pdf1} 
\includegraphics[width=7.0cm]{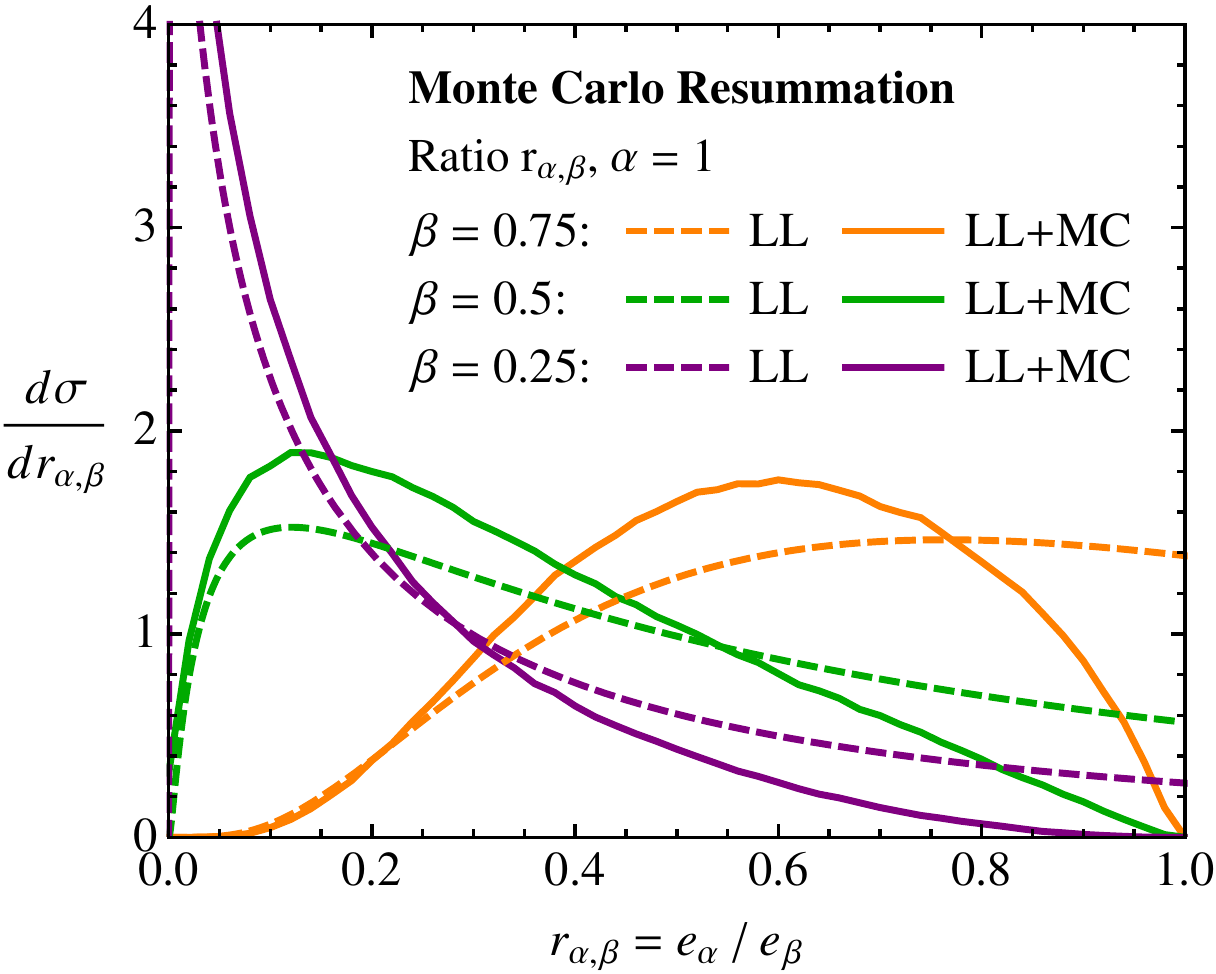}
}
\end{center}
\caption{Comparing the distribution of the ratio $r_{\alpha, \beta} \equiv e_\alpha/e_\beta$ between the analytic LL formula in \Eq{eq:resumrLL} (dashed) and the LL+MC shower with fixed $\alpha_s$ (solid).   Shown are $\alpha=2$ (left) and $\alpha=1$ (right), sweeping over $\beta$.   At small values of $r_{\alpha, \beta}$ there is good agreement between the two methods, but LL+MC includes multiple emissions which dramatically changes the shape of the distribution near $r_{\alpha, \beta} =1$.
}
\label{fig:angrat_MC_plot}
\end{figure}

With good agreement established between the parton shower and the analytic formulas for the angularities $e _\alpha$, we now consider the distribution of the ratio observable $r_{\alpha,\beta}$.  In \Fig{fig:angrat_MC_plot}, we compare the parton shower resummation to the analytic expression in \Eq{eq:resumrLL}.  There is very good agreement between the two methods at small values of $r_{\alpha,\beta}$, which, from the discussion in \Sec{sec:LL_double}, is where we expect the LL resummation for the ratio observable to be accurate.   

Near $r_{\alpha,\beta} = 1$, though, there are dramatic differences between the LL and LL+MC curves, due to the effect of multiple emissions.  For the ratio of $e_\alpha$ to $e_\beta$ to equal 1, there must be a single emission with arbitrary energy fraction and splitting angle equal to $\theta = 1$.  The effect of any subsequent emissions would be to reduce the value of $r_{\alpha,\beta}$.  The parton shower includes the effect of subsequent emissions on the value of the ratio, and up to cutoff effects, there is zero probability for there to be no emissions after the first one.  Thus, the fact that the ratio observable distribution vanishes at $r_{\alpha,\beta} =1$ is physical.  As discussed earlier, multiple emissions effects arise strictly beyond LL order.   An emission which results in $r_{\alpha,\beta} =1$ is necessarily a wide-angle emission, implying that the multiple emissions effect is at least one collinear logarithm down from LL order.\footnote{Also, instead of being a $\log r$ effect, this is a $\log(1-r)$ effect.}  That said, accounting for multiple emissions is crucial for obtaining the qualitatively correct distributions.

\begin{figure}
\begin{center}
\subfloat[]{\label{fig:angrat_MCrunning_pdf2}
\includegraphics[width=7.0cm]{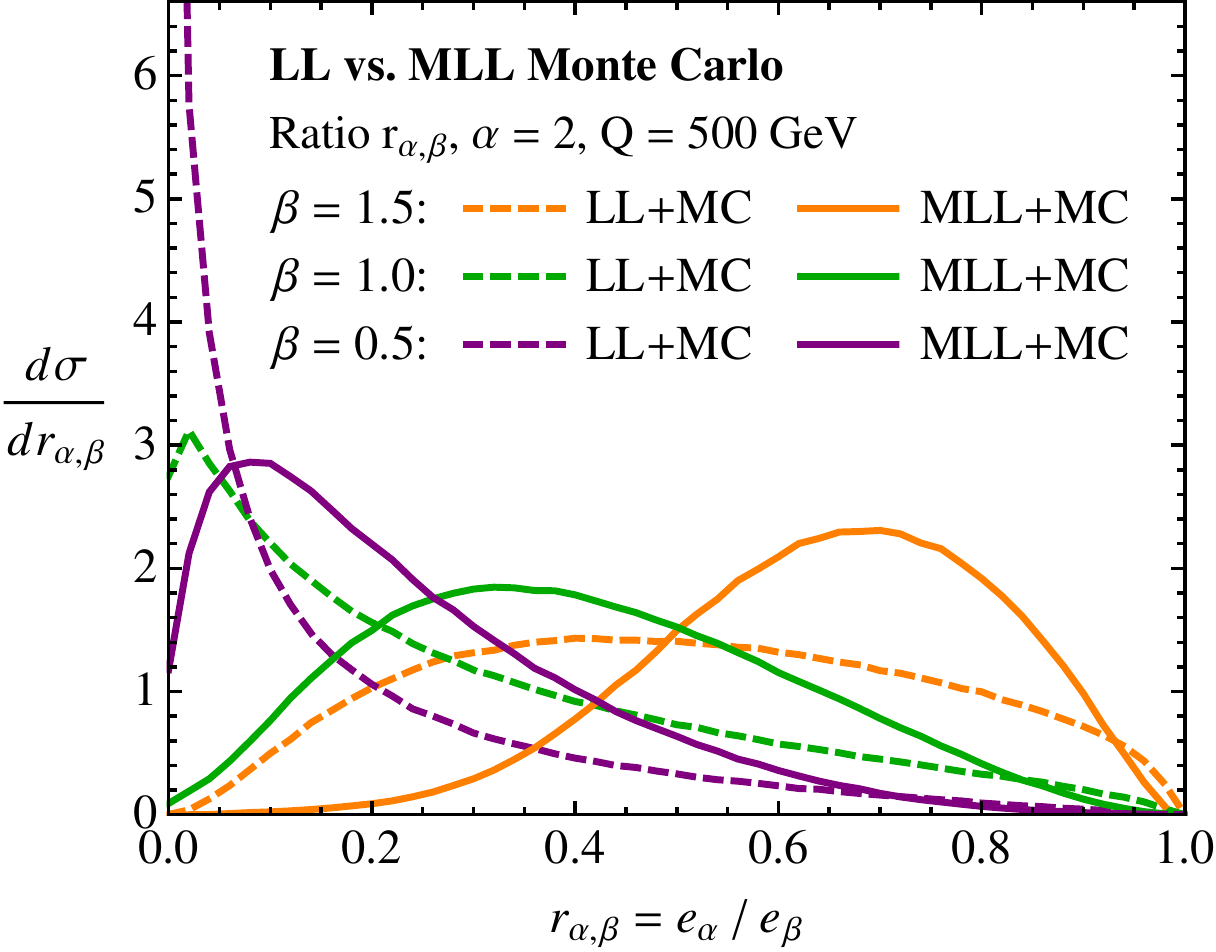}
}
$\quad$
\subfloat[]{\label{fig:angrat_MCrunning_pdf1} 
\includegraphics[width=7.0cm]{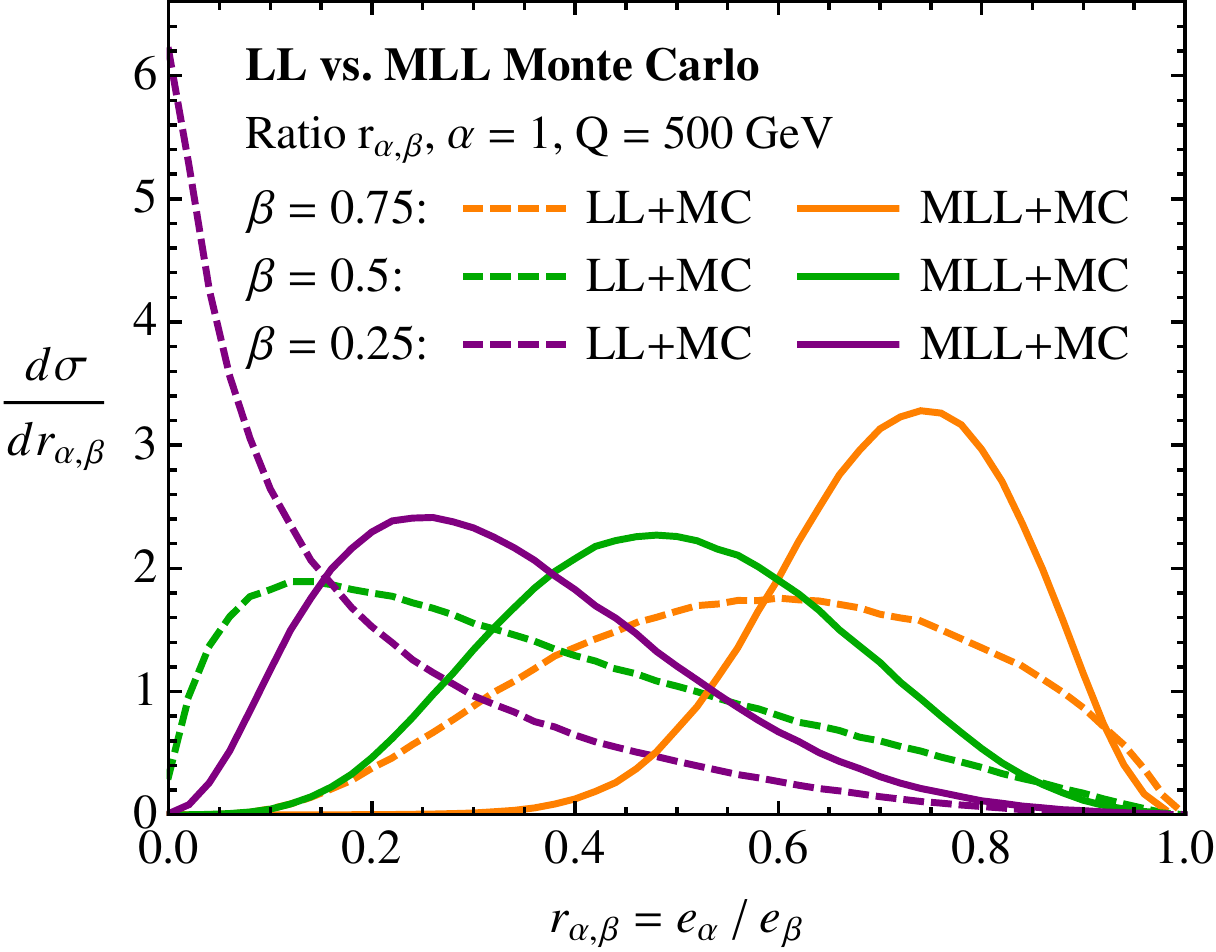}
}
\end{center}
\caption{Comparing the distribution of the ratio $r_{\alpha, \beta} \equiv e_\alpha/e_\beta$ between the LL+MC shower with fixed $\alpha_s$ (dashed) and the MLL+MC shower with running $\alpha_s$ (solid).   Shown are $\alpha=2$ (left) and $\alpha=1$ (right), sweeping over $\beta$.   Running $\alpha_s$ increases the Sudakov suppression, pushing the distributions away from $r_{\alpha,\beta} = 0$.  Because of multiple emissions, the Sudakov supression is even more enhanced than in the MLL results from \Fig{fig:mll_plot}.
}
\label{fig:ang_MCrunning_plot}
\end{figure}

Another qualitatively important effect is running $\alpha_s$, already seen in the difference between the LL and MLL curves in \Fig{fig:mll_plot}.  In our MLL+MC shower, we include both running $\alpha_s$ and subleading terms in the splitting function.  We compare the LL+MC and MLL+MC showers in \Fig{fig:ang_MCrunning_plot}, where the additional Sudakov suppression from the running coupling is apparent.  While not fully accurate to NLL level, the MLL+MC shower does include three key effects that show up at this order---running $\alpha_s$, subleading terms in the splitting function, and multiple emissions---and should give a good description of the qualitiative behavior of the ratio observable.  We stress that most publicly-available Monte Carlo programs include all of these effects by default.  \Fig{fig:ang_MCrunning_plot} illustrates the necessity of higher-order resummation for the accuracy of the distributions.

\section{Sensitivity to Non-Perturbative Physics}
\label{sec:nonpert}

Thus far, our discussion has focused on perturbatively calculable aspects of the ratio observable, where Sudakov safety ensures sensible resummed distributions.  We now turn to the important question of non-perturbative effects.  Given a hard scattering at an energy $Q$, these effects can potentially be order one.  Here we show that at sufficiently high energies, non-perturbative corrections scale like $\Lambda/Q$ to some positive (possibly fractional) power, where $\Lambda \simeq 0.5~\text{GeV}$ is a characteristic non-perturbative scale.  As long as these corrections fall sufficiently fast with respect to $Q$, then one can say that perturbatively calculated distributions will be robust to non-perturbative corrections.  

To prove that non-perturbative effects are suppressed by inverse powers of the energy $Q$ requires the existence of a factorization theorem.  To date, all such factorization theorems are formulated for IRC-safe observables where fixed-order cross sections exist.  Because the ratio observable is not IRC safe, though, one might worry that it is not only sensitive to non-perturbative physics, but in such a way that the corrections are independent of the energy $Q$ of the jet.  If this were to be the case, then even at arbitrarily high energies, non-perturbative effects could not be neglected for an accurate description of the observable.  As of yet, though, no factorization theorem exists to provide a definitive answer for how non-perturbative corrections will affect the ratio observable.\footnote{We suspect that the right strategy is to prove a factorization theorem for the double differential cross section and then project onto the ratio $r_{\alpha,\beta}$ appropriately.}  

Nevertheless, we will use some simple quantitative assumptions to argue that non-pertur\-bative corrections to the ratio distribution will be small.\footnote{We thank Iain Stewart,  Duff Neill, and Gavin Salam for extensive discussions of these points.} The strategy of our argument is to break up the integral in  \Eq{eq:ratio_dist_def}---which defines the cross section of the ratio observable from the double differential cross section---into two pieces.  For one piece, the non-perturbative effects manifest themselves as corrections to the perturbative cross section suppressed by (fractional) powers of $1/Q$.  This follows from the assumption that the double differential cross section has a valid operator product expansion (OPE) in this region of phase space.  The second piece is the direct contribution to the cross section from the non-perturbative region.  
At sufficiently high energies, we argue that the contribution from this region to the cross section of the ratio observable is exponentially suppressed, as a consequence of Sudakov safety.  These observations provide strong evidence that the ratio distribution has only power-suppressed dependence on non-perturbative physics at sufficiently high energies.

\subsection{The Shape Function}

We first review how non-perturbative physics affects IRC-safe observables like the jet angularities.  There are many strategies to gain a quantitative understanding of non-perturbative effects \cite{Akhoury:1995sp,Dokshitzer:1995zt,Dokshitzer:1995qm,Dokshitzer:1998pt,Gardi:1999dq,Gardi:2001ny,Gardi:2002bg,Lee:2006nr}, with perhaps the most general method being the shape function \cite{Korchemsky:1999kt,Korchemsky:2000kp}.  The shape function encodes non-perturbative physics contributions to an observable and is independent of the energy scale at which the observable is evaluated.  To find the cumulative distribution of an IRC-safe observable, one convolves the perturbative cumulative distribution with the shape function.  For recoil-free angularities $e_\alpha$ or other additive observables, this takes the form\footnote{\label{foot:pc}This form of the convolution is only valid for angularities with $\alpha>1$.  To get the correct scaling of the power corrections in \Eq{eq:deltaNP} for $\alpha\leq 1$, the form of the shape function and convolution is different.}
\begin{equation}
\Sigma(e_\alpha) = \int_0^{e_\alpha Q} d\epsilon \, f(\epsilon)\, \Sigma^\text{pert}\left( e_\alpha - \frac{\epsilon}{Q} \right) \ ,
\end{equation}
where $f(\epsilon)$ is the shape function (generically different for each observable) and $Q$ is the energy of the jet.  The differential cross section for $e_\alpha$ can then be computed by differentiating the cumulative distribution.

The shape function only has support in an energy range of order the QCD scale $\Lambda$.  For values of $e_\alpha$ such that $e_\alpha Q \gg \Lambda$, the OPE is valid and the cumulative distribution can therefore be expanded in derivatives of the perturbative cumulative distribution (see \Eq{eq:ang_powcorr} below). The precise scaling of the power corrections depends on whether the observable is most sensitive to non-perturbative energies or angles.  When a non-perturbative emission contributes to the angularity $e_\alpha$, this means that either the energy fraction $z$ or the splitting angle $\theta$ (or both) is non-perturbative:
\begin{equation}
z,\, \theta \lesssim \frac{\Lambda}{Q} \ .
\end{equation}
For values of the angular exponent $\alpha>1$, the angularities are most sensitive to the non-perturbative energy.  Because $e_\alpha = z\theta^\alpha$ with one emission, the power correction scales as $\Lambda/Q$.  For $\alpha<1$, the angularities are most sensitive to the non-perturbative splitting angle, and so it is expected that the power corrections scale as $(\Lambda / Q)^\alpha$ \cite{Manohar:1994kq,Banfi:2004yd}.\footnote{This would only be true for the recoil-free angularities measured about the broadening axis.  For recoil-sensitive angularities measured about the jet axis, the power corrections for $\alpha<1$ would have an extra factor of $\log \Lambda/Q$.}  When $\alpha = 1$, the angularities are equally sensitive to the angle and the energy of the emission; this introduces an extra logarithm in the power corrections which then scale like $(\Lambda/Q)\log \Lambda/Q$.  All of these scalings can be nicely packaged in the formula\footnote{This form is perhaps a bit unrealistic since it has a finite $\alpha \to 0$ limit.  Instead, one probably expects large (unbounded) non-perturbative corrections as $\alpha$ approaches zero since the strict $\alpha = 0$ limit is IRC unsafe.}
\begin{equation}
\label{eq:deltaNP}
\delta_\text{NP}^\alpha = \frac{\frac{\Lambda}{Q}-\left(\frac{\Lambda}{Q}\right)^\alpha}{\alpha - 1} \ .
\end{equation}
It should be stressed that \Eq{eq:deltaNP} has not been derived from any model of non-perturbative physics, but merely encodes the expected scaling with $\Lambda/Q$ of the non-perturbative corrections as a function of $\alpha$.

For values of $e_\alpha$ such that $e_\alpha Q \gg \Lambda$, we have the expansion\footnote{Using the form of the shape function from \Ref{Korchemsky:2000kp} introduces boundary terms which depend on a factorization scale defining the separation of the perturbative and non-perturbative regions.  The expression of \Eq{eq:ang_powcorr}, which has no boundary terms, follows from an $\overline{\text{MS}}$ shape function as formulated in \Ref{Hoang:2007vb}.}
\begin{equation}\label{eq:ang_powcorr}
\Sigma^\text{OPE}(e_\alpha) = \Sigma^\text{pert}\left( e_\alpha \right) + c_1 \delta_\text{NP}^\alpha \frac{\partial}{\partial e_\alpha} \Sigma^\text{pert}\left( e_\alpha \right) + {\cal O}\left( \left( \delta_\text{NP}^\alpha \right)^2 \right) \ ,
\end{equation}
where $c_1$ is a constant.  For example, angularities with $\alpha>1$, where $\alpha = 2-a$ in standard language, have the well-known expansion \cite{Dokshitzer:1995zt,Lee:2006nr}
\begin{equation}\label{eq:th_powcorr}
\Sigma^\text{OPE}(e_\alpha) = \Sigma^\text{pert}\left( e_\alpha \right) + \frac{1}{\alpha-1} \frac{\Omega_1}{Q} \frac{\partial}{\partial e_\alpha} \Sigma^\text{pert}\left( e_\alpha \right) + {\cal O}\left(  \frac{\Lambda^2}{Q^2} \right) \ ,
\end{equation}
where $\Omega_1 \simeq  \Lambda$ is a universal (observable-independent) constant.\footnote{Strictly speaking, the universality of $\Omega_1$ is only true if angularities are measured in the $E$-scheme.  See \Refs{Salam:2001bd,Mateu:2012nk}.}   For values of $e_\alpha$ such that $e_\alpha Q \lesssim \Lambda$, the OPE is no longer valid and the entire shape function must be used to determine the effect of non-perturbative physics.  Note, however, that both the non-perturbative corrections in the OPE regime as well as the range of $e_\alpha$ over which the full shape function must be used decrease as (fractional) powers of $\Lambda / Q$, formally vanishing at arbitrarily high energies.  This property is a consequence of IRC safety and guarantees that the angularities can be reliably computed in perturbative QCD. 

\subsection{Non-Perturbative Effects on the Ratio Observable}

We will now use similar shape function arguments to study the effect of non-perturbative physics on the ratio observable.  Our key assumption is that there exists a shape function for the double cumulative distribution of angularities $e_\alpha$ and $e_\beta$, which presumably encodes non-perturbative correlations between $e_\alpha$ and $e_\beta$.  Given a shape function $f(\epsilon_1,\epsilon_2)$, the non-perturbative effects would be included by convolution:\footnote{As discussed in footnote \ref{foot:pc}, this form of the convolution would only be valid for $\alpha,\beta>1$.  The form of the convolution must be different if $\alpha$ or $\beta$ is less than 1 for the power corrections to have the scaling in \Eq{eq:deltaNP}.}
\begin{equation}\label{eq:dub_np}
\Sigma(e_\alpha,e_\beta) = \int_0^{e_\alpha Q} d\epsilon_1 \int_0^{e_\beta Q} d\epsilon_2 \, f(\epsilon_1,\epsilon_2)\,\Sigma^\text{pert}\left(e_\alpha-\frac{\epsilon_1}{Q},e_\beta-\frac{\epsilon_2}{Q}\right) \ .
\end{equation}
For $e_\alpha$ and $e_\beta$ sufficiently large such that $e_\alpha Q \gg \Lambda$ and $e_\beta Q \gg \Lambda$, the OPE is appropriate.  The double cumulative distribution can be written in the OPE regime as
\begin{align}\label{eq:ddist_ope}
\Sigma^\text{OPE}(e_\alpha,e_\beta)=& \ \Sigma^\text{pert}(e_\alpha,e_\beta) + c_{1,0}\delta_\text{NP}^\alpha\frac{\partial}{\partial e_\alpha}\Sigma^\text{pert}(e_\alpha,e_\beta)+ c_{0,1}\delta_\text{NP}^\beta\frac{\partial}{\partial e_\beta}\Sigma^\text{pert}(e_\alpha,e_\beta) \nonumber \\
&+{\cal O}\left( \left( \delta_\text{NP}^\alpha \right)^2,\left( \delta_\text{NP}^\beta \right)^2 \right) \ ,
\end{align}
where the constants $c_{1,0}$ and $c_{0,1}$ depend on the corresponding observables.  In \App{app:powcorr}, we discuss the simple relationship of these coefficients to the power corrections of the individual angularities.  For $\alpha > \beta$, one generically expects the $\delta_\text{NP}^\beta$ term to dominate the power correction at large $Q$ (see \Tab{tab:exponents} below).

Assuming that \Eq{eq:dub_np} is valid, the double differential cross section $d^2 \sigma/ d e_\alpha d e_\beta$ can be found by differentiating $\Sigma(e_\alpha,e_\beta)$ with respect to $e_\alpha$ and $e_\beta$.  Then, the differential cross section for the ratio observable  $d\sigma/dr_{\alpha,\beta}$ can be computed by marginalizing according to \Eq{eq:ratio_dist_def}, automatically accounting for non-perturbative effects from the shape function $f(\epsilon_1,\epsilon_2)$.  Concretely,
\begin{align}\label{eq:np_rdist}
\frac{d\sigma}{dr} & = \int_0^1 de_\beta \,\left. e_\beta \left( \frac{\partial}{\partial e_\alpha}\frac{\partial}{\partial e_\beta}\Sigma(e_\alpha,e_\beta)  \right) \right|_{e_\alpha = re_\beta} \ .
\end{align}
Here, we leave the phase space boundaries implicit in the double differential cross section such that $e_\beta$ is integrated over its entire range.  We emphasize again that the only assumption we have made thus far is that the shape function $f(\epsilon_1,\epsilon_2)$ describes the dominant power corrections.

We now study \Eq{eq:np_rdist} to argue that the non-perturbative corrections to the cross section for the ratio observable are suppressed by (fractional) powers of $\Lambda/Q$ for sufficiently large $Q$.   We can break this integral into two parts: one part in which the OPE expansion is valid and another where the full shape function must be used:
\begin{align}\label{eq:rnp_exp}
\frac{d\sigma}{dr}=&\int_{\eta /Q}^1 de_\beta \,\left. e_\beta \left( \frac{\partial}{\partial e_\alpha}\frac{\partial}{\partial e_\beta}\Sigma^\text{OPE}(e_\alpha,e_\beta) \right) \right|_{e_\alpha = re_\beta} \nonumber \\
&+\int_0^{\eta /Q} de_\beta \,\left. e_\beta \left( \frac{\partial}{\partial e_\alpha}\frac{\partial}{\partial e_\beta}\Sigma(e_\alpha,e_\beta) \right) \right|_{e_\alpha = re_\beta} \ .
\end{align}
Here, $\eta$ is a fixed energy scale above which the OPE is valid: $Q\gg\eta \gg \Lambda$.  
From \Eq{eq:ddist_ope}, it is clear that the contribution from the first line to the differential cross section of $r$ (the ``OPE'' contribution) has only power-suppressed dependence on non-perturbative physics.  Because there is an explicit non-zero lower bound of this integral, one can think of this term as corresponding to the IRC-safe modification of $r_{\alpha,\beta}$ in \Eq{eq:ircsaferwithepsilon}.  Like with an ordinary IRC-safe observable, this term has a valid Taylor expansion in $\alpha_s$. 

The second line in \Eq{eq:rnp_exp} probes the non-perturbative region directly and so its contribution to the cross section is more subtle.  From Sudakov safety, however, we know that the perturbative double cumulative distribution is exponentially suppressed at small values of $e_\beta$ due to the Sudakov factor.  Regardless of the shape function $f(\epsilon_1,\epsilon_2)$, the second line of \Eq{eq:rnp_exp} will be exponentially suppressed at sufficiently high energies $Q$.  Because this suppression requires the existence of the Sudakov factor, there is not a valid Taylor series in $\alpha_s$ here.   As $\alpha_s\to 0$ and the Sudakov factor weakens, the contribution from the singular region becomes important over an increasingly large energy range.  Indeed, if one tries to perform an expansion in $\alpha_s$, the non-perturbative corrections will have an essential singularity in $\alpha_s$.  At exactly $\alpha_s = 0$, the Sudakov factor provides no suppression and the contribution from the singular region dominates the observable, rendering the perturbative calculation irrelevant.  For any finite value of $\alpha_s$, though, the Sudakov factor exponentially suppresses any non-perturbative effects in the second term in \Eq{eq:rnp_exp}.  

Of course, for a fixed value of $Q$, one can always find a value of $r$ where the physics is entirely non-perturbative (just as one can always find a value of the angularity $e_\alpha \lesssim \delta_\text{NP}^\alpha$ where non-perturbative effects dominate).  But for a fixed value of $r$ (or a fixed value of $e_\alpha$), Sudakov safety ensures that one can always go to a high enough energy $Q$ such that the OPE expansion dominates the non-perturbative description.

We can estimate the energy scale $Q_\text{Sud}$ above which Sudakov safety controls non-pertur\-bative effects.  To do this, we will compare the relative size of the two terms from \Eq{eq:rnp_exp}.  This requires knowledge of the shape function describing the non-perturbative physics; however, all we want to show is that the perturbative Sudakov factor is sufficient to exponentially damp the contribution from the term integrated over the region $e_\beta \in [0,\eta/Q]$.  Thus, we will study the ratio of the two terms computed with the perturbative LL double differential cross section, as defined in \Eq{eq:resum_double}.  We find
\begin{equation}
\frac{\int_0^{\eta /Q} de_\beta \,\left. e_\beta \left( \frac{\partial}{\partial e_\alpha}\frac{\partial}{\partial e_\beta}\Sigma^\text{LL}(e_\alpha,e_\beta) \right) \right|_{e_\alpha = re_\beta}}{\int_{\eta /Q}^{r^{\frac{\beta}{\alpha-\beta}}} de_\beta \,\left. e_\beta \left( \frac{\partial}{\partial e_\alpha}\frac{\partial}{\partial e_\beta}\Sigma^\text{LL}(e_\alpha,e_\beta) \right) \right|_{e_\alpha = re_\beta}} = \frac{\pi \sqrt{\beta}}{2\sqrt{\alpha_s}\sqrt{C_F}\left(\log r^{\frac{\beta}{\alpha-\beta}} - \log\frac{\eta}{Q}\right)} +{\cal O}(\alpha_s^0) \ .
\end{equation}
Assuming that $\log r \simeq 0$, the ratio is small when
\begin{equation}\label{eq:Qsud}
Q\gg \eta \,e^{\frac{\pi \sqrt{\beta}}{2\sqrt{\alpha_s}\sqrt{C_F}}} \equiv Q_\text{Sud} \ .
\end{equation}
For values of energy $Q\gg Q_\text{Sud}$, the contribution from the non-perturbative region is exponentially suppressed by the perturbative Sudakov factor and the non-perturbative corrections are dominantly described by the OPE.

The form of $Q_\text{Sud}$ in \Eq{eq:Qsud} manifests the essential singularity in $\alpha_s$ mentioned earlier.  As $\alpha_s\to 0$, the energy scale $Q_\text{Sud}$ becomes arbitrarily large, but for finite $\alpha_s$, $Q_\text{Sud}$ can take on reasonable values.  Because the LL expression has fixed $\alpha_s$ and does not include multiple emissions, the resulting estimate for $Q_\text{Sud}$ is somewhat conservative.  For $\beta=1$, $\alpha_s=0.12$, and energy scale $\eta = 5$ GeV, $Q_\text{Sud}$ is approximately 250 GeV.  Nevertheless, we must assume that the LL expression dominates at sufficiently small values of $e_\beta$ to guarantee that this contribution is exponentially suppressed at high energies.  In the numerical estimates below, we find that Sudakov safety is effective typically at $Q_\text{Sud} \simeq 100\text{--}1000~\text{GeV}$ depending on the choice of $\alpha$ and $\beta$.

\subsection{Numerical Analysis}
\label{subsec:numericalNP}

To test the suppression of non-perturbative physics on the ratio observable, we implement a simple shape function which just shifts each of the angularities by an amount $\delta_\text{NP}^\alpha$ from \Eq{eq:deltaNP}.  That is, for $e_\alpha \gg \delta_\text{NP}^\alpha$ and $e_\beta \gg \delta_\text{NP}^\beta$, 
we take the effect of this shape function on the perturbative cumulative double distribution to be
\begin{equation}
\label{eq:toyshape}
\Sigma\left( e_\alpha,e_\beta \right) = \Sigma^\text{pert}\left( e_\alpha- \delta_\text{NP}^\alpha,e_\beta - \delta_\text{NP}^\beta \right) \ . 
\end{equation}
A more realistic shape function would be a full distribution that accounts for correlations between $e_\alpha$ and $e_\beta$, but \Eq{eq:toyshape} is sufficient to understand the scaling of the non-perturbative effects with $Q$.

For an accurate modeling of higher-order physics not included in our LL analysis, we generate the perturbative distribution with our MLL+MC parton shower from \Sec{sec:MC_resum}.  In particular, one-loop running of $\alpha_s$ is needed to make sure the scaling with energy is realistic.  We terminate the Monte Carlo when the scale of an emission falls below $1$ GeV and we set the scale of non-perturbative physics to be $\Lambda = 0.5$ GeV.  At the end of the shower, an amount $\delta_\text{NP}^\alpha$ is added to the perturbative value of the angularity $e_\alpha$ to account for non-perturbative physics.  

\begin{figure}[]
\begin{center}
\subfloat[]{\label{fig:llshape_500_2}
\includegraphics[width=7.0cm]{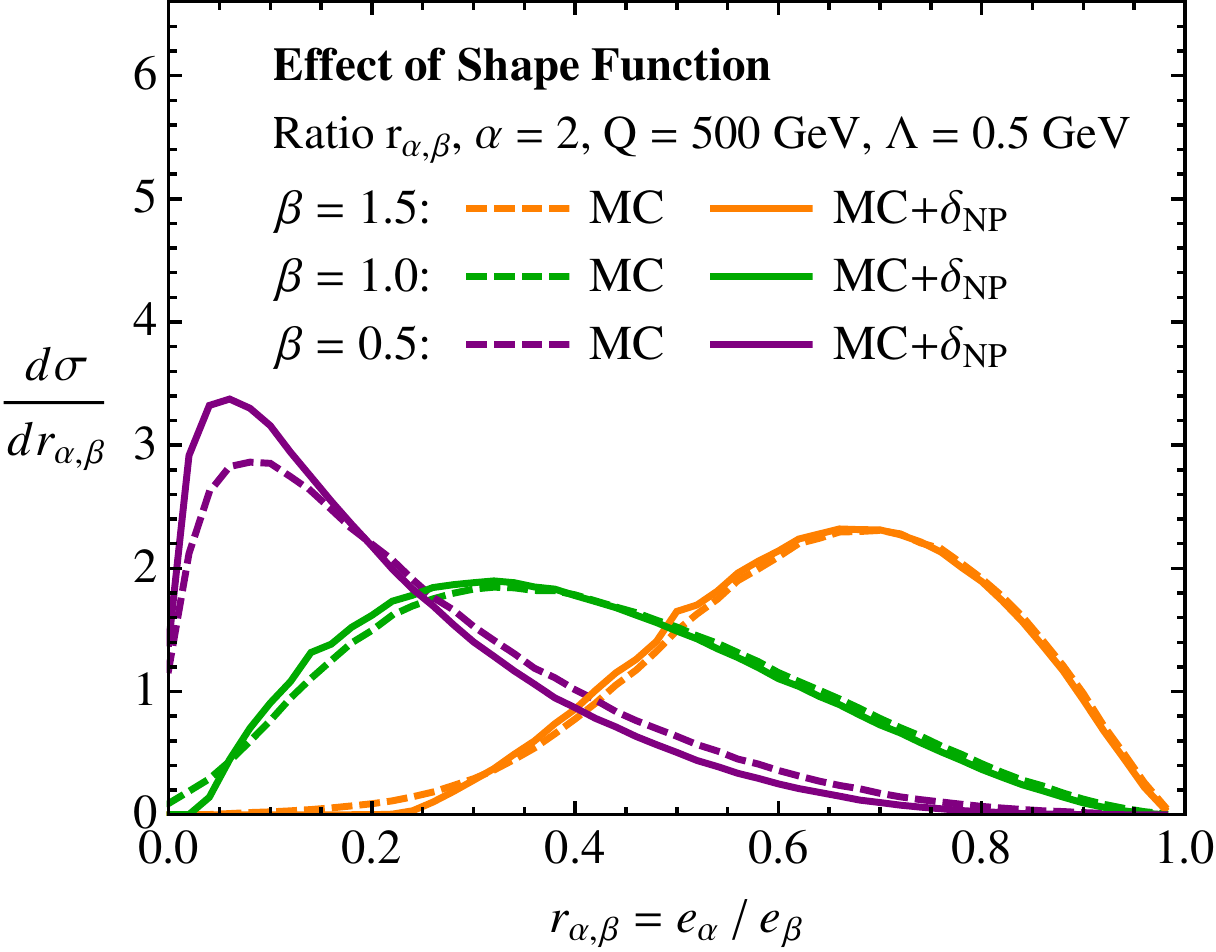}
}
$\quad$
\subfloat[]{\label{fig:llshape_500_1} 
\includegraphics[width=7.0cm]{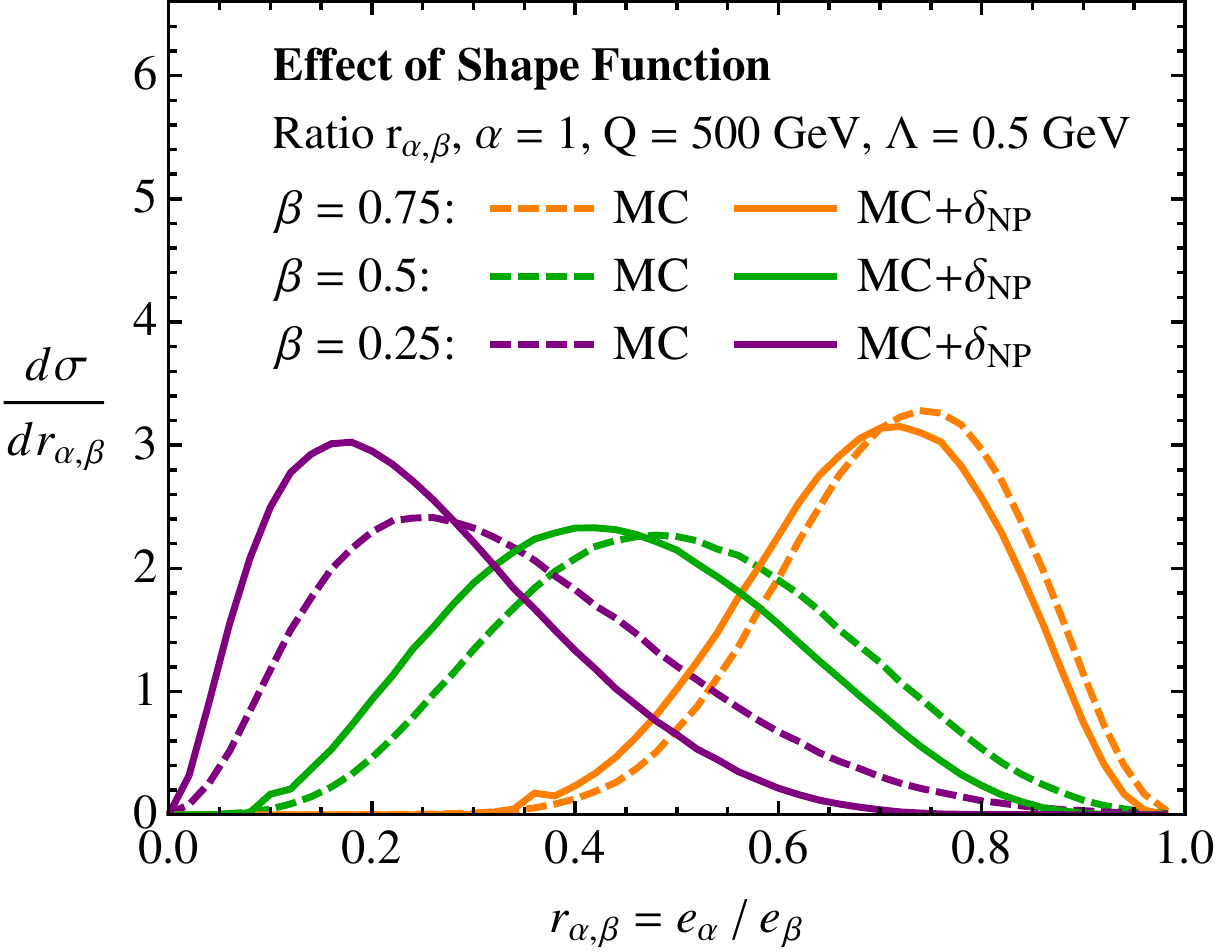}
}\\
\subfloat[]{\label{fig:llshape_5000_2} 
\includegraphics[width=7.0cm]{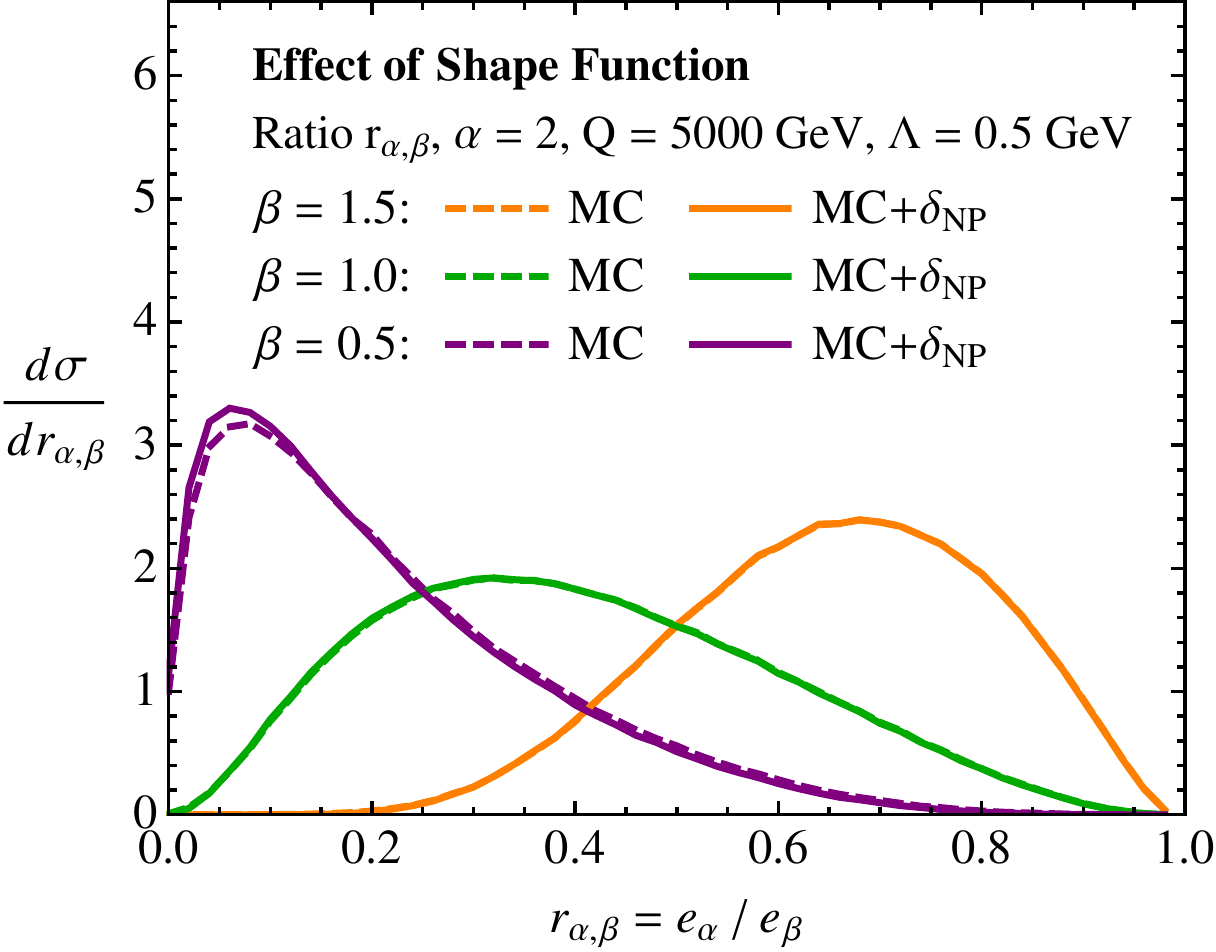}
}
$\quad$
\subfloat[]{\label{fig:llshape_5000_1} 
\includegraphics[width=7.0cm]{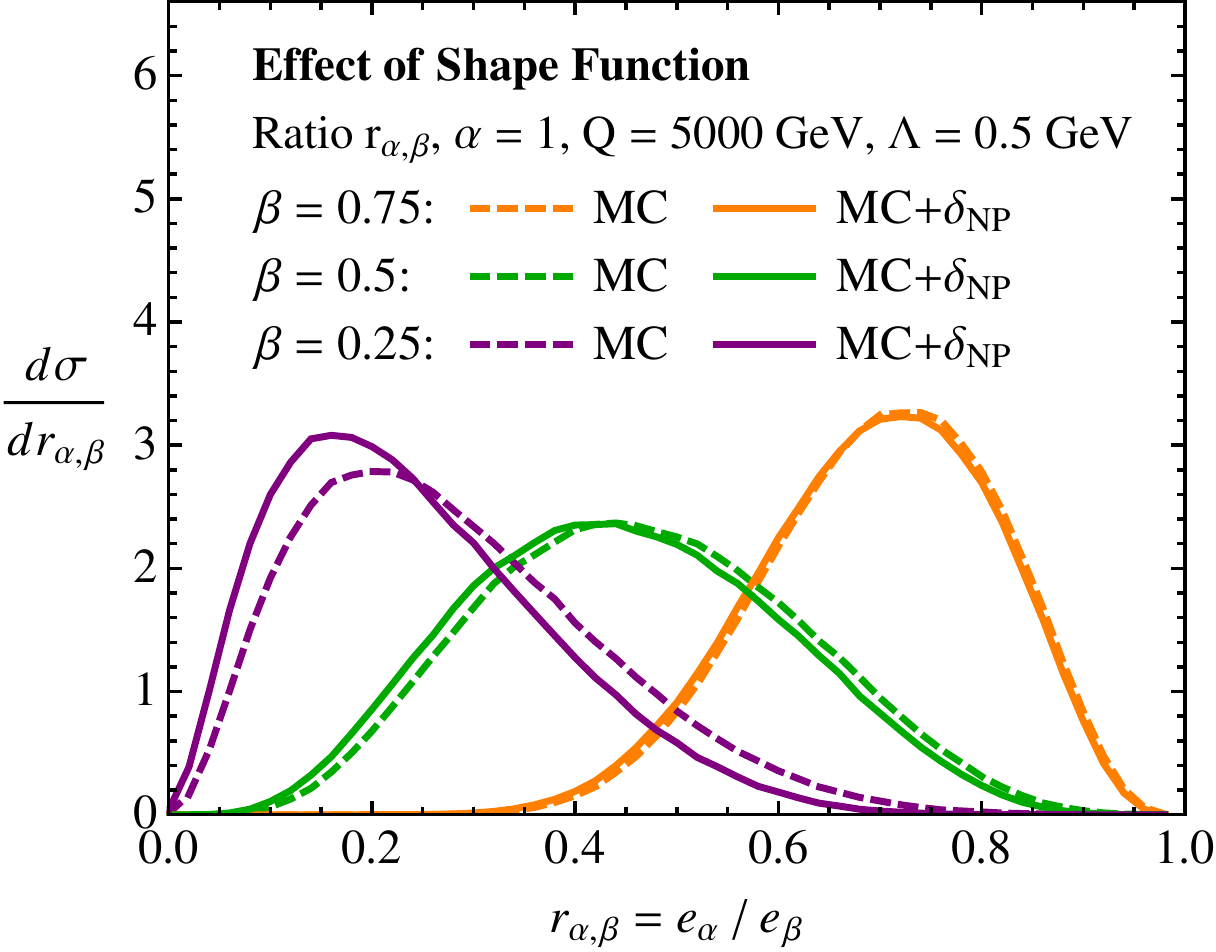}
}
\end{center}
\caption{Convolving the MLL+MC results for the ratio $r_{\alpha,\beta}$ in \Sec{sec:MC_resum} with the toy shape function in \Eq{eq:toyshape} with $\Lambda = 0.5$ GeV.  Shown are $Q = 500$ GeV (top row) and $Q = 5000$ GeV (bottom row), with either $\alpha=2$ (left column) or $\alpha = 1$ (right column), sweeping $\beta$.  As expected, the non-perturbative corrections fall off at sufficiently high energies.  In these plots, ``MC'' refers to the MLL+MC shower.
}
\label{fig:llshape_plot}
\end{figure}

We show the effects of adding non-perturbative corrections in \Fig{fig:llshape_plot}, plotting two different jet energies, $Q=500$ GeV and $Q=5000$ GeV.  Because the scaling of the power corrections in \Eq{eq:deltaNP} depends on the angular exponents, the different ratio curves $r_{\alpha,\beta}$ are sensitive to different power corrections. Overall the non-perturbative corrections to the ratio are small and decrease significantly as the energy of the jet increases.

\begin{figure}[]
\begin{center}
\subfloat[]{\label{fig:angrat500_py_2}
\includegraphics[width=7.0cm]{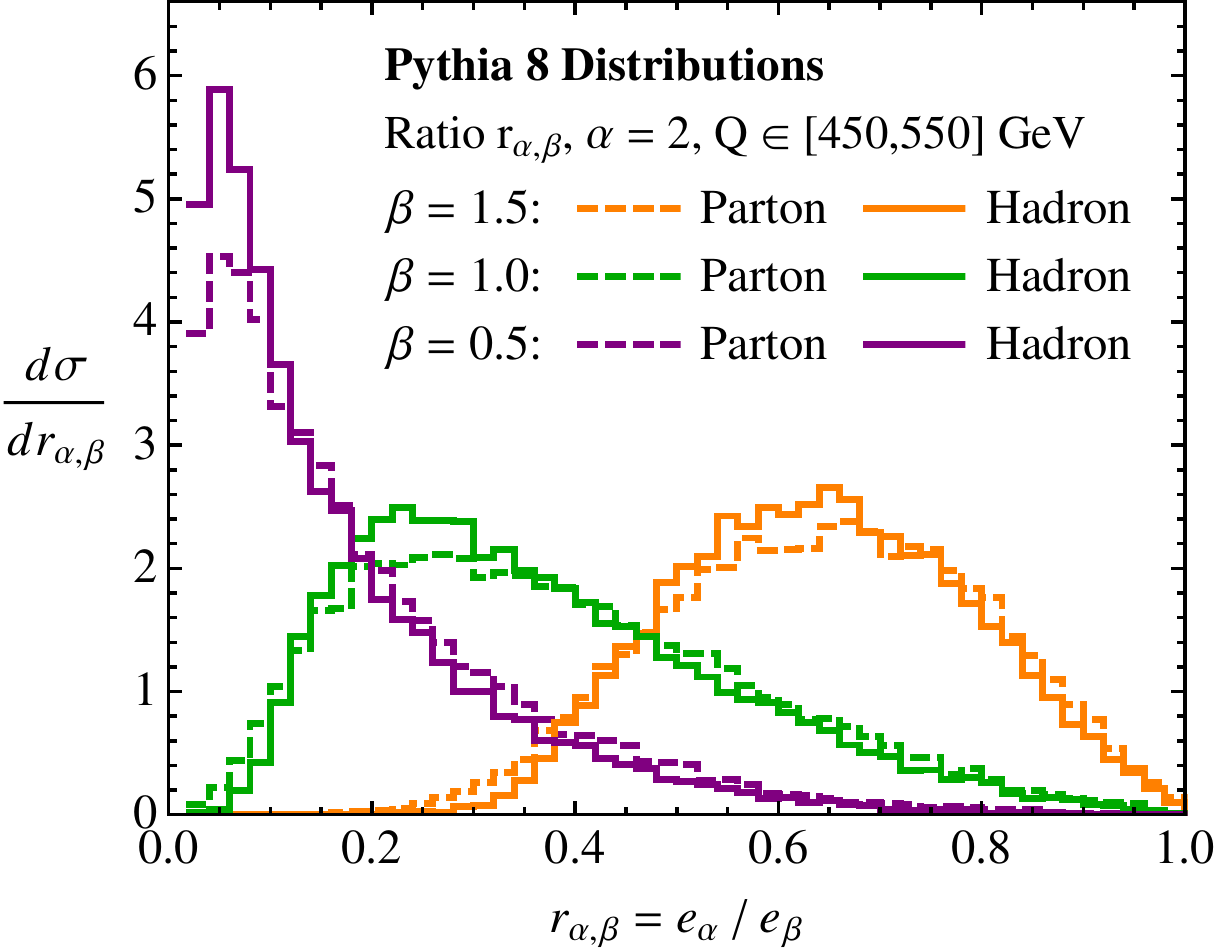}
}
$\quad$
\subfloat[]{\label{fig:angrat500_py_1} 
\includegraphics[width=7.0cm]{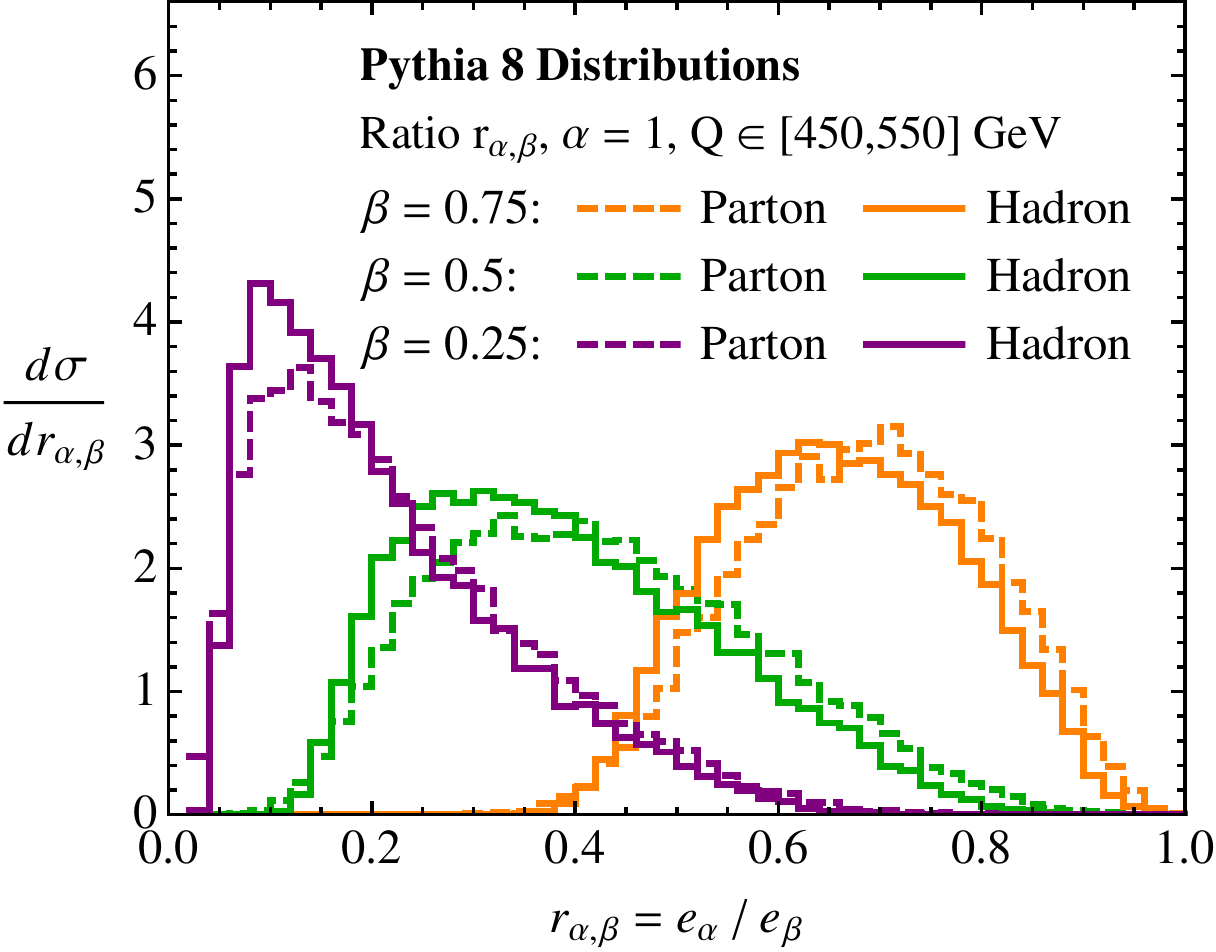}
}\\
\subfloat[]{\label{fig:angrat5000_py_2} 
\includegraphics[width=7.0cm]{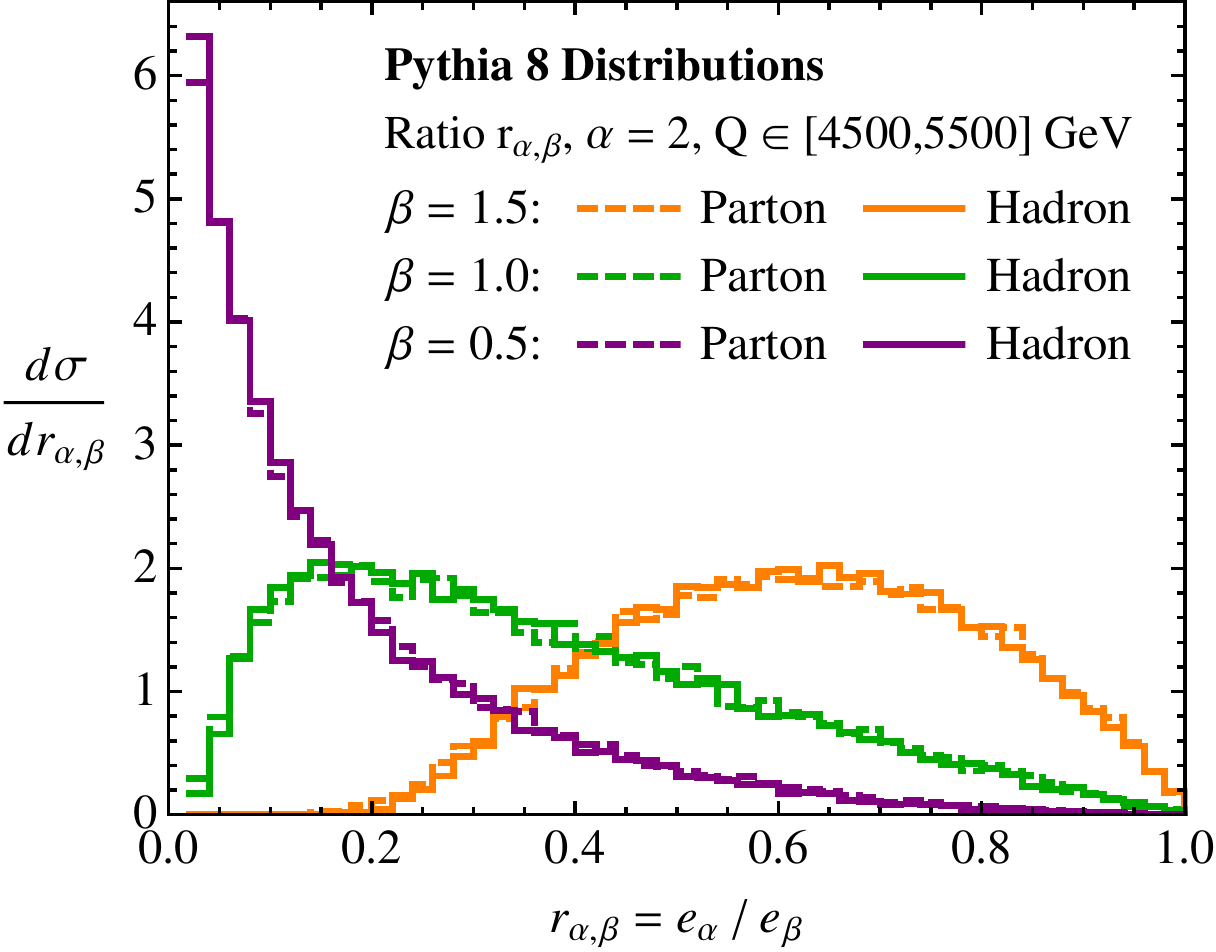}
}
$\quad$
\subfloat[]{\label{fig:angrat5000_py_1} 
\includegraphics[width=7.0cm]{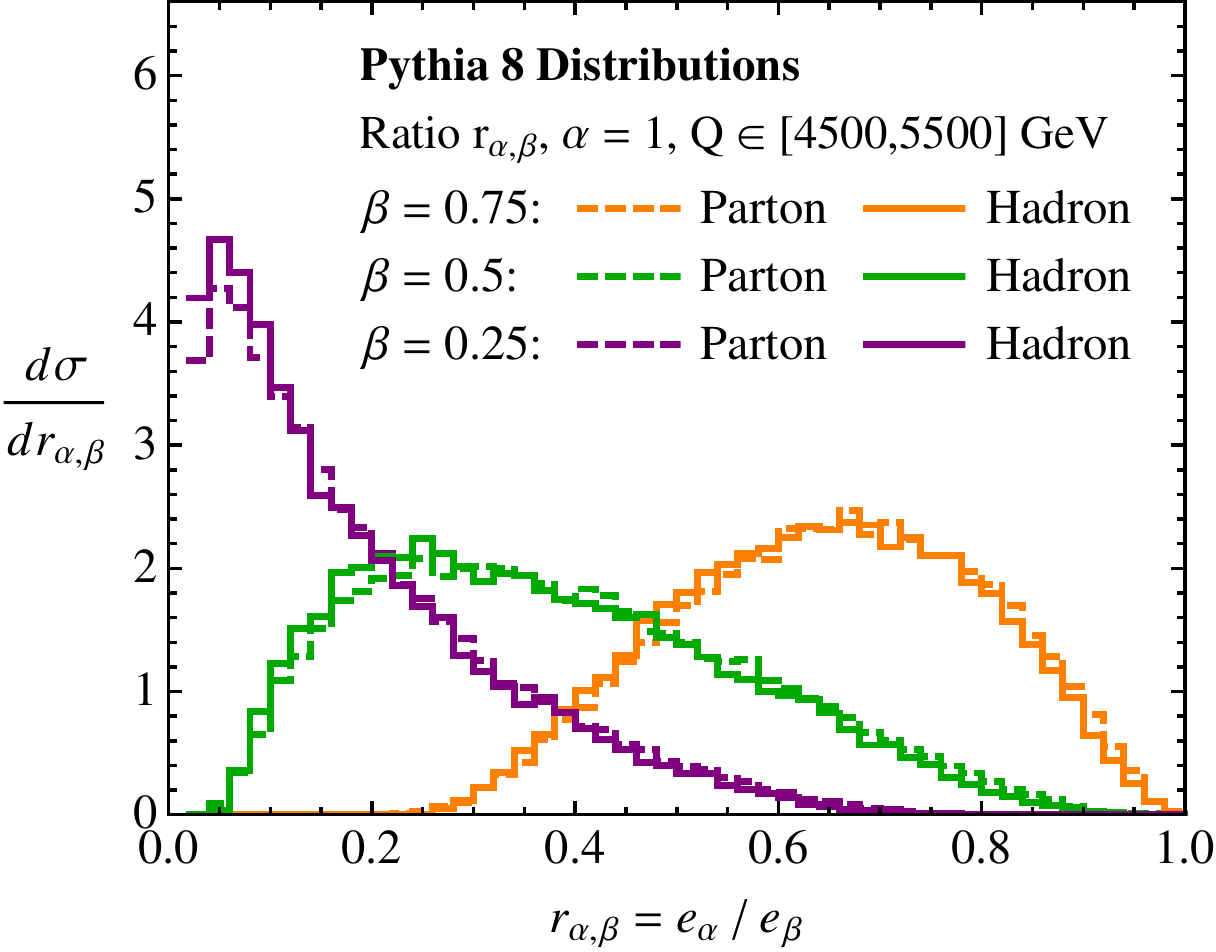}
}
\end{center}
\caption{Effect of hadronization on the ratio $r_{\alpha,\beta}$ as simulated in \pythia{8.165}.  The sample consists of the hardest jet from $e^+e^- \to q\bar{q}$ events found by the anti-$k_T$ algorithm, keeping all particles that lie within a radius $R_0=1.0$ of the broadening axis.  Shown are jets that lie in the energy range $Q = [450,550]$ GeV (top row) and $Q = [4500,5500]$ GeV (bottom row), with either $\alpha=2$ (left column) or $\alpha = 1$ (right column), sweeping $\beta$.  The dashed (solid) curves are the distribution at parton (hadron) level.  These results qualitatively agree with the MLL+MC analysis in \Fig{fig:llshape_plot}.
}
\label{fig:ang_MC}
\end{figure}

As a cross check of our analysis, we can study the effect of non-perturbative physics in a  full Monte Carlo simulation by observing the sensitivity of the cross section to hadronization.   We generate $e^+e^- \to q \bar{q}$ events simulated with \pythia{8.165} \cite{Sjostrand:2006za,Sjostrand:2007gs} at center-of-mass energies of 1 and 10 TeV with hadronization turned on and off.\footnote{The quarks that are produced are only $u$, $d$, or $s$, so mass effects should be minimal.  Apart from the turning hadronization on and off, we use the default \pythia{8} settings.}  To analyze the jets, we cluster jets with the $e^+e^-$ anti-$k_T$ algorithm \cite{Cacciari:2008gp} with \fastjet{3.0.3} \cite{Cacciari:2011ma} with a fat jet radius $R_0 = 1.5$.  We analyze only the hardest jet in the event, requiring that the cosine of the angle between the jet axis and the initiating hard parton be greater than $0.9$.  In keeping with the discussion in \Sec{sec:broadening}, we only include particles that lie within an angle $R_0=1.0$ from the broadening axis of the hardest jet.  The energy of the jets is required to be in the range of $Q\in [450,550]$ GeV for the 1 TeV sample and $Q\in [4500,5500]$ GeV for the 10 TeV sample.  We then measure the recoil-free angularities for various values of the angular exponent $\alpha$ of the jets in the sample.  The \pythia{8} results are shown in \Fig{fig:ang_MC}, which agree qualitatively with our MLL+MC study in \Fig{fig:llshape_plot}.

Finally, we can study the energy dependence of the non-perturbative corrections directly.  To do this, we compare moments of the distribution of the ratio observable with and without the inclusion of a shape function.  The moments of the ratio observable are defined as
\begin{equation}
\langle r^n \rangle = \frac{1}{\sigma}\int_0^1 dr \, r^n \, \frac{d\sigma}{dr} \ .
\end{equation}
In our analysis, we will focus on the mean $\mu=\langle r \rangle$ and the variance $\sigma^2=\langle r^2 \rangle - \langle r \rangle^2$.  These combinations of moments provide a probe into power corrections at different orders in $\Lambda/Q$.  For additive IRC-safe observables like thrust, it was noted in \Ref{Abbate:2012jh} that while the mean value of thrust receives power corrections starting at order $\Lambda/Q$, the variance of the thrust distribution first receives power corrections at order $\Lambda^2/Q^2$.  This is a consequence of the form of the leading power corrections to thrust.\footnote{For the simple shape function we use here, the mean value of angularities $e_\alpha$ is shifted by $\delta_\text{NP}^\alpha$, while the variance is unchanged because the shape function merely translates the angularity distribution.  For a more complicated shape function, the variance will indeed receive corrections beginning at $\left(\delta_\text{NP}^\alpha\right)^2$ order.  In addition, there exists power corrections suppressed by $\alpha_s$ starting at $\alpha_s \delta_\text{NP}^\alpha$.}  In contrast, the ratio observable is not additive, so we do not expect $\mu$ or $\sigma^2$ to have dramatically different dependence on $\Lambda/Q$.

\begin{figure}[]
\begin{center}
\subfloat[]{\label{fig:llshape_qvar2}
\includegraphics[width=7.0cm]{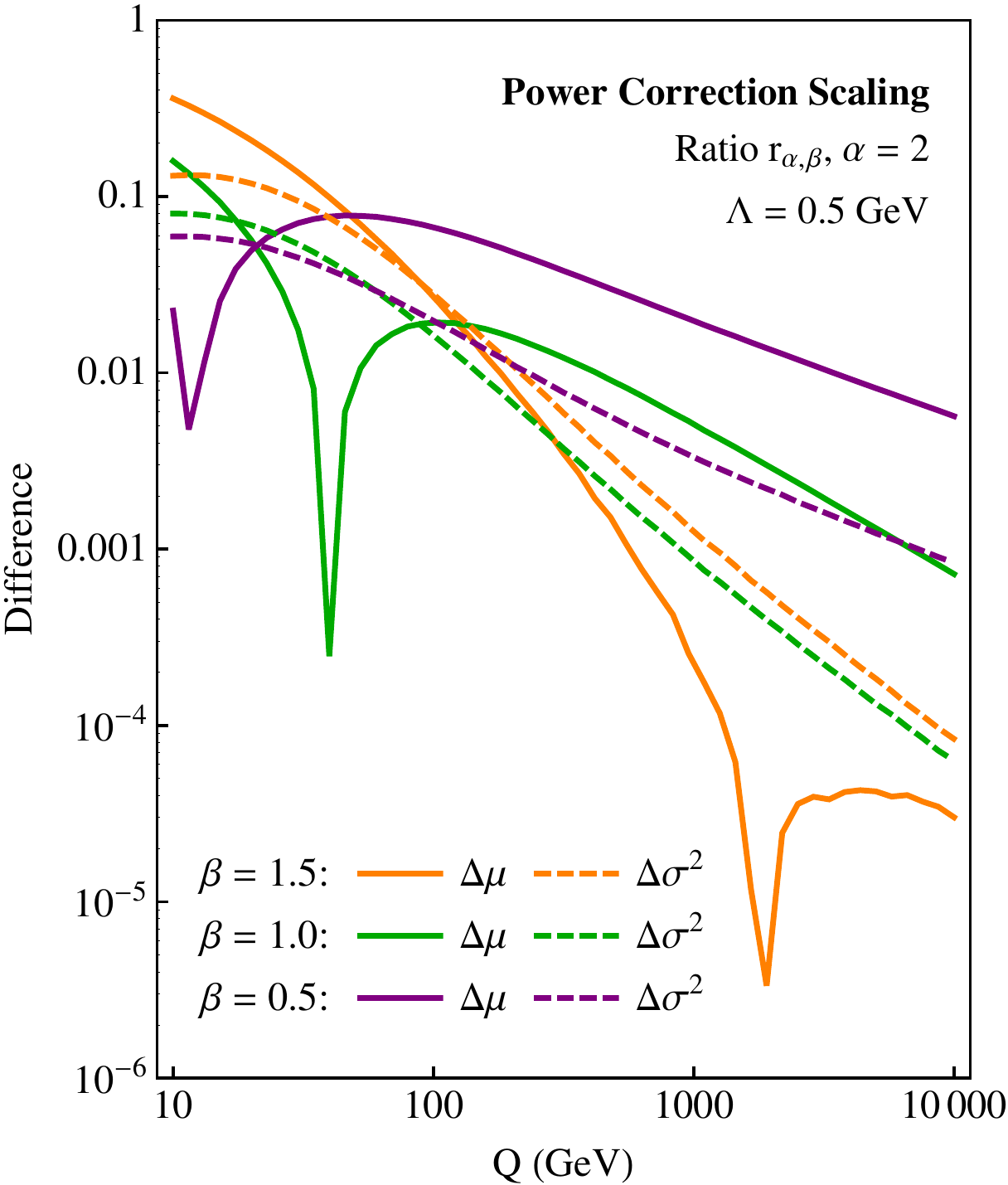}
}
$\quad$
\subfloat[]{\label{fig:llshape_qvar1} 
\includegraphics[width=7.0cm]{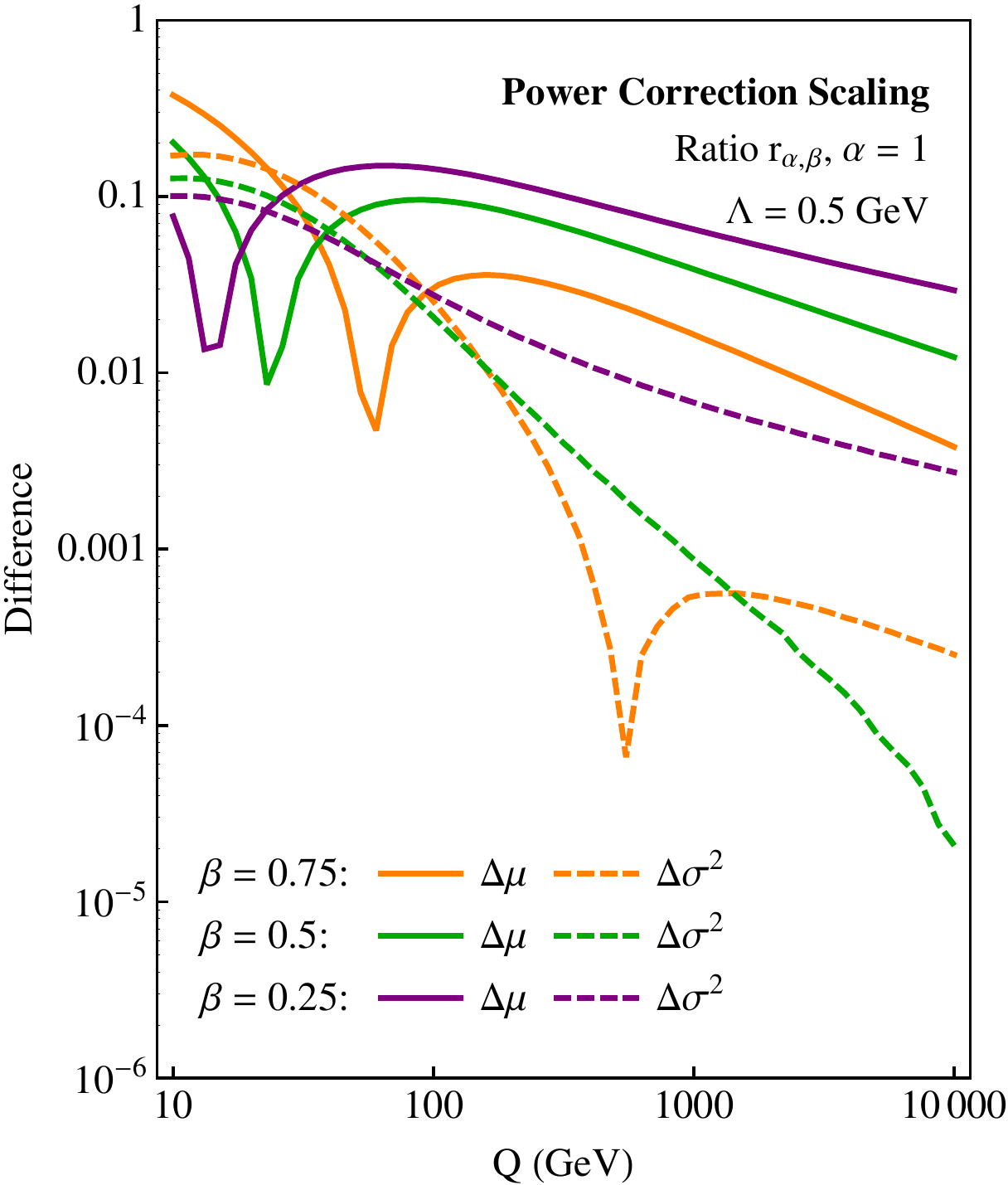}
}
\end{center}
\caption{Energy dependence of the non-perturbative corrections to the mean $\mu=\langle r \rangle$ and the variance $\sigma^2=\langle r^2 \rangle - \langle r \rangle^2$ of the MLL+MC ratio distribution.  Shown is difference between the value with and without the inclusion of the shape function as a function of the energy $Q$ of the jet, with $\alpha = 2$ (left) and $\alpha = 1$ (right), sweeping $\beta$.  The scale of non-perturbative physics is set to $\Lambda = 0.5$ GeV.  The sharp dips correspond to places where the non-perturbative corrections are accidentally small.  The power corrections to $\mu$ and $\sigma^2$ indeed scale like a power of $\Lambda/Q$ (i.e. linear dependence on a log-log plot).}
\label{fig:llshape_qvar}
\end{figure}

\begin{table}
\begin{center}
\begin{tabular}{|r||ccc|ccc|}
\hline
& \multicolumn{3}{|c|}{$\alpha = 2$}  & \multicolumn{3}{|c|}{$\alpha = 1$}\\
 & $\beta = 1.5$ & $\beta = 1.0$ & $\beta = 0.5$ & $\beta = 0.75$ & $\beta = 0.5$ & $\beta = 0.25$\\
 \hline \hline
$\gamma_{\Delta \mu}$ & --- &  0.86 & 0.53 & 0.64 &0.50 & 0.33 \\
$\gamma_{\Delta \sigma^2}$ & 1.15 & 1.10 & 0.58 &   0.50 & --- & 0.35 \\
\hline
\end{tabular}
\end{center}

\caption{Scaling exponent of the $(\Lambda/Q)^\gamma$ power correction to the mean $\mu$ and variance $\sigma^2$, estimated from the large $Q$ behavior of the curves in \Fig{fig:llshape_qvar}.  Omitted entries correspond to ambiguous situations where there a sharp dip in the corresponding curve due to an accidental cancellation.  We see that the exponent $\gamma$ roughly follows the scaling of $\delta_\text{NP}^\beta$ (and not $\delta_\text{NP}^\alpha$), as expected from the discussion below \Eq{eq:ddist_ope}.}
\label{tab:exponents}
\end{table}

We illustrate the energy dependence of the non-perturbative corrections in \Fig{fig:llshape_qvar}.  Here, we plot the difference of the means $\Delta \mu$ and variances $\Delta \sigma^2$ between the MLL+MC distributions with and without the inclusion of the shape function as a function of the jet energy.  As shown in \Tab{tab:exponents}, all of the differences fall like a (fractional) power of $\Lambda/Q$ at very high energies, as expected from our earlier arguments.  Unlike for observables like thrust, non-perturbative corrections to the variance do not typically scale away more quickly with $Q$ than corrections to the mean, so the power corrections to the ratio observable do not manifest themselves as simple translations of the cross section by an amount $\Lambda/Q$.  Depending on the choice of $\alpha$ and $\beta$, corrections to the mean and variance are less than 10\% for $Q \gtrsim 100\text{--}1000~\text{GeV}$.  Taken together, the observations made in this section are strong evidence that the non-perturbative corrections to the ratio observable are small and decrease as inverse powers of energy.  This behavior, familiar from IRC-safe observables, follows from Sudakov safety.

\section{Conclusions}\label{sec:conc}

By explicitly computing the LL resummed double differential cross section of two angularities and marginalizing, we have shown that the cross section of the ratio $r_{\alpha,\beta} \equiv e_\alpha/e_\beta$ is well-defined in perturbative QCD.  This is in spite of the fact that the ratio is not IRC safe and so is undefined at any fixed-order in $\alpha_s$.  Instead, the ratio observable is ``Sudakov safe'', where logartihmic resummation suppresses the singular regions of phase space.  We have found that Monte Carlo parton showers resum the leading logarithms of the ratio cross section correctly, while also incorporating the important effect of multiple emissions which first arise at NLL order.  Multiple emissions enhance the Sudakov suppression as $r_{\alpha,\beta} \to 0$ and suppress the cross section near $r_{\alpha,\beta}=1$.  The accuracy of the cross section of the ratio observable can be systematically improved by computing the double differential cross section to higher orders in resummed perturbation theory and matching to fixed-order results.

Because the ratio observable $r_{\alpha,\beta}$ is not IRC safe, one has to check that non-perturbative corrections vanish sufficiently fast as the energy increases.  After all, even though we are able to compute the differential cross section of $r_{\alpha,\beta}$ in resummed perturbation theory, this result would be meaningless if non-perturbative physics dominated at arbitrarily high energies.  
By assuming the existence of a shape function for the double differential cross section of angularities for incorporating non-perturbative physics, we have shown that, at sufficiently high energies, the corrections decrease as inverse powers of the energy of the jet.  
This follows from Sudakov safety, where potentially large non-perturbative effects are exponentially suppressed at high energies because of the perturbative Sudakov factor.   We conjecture that any observable that has a Sudakov factor that suppresses the non-perturbative regime is Sudakov safe and can be computed reliably in resummed perturbation theory.
We leave a proof of a factorization theorem for the double differential cross section for later work.

We chose to study the ratio of angularities because of their simplicity, but we anticipate extending the discussion and considerations here to more phenomenologically motivated ratio-type observables.  Dimensionless ratios are ubiquitous in the study of jet substructure, including $N$-subjettiness \cite{Thaler:2010tr,Thaler:2011gf}, energy correlation function ratios \cite{Larkoski:2013eya}, planar flow \cite{Almeida:2008yp,Field:2012rw}, and angular correlation and structure functions \cite{Jankowiak:2011qa}.  
These interesting ratio observables are more complicated to study than the angularities, though, since they typically involve the ratio of two observables that are first non-zero at different orders in perturbation theory.  This results in complicated phase space considerations and relatively high-order calculations to determine the leading contributions to the observable.  To date, most studies of ratios observables rely on Monte Carlo predictions, so it is crucial to understand analytically to what extent these predictions can be trusted. 

As an example, we will outline the calculation of the ratio of $N$-subjettiness jet observables $\Nsub{2,1}{\beta}=\Nsub{2}{\beta}/\Nsub{1}{\beta}$ for arbitrary values of $\beta$.  $N$-subjettiness $\tau_N^{(\beta)}$ is defined as 
\begin{equation}
\label{eq:Nsubdef}
\Nsub{N}{\beta} =\sum_{i} p_{T i}\min\left\{  R_{1,i}^\beta, R_{2,i}^\beta,\dotsc, R_{N,i}^\beta   \right\} \ ,
\end{equation}
where the sum runs over all particles in the jet and $R_{A,i}$ is the distance from axis $A$ to particle $i$.\footnote{There are various ways to choose subjet axes, including minimizing $\Nsub{N}{\beta}$ over all possible axes directions.}
With a cut on the mass of a jet, $\tau_{2,1}^{(\beta)}$ is IRC safe.  A calculation of $\tau_{2,1}^{(2)}$ at fixed mass for boosted $Z$ bosons was presented in \Ref{Feige:2012vc}.  We are interested, however, in determining the double differential cross section of $\tau_1^{(\beta)}$ and $\tau_2^{(\beta)}$ with no mass cut, especially since this variable was measured by the ATLAS experiment in \Ref{ATLAS:2012am}.\footnote{Ultimately, we would like to understand the behavior of $\tau_{3,2}^{(\beta)}$ which is relevant for boosted top identification.  In that case, a cut on the jet mass does not regulate the denominator $\Nsub{2}{\beta}$, as pointed out in \Ref{Soyez:2012hv}.}

At fixed-order in perturbation theory, computing the double differential cross section requires an ${\cal O}(\alpha_s^2)$ matrix element because the observable $\tau_2^{(\beta)}$ is first non-zero for a jet with three constituents.  While this calculation would be more challenging than the analysis of angularities presented here, it is in principle straightforward.  The lowest-order distribution for planar flow (at fixed mass)---another observable that is first non-zero at ${\cal O}(\alpha_s^2)$---was presented in \Ref{Field:2012rw}.

To obtain the resummed double differential cross section presents different challenges.  Ideally, a resummation analysis would resum up through NLL order for sufficient accuracy; however, even the LL resummation would be interesting.   In the case of angularities, we found that there were two possible phase space regions which contributed to the LL distribution: when one emission dominated both angularities and when two different emissions set the two angularities.  For the case of $N$-subjettiness, the phase space regions that contribute at LL to the double differential distribution are more complicated.  Possible phase space regions include:
\begin{enumerate}
\item One emission dominating $\tau_1^{(\beta)}$; a second emission dominating $\tau_2^{(\beta)}$.  This emission can come from the hard jet or off of the emission that dominates $\tau_1^{(\beta)}$.
\item One emission dominating $\tau_1^{(\beta)}$; two separate emissions dominating $\tau_2^{(\beta)}$.
\end{enumerate}
Thus, to compute the LL (or MLL) resummed double differential distribution of $\tau_1^{(\beta)}$ and $\tau_2^{(\beta)}$ requires the consideration of up to three emissions in the jet. Of course, restricting the phase space (i.e.~with a cut on $\tau_1^{(\beta)}$) simplifies the emission structure, but for phenomenology, we would like to compute the cross section for arbitrary values of $\tau_1^{(\beta)}$ and $\tau_2^{(\beta)}$.  

A resummed calculation like the one sketched above would shed significant light on the performance of $N$-subjettiness as a discrimination observable and provide new insight into other powerful observables that could be constructed. A full NLL calculation may require developing new tools for resummation.  For example, because of the numerous phase space constraints in the double differential cross section, the use of soft-collinear effective theory \cite{Bauer:2000ew,Bauer:2000yr,Bauer:2001ct,Bauer:2001yt,Bauer:2002nz} for resummation 
would require identifying modes in regions of the phase space that have not yet been studied in detail 
(see, e.g., SCET$_+$ of \Ref{Bauer:2011uc} as a first step in this direction).  In addition, the ratio observable would receive contributions from many such regions, and some kind of interpolation would be needed to obtain the ratio cross section.  As ratio observables are becoming more widely used in studies of jets at the Large Hadron Collider \cite{Aad:2012meb,ATLAS:2012am,ATLAS:2012dp,ATLAS:2012jla,ATLAS:2012kla,Aad:2013gja}, it is important to further develop the analysis of double differential cross sections and ratio observables.  This will put ratio observables on a firm theoretical footing and provide robust predictions which can be compared to the growing experimental results.

\begin{acknowledgments}
We thank Iain Stewart, Duff Neill, Gavin Salam, Gregory Soyez, and Gregory Korchemsky for helpful discussions.  A.L.\ and J.T.\ are supported by the U.S. Department of Energy (DOE) under cooperative research agreement DE-FG02-05ER-41360.  J.T.\ is supported by the DOE Early Career research program DE-FG02-11ER-41741.
\end{acknowledgments}

\appendix

\section{Matching Fixed-Order to Resummation}
\label{app:match}

In this appendix, we introduce a procedure for matching the fixed-order double differential cross section to the LL resummed double differential cross section.  The method we consider is based on Log-R matching \cite{Catani:1992ua}, and is the natural generalization to double differential distributions of the known results for their single differential counterparts.  The matched double differential cross section also allows for a systematic improvement in the accuracy of the cross section for the ratio $r_{\alpha,\beta}$, by including fixed-order corrections to the resummed cross section to any order in $\alpha_s$.

The Log-R matching procedure requires exponentiating the fixed-order cumulative distribution and eliminating double counting of the logarithms that occur in both the resummation and the fixed-order expression.  To do this, we must first compute the fixed-order cumulative distribution $\Sigma(e_\alpha,e_\beta)$ from \Eq{doubledist} on the physical phase space region of $e_\beta > e_\alpha$ and $e_\alpha^\beta > e_\beta^\alpha$.  We find
\begin{align}\label{cumulant}
\Sigma^{\rm LO}(e_\alpha,e_\beta)&=\int_0^{e_\alpha} de_\alpha' \int_0^{e_\beta} de_\beta \, \frac{d^2\sigma}{de_\alpha \, de_\beta} \nonumber \\
&=  1- \frac{\alpha_s}{\pi}C_F \left\{ \frac{7}{4\beta} + \frac{3}{2}\frac{\log e_\beta}{\beta} +\frac{\log^2 e_\beta}{\beta}  -\frac{2}{\alpha}e_\alpha + \frac{e_\alpha^2}{4\alpha} -\frac{2(\alpha-\beta)}{\alpha \beta} e_\alpha^{-\frac{\beta}{\alpha-\beta}}e_\beta^{\frac{\alpha}{\alpha-\beta}}  \right.  \nonumber\\
&\left.  \qquad  + \ \frac{\alpha-\beta}{4\alpha\beta}e_\alpha^{-\frac{2\beta}{\alpha-\beta}}e_\beta^{\frac{2\alpha}{\alpha-\beta}} +\frac{\log^2\frac{e_\alpha}{e_\beta}}{\alpha-\beta}   \right\}\Theta\left( e_\alpha^\beta - e_\beta^\alpha  \right) \Theta\left( e_\beta - e_\alpha  \right) \ .
\end{align}
Similarly, the resummed cumulative distribution is just the Sudakov factor from \Eq{eq:sudakov}:
\begin{equation}
\Delta(e_\alpha,e_\beta) =e^{- \frac{\alpha_s}{\pi}C_F\left( \frac{1}{\beta}\log^2 e_\beta +\frac{1}{\alpha-\beta}\log^2\frac{e_\alpha}{e_\beta} \right)}\Theta\left( e_\alpha^\beta - e_\beta^\alpha  \right) \Theta\left( e_\beta - e_\alpha  \right) \ .
\end{equation}

Defining $R_1$ via
\begin{equation}
\Sigma^\text{LO}(e_\alpha,e_\beta) = 1-\frac{\alpha_s}{\pi}C_F R_1 \ ,
\end{equation}
the LL resummed matched to ${\cal O}(\alpha_s)$ fixed-order double cumulative distribution is then
\begin{equation}\label{eq:logrmatch}
\Sigma^\text{LL+LO} = e^{-\frac{\alpha_s}{\pi}C_F R_1} \ .
\end{equation}
This cumulative distribution has the correct limits: as $e_\alpha,e_\beta\to 0$, it reduces to the Sudakov factor; as $e_\alpha,e_\beta\to 1$, the fixed-order distribution dominates.  The double differential cross section that follows is just the double derivative of \Eq{eq:logrmatch}.  We find
\begin{align}
\frac{d^2\sigma^\text{LL+LO}}{de_\alpha \, de_\beta} =& \ \frac{\partial}{\partial e_\alpha}\frac{\partial}{\partial e_\beta} e^{-\frac{\alpha_s}{\pi}C_F R_1}  \label{eq:logRresult} \\
=&\ \frac{2\alpha_s}{\pi} \frac{C_F}{\alpha-\beta}\left(\frac{1}{e_\alpha e_\beta} -e_\alpha^{-\frac{\alpha}{\alpha-\beta}}e_\beta^{\frac{\beta}{\alpha-\beta}} + \ \frac{e_\alpha^{-\frac{\alpha+\beta}{\alpha-\beta}}  e_\beta^{\frac{\alpha+\beta}{\alpha-\beta}}}{2}  \right) e^{-\frac{\alpha_s}{\pi}C_FR_1}\nonumber \\
&+ \ 4\frac{\alpha_s^2}{\pi^2}C_F^2\left[
\left(\frac{1}{\alpha }-\frac{ e_{\beta }^{\frac{\alpha
   }{\alpha -\beta }} e_{\alpha }^{-\frac{\alpha }{\alpha
   -\beta }}}{\alpha }+\frac{e_{\beta }^{\frac{2 \alpha
   }{\alpha -\beta }} e_{\alpha }^{-\frac{\alpha +\beta
   }{\alpha -\beta }}}{4 \alpha }-\frac{ \log
   \frac{e_{\alpha }}{e_{\beta }}}{(\alpha -\beta
   ) e_{\alpha }}-\frac{e_{\alpha }}{4 \alpha }\right) \right.\nonumber \\
   &\times
  \left. \left(-\frac{e_{\beta }^{\frac{\alpha +\beta }{\alpha
   -\beta }} e_{\alpha }^{-\frac{2 \beta }{\alpha -\beta }}}{4
   \beta }+\frac{ e_{\beta }^{\frac{\beta }{\alpha -\beta }}
   e_{\alpha }^{-\frac{\beta }{\alpha -\beta }}}{\beta
   }+\frac{ \log \frac{e_{\alpha }}{e_{\beta
   }}}{(\alpha -\beta ) e_{\beta }}-\frac{3}{4 \beta 
   e_{\beta }}-\frac{ \log e_{\beta }}{\beta 
   e_{\beta }}\right)
\right]e^{-\frac{\alpha_s}{\pi}C_FR_1}, \nonumber
\end{align}
which is defined on the physical phase space $e_\beta > e_\alpha$, $e_\alpha^\beta > e_\beta^\alpha$.  The differential cross section for the ratio observable $r_{\alpha,\beta}$ can then be computed by marginalizing according to \Eq{eq:ratio_dist_def}:
\begin{equation}
\frac{d\sigma^{\text{LL+LO}}}{dr} = \int_0^{r^{\frac{\beta}{\alpha-\beta}}} de_\beta \, e_\beta \, \left. \frac{d^2\sigma^\text{LL+LO}}{de_\alpha \, de_\beta}\right|_{e_\alpha = r e_\beta} \ .
\end{equation}
Results from this LL+LO distribution are shown in \Fig{fig:match_plot}.

\section{Modified Leading Logarithmic Resummation}
\label{app:mll}

In this appendix, we compute the MLL Sudakov factor for the double differential cross section of angularities $e_\alpha$ and $e_\beta$.  To do this, we need the expression for the one-loop running coupling and the subleading terms in the splitting function.  The running coupling is
\begin{equation}
\alpha_s(\mu) = \frac{\alpha_s(Q)}{1+\alpha_s(Q)\beta_0 \log\frac{\mu}{Q}} \ ,
\end{equation}
where $\alpha_s(Q)$ is the coupling evaluated at the scale $Q$ and $\beta_0$ is the coefficient of the one-loop $\beta$-function:
\begin{equation}
\beta_0 = \frac{11 C_A - 2n_F}{6\pi} \ .
\end{equation}
For 5 flavors of quarks ($n_F = 5$), $\beta_0 = \frac{23}{6\pi}$.  The quark splitting function is
\begin{equation}\label{eq:splitfull}
P_q(z)=C_F \frac{1+(1-z)^2}{z} \ .
\end{equation}
To MLL accuracy, we can replace the non-singular component of the splitting function by its average on $z\in[0,1]$ which gives
\begin{equation}\label{eq:splitave}
P_q(z)^\text{MLL}=C_F \left( \frac{2}{z} - \frac{3}{2}  \right) \ .
\end{equation}
In the language used in this paper, using the averaged splitting function of \Eq{eq:splitave} gives MLL accuracy, while using the full splitting function from \Eq{eq:splitfull} gives MLL+LO accuracy.

To compute the MLL Sudakov factor, we follow a similar procedure to that used in \Sec{sec:strongorder} for the LL Sudakov.  The value of the exponent in the Sudakov factor is found by integrating over the same region as defined in \Fig{fig:ps_emis}, albeit with non-trivial dependence on $z$ and $\theta$.  To MLL accuracy, this is
\begin{align}
R(e_\alpha,e_\beta) &= \frac{C_F}{\pi}\int_0^1\frac{d\theta}{\theta}\int_0^1 dz \, P_q(z)^\text{MLL} \, \alpha_s(z\theta Q) \left[  \Theta\left( z\theta^\beta - e_\beta  \right) + \Theta\left(e_\beta - z\theta^\beta  \right)\Theta\left( z\theta^\alpha - e_\alpha \right)\right] \nonumber \\
&= \frac{C_F}{\pi}  \left[  \int_{e_\beta^{1/\beta}}^1 \frac{d\theta}{\theta}\int_{e_\beta/\theta^\beta}^1 dz \, \left( \frac{2}{z} -\frac{3}{2} \right) \frac{\alpha_s(Q)}{1+\alpha_s(Q) \beta_0 \log z\theta} 
\right.
\nonumber \\
&\left.
\qquad\qquad +\int_{\left( \frac{e_\alpha}{e_\beta}  \right)^{\frac{1}{\alpha-\beta}}}^1 \frac{d\theta}{\theta}\int_{e_\alpha/\theta^\alpha}^{e_\beta/\theta^\beta} dz \, \left( \frac{2}{z} -\frac{3}{2} \right) \frac{\alpha_s(Q)}{1+\alpha_s(Q) \beta_0 \log z\theta} 
\right] .
\end{align}
Note that we evaluate the coupling at the scale defined by the relative transverse momentum of the emitted gluon, $k_T = z\theta Q$.  Keeping all terms of the form $\alpha_s^n L^m$ where $L$ is the logarithm of $e_\alpha$ or $e_\beta$ with $n\leq m$, we find
\begin{align}\label{eq:radMLL}
R(e_\alpha,e_\beta) &=\frac{C_F}{\pi\beta_0} \left[ 
-\frac{2}{\alpha_s \beta_0}\frac{\beta}{\beta-1}\,
U\left( 1+\alpha_s\beta_0 \frac{\log e_\beta}{\beta}  \right) 
+ \frac{2}{\alpha_s \beta_0}\frac{1}{\alpha-1}\,
U\left( 1+\alpha_s\beta_0\log e_\alpha  \right) 
\right.
\nonumber \\
&\qquad\qquad +  \frac{2}{\alpha_s\beta_0}\frac{\alpha-\beta}{(\alpha-1)(\beta - 1)}\,
U\left( 1+\alpha_s \beta_0 \frac{\log e_\alpha^{1-\beta}e_\beta^{\alpha-1}}{\alpha-\beta}  \right)
\nonumber \\
&\qquad\qquad \left.  + \, \frac{3}{2}\log\left( 1+\alpha_s \beta_0\frac{\log e_\beta}{\beta} \right) \right]\ ,
\end{align}
where  $\alpha_s \equiv \alpha_s(Q)$ and the function $U(x)$ is the logarithm of the inverse of the Lambert $W$-function: $U(x)=x\log x$.  The MLL Sudakov factor for $e_\alpha$ and $e_\beta$ is thus
\begin{equation}\label{eq:sudMLL}
\Delta(e_\alpha,e_\beta)^\text{MLL} = e^{-R(e_\alpha,e_\beta)} \ .
\end{equation}
It is easy to check that the terms at $\alpha_s L^2$ order in \Eq{eq:sudMLL} agree with the LL expression for the Sudakov factor as computed in \Eq{eq:sudakov} in the $\beta_0\to 0$ limit. 

While it might appear that the MLL Sudakov is singular for $\alpha$ or $\beta$ equal to 1, \Eq{eq:radMLL} has a finite limit in each case.  For $\alpha = 1$, \Eq{eq:radMLL} becomes
\begin{align}
R(e_1,e_\beta) &= \frac{C_F}{\pi \beta_0} \left\{
\frac{2}{1-\beta}\log\frac{e_1}{e_\beta}- \frac{2}{\alpha_s \beta_0}\frac{1}{1-\beta}\left(1 + \alpha_s \beta_0 \log e_\beta\right)\log\left( 1+\alpha_s \beta_0 \log e_1  \right)
\right.
\nonumber \\
&\qquad\qquad +\frac{2}{\alpha_s \beta_0}\frac{\beta}{1-\beta}\left( 1+\alpha_s\beta_0 \frac{\log e_\beta}{\beta}  \right)\log\left( 1+\alpha_s\beta_0 \frac{\log e_\beta}{\beta}  \right) 
\nonumber \\
&\qquad\qquad \left.+\, \frac{3}{2} \log\left( 1+\alpha_s\beta_0 \frac{\log e_\beta}{\beta}  \right) \right\} \ .
\end{align}
For $\beta = 1$, \Eq{eq:radMLL} becomes
\begin{align}
R(e_\alpha,e_1)&=\frac{C_F}{\pi \beta_0}\left\{
\frac{2}{\alpha_s \beta_0}\frac{1}{\alpha-1}\left(1 + \alpha_s \beta_0 \log e_\alpha\right)\log\left( 1+\alpha_s \beta_0 \log e_\alpha  \right)
\right.
\nonumber \\
&\qquad\qquad -\frac{2}{\alpha_s \beta_0}\frac{\alpha}{\alpha-1}\left(1+\alpha_s\beta_0 \frac{\log e_\alpha}{\alpha} \right)\log\left( 1+\alpha_s\beta_0 \log e_1   \right)
\nonumber \\
&\qquad\qquad \left.
-\, \frac{2}{\alpha-1}\log e_\alpha + 2\frac{\alpha}{\alpha - 1}\log e_1 + \frac{3}{2}\log\left( 1+\alpha_s \beta_0 \log e_1  \right)\right\} \ .
\end{align}
The double differential cross section of $e_\alpha$ and $e_\beta$ and the cross section for the ratio observable can be found from the Sudakov factor as described in \Sec{sec:LL_double}.

Results for the MLL distribution are shown in \Fig{fig:mll_plot}.  Because MLL includes some subleading terms in the splitting function, this distribution formally includes all logarithmically-enhanced terms of LL+LO in \App{app:match}.  The MLL+LO distributions shown in \Fig{fig:mlllo_plot} uses the full splitting functions, and includes all of the physics of LL+LO.

\section{Power Corrections to the Double Cumulative Distribution}\label{app:powcorr}

In this appendix, we relate the leading power corrections of the double cumulative distribution of angularities to those of a single angularity.  In the OPE region, the power corrections to the double cumulative distribution are
\begin{align}\label{eq:dub_ope}
\Sigma^\text{OPE}(e_\alpha,e_\beta)=& \ \Sigma^\text{pert}(e_\alpha,e_\beta) + c_{1,0}\delta_\text{NP}^\alpha\frac{\partial}{\partial e_\alpha}\Sigma^\text{pert}(e_\alpha,e_\beta)+ c_{0,1}\delta_\text{NP}^\beta\frac{\partial}{\partial e_\beta}\Sigma^\text{pert}(e_\alpha,e_\beta) \nonumber \\
&+{\cal O}\left( \left( \delta_\text{NP}^\alpha \right)^2,\left( \delta_\text{NP}^\beta \right)^2 \right) \ ,
\end{align}
while the power corrections to the single cumulative distribution are
\begin{equation}
\Sigma^\text{OPE}(e_\alpha) = \Sigma^\text{pert}\left( e_\alpha \right) + c_1 \delta_\text{NP}^\alpha \frac{\partial}{\partial e_\alpha} \Sigma^\text{pert}\left( e_\alpha \right) + {\cal O}\left( \left( \delta_\text{NP}^\alpha \right)^2 \right) \ .
\end{equation}
Integrating out the angularity $e_\beta$ in the double cumulative distribution necessarily results in the single cumulative distribution.  To integrate out $e_\beta$, we set $e_\beta=1$:
\begin{equation}
\Sigma(e_\alpha) = \left.\Sigma(e_\alpha,e_\beta)\right|_{e_\beta = 1} \ .
\end{equation}
Applying this to \Eq{eq:dub_ope} and associating terms at $\Lambda/Q$ order, we find
\begin{align}
\Sigma^\text{OPE}(e_\alpha,1)&= \ \Sigma^\text{pert}(e_\alpha,1) + c_{1,0}\delta_\text{NP}^\alpha\frac{\partial}{\partial e_\alpha}\Sigma^\text{pert}(e_\alpha,1)+\left. c_{0,1}\delta_\text{NP}^\beta\frac{\partial}{\partial e_\beta}\Sigma^\text{pert}(e_\alpha,e_\beta)\right|_{e_\beta = 1} \nonumber \\
&=\Sigma^\text{pert}\left( e_\alpha \right) + c_1 \delta_\text{NP}^\alpha \frac{\partial}{\partial e_\alpha} \Sigma^\text{pert}\left( e_\alpha \right) \ .
\end{align}
We assume that the double differential cross section vanishes when either $e_\alpha$ or $e_\beta$ equal 1, which implies that 
\begin{equation}
\left.\frac{\partial}{\partial e_\beta}\Sigma^\text{pert}(e_\alpha,e_\beta)\right|_{e_\beta = 1} = 0 \ .
\end{equation}
Then, to first order in $\delta_\text{NP}^\alpha$, $c_{1,0}=c_1$.  A similar result holds for $e_\beta$.  Correlations between the power corrections of the different angularities first arise at higher orders in powers of $\Lambda/Q$.

\bibliography{RatioDraft}

\end{document}